\documentstyle[epsfig]{elsart}
\input rotate
\newcommand{\re}{\mathop{\mathrm{Re}}}
\newcommand{\im}{\mathop{\mathrm{Im}}}
\begin{document}

\begin{frontmatter}
\title{Study of the reaction $\pi^- p_{\uparrow} \to \pi^- \pi^+ n$ 
on the polarized proton target at 1.78 GeV/c. Experiment and amplitude 
analysis.}

\author{I.G.~Alekseev}, 
\author{P.E.~Budkovsky},
\author{V.P.~Kanavets}, 
\author{L.I.~Koroleva},
\author{I.I.~Levintov}, 
\author{V.I.~Martynov}, 
\author{B.V.~Morozov}, 
\author{V.M.~Nesterov}, 
\author{V.V.~Ryltsov},
\author{D.N.~Svirida}, 
\author{A.D.~Sulimov}, 
\author{V.V.~Zhurkin}
\address{Institute for Theoretical and Experimental Physics, \\ 
         B.~Cheremushkinskaya 25, Moscow, 117259, Russia \\
         Tel: 7(095)123-80-72, Fax: 7(095)127-08-33 \\
         E-mail : kanavets@vitep5.itep.ru}
\author{V.V.~Sumachev}  
\address{Petersburg Nuclear Physics Institute, \\             
         Gatchina, Leningrad district, 188350, Russia}
\begin{abstract}
We present the results of the experimental study of the reaction 
of dipion production by the beam of the negative pions with momentum
1.78~GeV/c on the polarized proton and liquid hydrogen targets. The 
experiment covers the region of dipion masses near the mass of 
$\rho$-meson and small momenta transferred $|t|<0.2$~(GeV/c)$^2$. 
The whole set of spin density matrix elements was reconstructed 
and model-independent and model-dependent amplitude analyses of 
the reaction were performed. The experiment allows to exclude 
the ambiguity of the amplitude analysis existing at high energies. 
The results contain evidences in favour of existing of narrow
$\sigma(750)$ $00^{++}$.
The data also allows to estimate intercept of $a_1$-meson Regge
trajectory. 

The experiment was performed at the ITEP proton synchrotron, Moscow.

PACS number(s) : 13.75.Gx, 13.88.+e.

\end{abstract}

\end{frontmatter}

\section{Introduction}
Exclusive reactions of pion production are an important source of
information on the nature of the strong interaction. They allow
to study the dynamics of the processes as well as the
pion-pion interaction.
Up to now a number of experiments which study production of pions 
by pions were fulfilled on unpolarized proton targets 
\cite{Jacobs,Haber,Wicklund,Leksin,Charlesworth,Reynolds,Bacon} 
in the 	intermediate energy 
region. Their results showed the predominant role of the one pion
exchange (OPE) and noticeable effects of absorption. The whole
picture was satisfactorily described by models of the one pion exchange 
with absorption (OPEA) \cite{Ochs} and Regge-model with moving 
branchings
\cite{Kaydalov}. Nevertheless the absence of information on the spin
dependence of the reaction at the intermediate energies (except the 
preliminary results of this experiment \cite{RhoOld}) did not allow 
to go
further in the understanding of the reaction mechanism, particularly
it was not possible to estimate contribution of the axial-vector 
exchange.

Here we should mention two experiments with transversely polarized 
targets performed at high energies. Authors of the first experiment 
\cite{Becker} used a polarized proton target to study the reaction
\begin{equation}
\label{eq:pip}
\pi^- p \to \pi^+ \pi^- n
\end{equation}
at the incident pion beam momentum 17.2~GeV/c. In the other 
\cite{Lesquen} the reaction
\begin{equation}
\label{eq:pid}
\pi^+ n \to \pi^+ \pi^- p
\end{equation}
was explored at 5.98 and 11.85~GeV/c on a polarized deutron target. In 
both experiments large asymmetries (up to 50\%) in the $\rho$-meson 
production were observed. This was explained as a result of the essential 
contribution of the axial-vector exchange (a$_1$-meson), that caused 
doubts about the results of partial wave analyses, which neglected this 
mechanism. Untrivial results of these experiments put forward the 
question of the energy dependence of the asymmetries. The latter would 
allow to estimate intercept of axial-vector Regge trajectory.

The angular distributions of the products of the reaction (\ref{eq:pip})
at the transversely polarized target allow to fulfill a nearly full
model independent amplitude analysis of the reaction \cite{Lutz}. 
So in the region of the dipion masses below 1~GeV where contribution
of waves with $L \ge 2$ is negligible, one can directly reconstruct 
modules of all the transversity amplitudes and most of the phases.
Besides all, this makes possible to separate the intensities of the 
dipions produced in the $S$- and $P$-wave states without additional 
assumptions and thus study the mass dependence of $S$-wave which is
important for the scalar meson spectroscopy.

Up to now the main source of information about light quark meson
spectroscopy is the partial wave analyses of the pion-pion scattering.
Old analyses were based on the data obtained on unpolarized targets.
In order to study the pion-pion interaction authors of these works
separated the diagram of the  one pion exchange and then made a 
transition from the scattering on the virtual pion to the scattering 
on the real one
\cite{Muhin}. The other method is to perform a partial wave analysis
basing on the amplitudes of the reaction. But the model independent
amplitude reconstruction is impossible without measurements on 
polarized targets. In the absence of such data the assumption 
was used that
in the helicity system one could disregard amplitudes without the 
nucleon helicity 
flip \cite{Williams}. In addition the partial wave analyses
used strong but not well proven assumptions of "spin coherence", 
that is the absence of $a_1$ exchanges, and "phase coherence", 
that is the equality
of phases of $P$-waves with zero and unit helicities. On the other
hand the analysis of the situation using Roy's equations \cite{Roy},
incorporating conditions of unitarity, analyticity and crossing 
symmetry, showed that solutions of partial wave analysis are 
self content and generally correct \cite{Patarakin}, though individual 
analyses had noticeable difference in the $S$-wave phases.

The results of partial wave analyses had an ambiguity in the behavior
of the $\delta^0_0$ phase ($S$-wave with zero isospin) at dipion masses
above 700~MeV, so called "UP-DOWN" ambiguity. The "UP" solution
revealed more quick change of the $\delta^0_0$ phase under the peak of
$\rho$-meson and was considered as resonance opposite to the "DOWN"
solution. The direct measurements of the mass dependence of the
pion-pion interaction intensity
with two $\pi^0$-mesons in the final state did not
lead to a clear conclusion about the existence of the relatively narrow
scalar resonance \cite{Cason,Amsler}. The experiments at high energies
with polarized targets \cite{Becker,Lesquen} showed large polarization
effects that contradicted to the assumptions on which partial wave
analyses were based. The model-independent amplitude analysis of the
data from \cite{Becker,Lesquen} performed by M.~Svec in \cite{Svec}
gave 4-fold ambiguity in the solution, but all the variants corresponded 
to the resonant behaviour of the $S$-wave in the $\rho$-meson region.
The important additional argument in favour of the existence of the
resonance, obtained in this analysis, is the weak mass dependence
of the relative phase between $S$-wave and $P$-wave with zero helicity.
The simultaneous amplitude analysis of the reaction 
(\ref{eq:pip}) together
with the reaction $\pi^+ p \to \pi^+ \pi^- \Delta^{++}$ \cite{Donohue}
gave two branches of the solution.
One corresponded to a relatively narrow resonance ($M=750$~MeV and
$\Gamma=200$~MeV) and the other --- to a wide one ($M=600$~MeV and
$\Gamma=450$~MeV). The recent partial wave analysis of the data 
\cite{Becker},
which took into account the axial-vector exchange, gave also two branches 
of the solution for the phase $\delta^0_0$ \cite{Kaminski}. So the 
current amount 
of the experimental data over peripheral meson production is not 
sufficient to solve the "UP-DOWN" ambiguity, which preserves the open
problem in the meson spectroscopy. Nevertheless this data points out to
the existence of the scalar-isoscalar resonance in the mass region below
1~GeV (wide sigma-meson $f_0(400-1200)$ and/or narrow $f_0(750)$).
The history of the search of the scalar-isoscalar meson is long and
contradictory. The $\sigma$-meson was for the first time proposed
in the work \cite{Gell}. Up to 1974 this resonance was present in
the PDG tables as $\sigma$, $\epsilon$ or $\delta_{0+}$. In 1996
it again appeared in PDG tables \cite{PDG} as $f_0(400-1200)$.

Thus the modern state of the problem of light scalar-isoscalar meson
requires additional experimental information and to provide such
information was one of the goals of
this work. We present here a new experimental data on the reaction
(\ref{eq:pip}) on the polarized proton and liquid hydrogen targets
in the region of small momentum transfered at 
$p_{\mathrm{beam}}=1.78$~GeV/c.
The results are compared with the ones at high energies 
($p_{\mathrm{beam}}=17.2$~GeV/c \cite{Becker,Svec}). The results are
analyzed with the help of the Regge phenomenology, which gives the
natural connection between the high and intermediate energy regions.

\section{Basic formalism}
Here we only want to review the basic formalism which could
be found elsewhere in more details 
\cite{Svec,Byckling,Jackson,Kimel,Lutz}.
At fixed beam momentum the reaction (\ref{eq:pip}) is described by 5 
kinematic
variables. We use two energy variables (the squared momentum transferred 
$t$\footnote{or $t'= t - t_{\mathrm{min}}$ instead of $t$}
and the dipion invariant mass $M_{\pi\pi}$) and three angular variables.
One angle $\psi$ describes the reaction plane and is defined
as the angle between the normal to the reaction plane and the
target polarization $P_t$. The other two angles describe the dipion 
decay in its rest frame and are defined as angles $\theta$ 
(Gotfried-Jackson angle)
and $\phi$ (Treiman-Yang angle) of the negative pion in the
Jackson coordinate system (helicity system of the t-channel,
axis $z$ along the beam and $y$ perpendicular to the reaction plane).

The dynamics of the reaction (\ref{eq:pip}) could be described by the 
set of the helicity amplitudes 
$\langle j,m,\chi |T| \lambda \rangle (s,t,M_{\pi\pi})$, where
$j$ and $m$ are the spin and the helicity of the dimeson, while $\chi$
and $\lambda$ are the neutron and proton helicities correspondingly.
At $M_{\pi\pi}<1$~GeV the dipion production with spins $j=0$ 
($S$-wave) and $j=1$ ($P$-wave) is dominant. Thus the reaction 
(\ref{eq:pip}) in the
energy region under consideration is described by 8 complex amplitudes:
two for dipion production in $S$-state (with and without neutron 
helicity flip) and 6 for dipion production in $P$-state (with dipion 
helicities +1,0,-1 with or without neutron helicity flip). An experiment
with a transversely
polarized target yield 15 elements of spin density matrix (SDME)
$\rho_{\alpha\beta}(s,t,M_{\pi\pi})$, defining the angular distributions
of the reaction products. Nine of them reflect the interference of the
helicity flip and non-flip amplitudes and can be measured only in 
polarized target experiment.

The normalized process intensity as function of the SDME and the angular 
variables could be written as follows:
\begin{equation}
\label{eq:i}
I(\theta,\phi,\psi) = I_0(\theta,\phi)+P_t \cdot \cos{\psi} \cdot
  I_Y(\theta,\phi)+ P_t \cdot \sin{\psi} \cdot I_X(\theta,\phi) , 
\label{eq:i0}
\end{equation}
where 
\begin{eqnarray}
I_0(\theta,\phi) &=& 
  1 + (\rho_{00}-\rho_{11})\cdot(3\cos^2\theta-1) - {} \nonumber \\
&&\rho_{1-1} \cdot 3\sin^2\theta\cos 2\phi - 
  \re\rho_{10} \cdot 3\sqrt{2}\sin 2\theta\cos\phi - {} \nonumber \\
&&\re\rho_{1S} \cdot 2\sqrt{6}\sin\theta\cos\phi +
  \re\rho_{0S} \cdot 2\sqrt{3}\cos\theta , \\
\label{eq:iY}
I_Y(\theta,\phi) &=& 
  A + (\rho^Y_{00}-\rho^Y_{11}) \cdot (3\cos^2\theta-1) - {} \nonumber \\
&&\rho^Y_{1-1} \cdot 3\sin^2\theta\cos 2\phi - 
  \re\rho^Y_{10} \cdot 3\sqrt{2}\sin 2\theta\cos\phi - {} \nonumber \\
&&\re\rho^Y_{1S} \cdot 2\sqrt{6}\sin\theta\cos\phi +
  \re\rho^Y_{0S} \cdot 2\sqrt{3}\cos\theta , \\
\label{eq:iX}
I_X(\theta,\phi) &=& 
  \im\rho^X_{1-1} \cdot 3\sin^2\theta\sin 2\phi +  
  \im\rho^X_{10}  \cdot 3\sqrt{2}\sin 2\theta \sin\phi + {} \nonumber \\
&&\im\rho^X_{1S}  \cdot 2\sqrt{6}\sin\theta\sin\phi \, . 
\end{eqnarray}
With two additional relations:
\begin{eqnarray}
\label{eq:norm1}
&&\rho_{SS}+\rho_{00}+2\rho_{11}=1 , \\
\label{eq:normA}
&&\rho^Y_{SS}+\rho^Y_{00}+2\rho^Y_{11}=A \, .
\end{eqnarray}
The first of them expresses the normalization and the second defines
$A$ --- conventional polarized target asymmetry.

The data on transversely polarized target is best analyzed in 
terms of the nucleon transversity amplitudes with definite t-channel 
exchange naturality :
$S$, $L$, $U$, $N$ and $\bar{S}$, $\bar{L}$, $\bar{U}$, $\bar{N}$. 
These amplitudes are linear combinations of helicity amplitudes and 
correspond to definite recoil nucleon transversity ("down" and "up" 
respectively). The amplitudes $S$, $\bar{S}$ and $L$, 
$\bar{L}$ describe production of the $S$-wave 
and $P$-wave dipions with zero helicity, respectively. Amplitudes 
$S$, $\bar{S}$, $L$, $\bar{L}$, $U$, $\bar{U}$
are dominated by unnatural exchange, while amplitudes 
$N$, $\bar{N}$ - by natural one.

\begin{eqnarray}
\label{eq:SLUN}
S &=& \frac{1}{\sqrt{2}}(\langle 0,0,+|T|+ \rangle + 
                 i \cdot \langle 0,0,+|T|- \rangle) , \\
\bar{S} &=& \frac{1}{\sqrt{2}}(\langle 0,0,+|T|+ \rangle -
                       i \cdot \langle 0,0,+|T|- \rangle) , \\
L &=& \frac{1}{\sqrt{2}}(\langle 1,0,+|T|+ \rangle + 
                 i \cdot \langle 1,0,+|T|- \rangle) , \\
\bar{L} &=& \frac{1}{\sqrt{2}}(\langle 1,0,+|T|+ \rangle - 
                       i \cdot \langle 1,0,+|T|- \rangle) , \\
U &=& \frac{1}{2}(\langle 1,+1,+|T|+ \rangle - 
                  \langle 1,-1,+|T|+ \rangle +  {} \nonumber \\
    &&    i \cdot \langle 1,+1,+|T|- \rangle - 
          i \cdot \langle 1,-1,+|T|- \rangle) , \\
\bar{U} &=& \frac{1}{2}(\langle 1,+1,+|T|+ \rangle - 
                        \langle 1,-1,+|T|+ \rangle - {} \nonumber \\
         &&    i \cdot \langle 1,+1,+|T|- \rangle + 
                i \cdot \langle 1,-1,+|T|- \rangle) , \\
N &=& \frac{1}{2}(\langle 1,+1,+|T|+ \rangle + 
                  \langle 1,-1,+|T|+ \rangle + {} \nonumber \\
    &&    i \cdot \langle 1,+1,+|T|- \rangle + 
          i \cdot \langle 1,-1,+|T|- \rangle) , \\
\bar{N} &=& \frac{1}{2}(\langle 1,+1,+|T|+ \rangle + 
                        \langle 1,-1,+|T|+ \rangle - {} \nonumber \\
          &&    i \cdot \langle 1,+1,+|T|- \rangle - 
                i \cdot \langle 1,-1,+|T|- \rangle \, .
\end{eqnarray}

These amplitudes are connected to SDME by the following equations:
\begin{eqnarray}
&\rho_{SS}+\rho_{00}+2\rho_{11} = \nonumber \\
&|S|^2+|\bar{S}|^2+|L|^2+|\bar{L}|^2+|U|^2+|\bar{U}|^2+|N|^2+|\bar{N}|^2 
= 1 , \label{eq:norm2} \\
&\rho_{00}-\rho_{11}=|L|^2 + |\bar{L}|^2 - \half (|N|^2 + |\bar{N}|^2 +
|U|^2 + |\bar{U}|^2) , \\
&\rho_{1-1} = \frac{1}{2} (|N|^2 + |\bar{N}|^2 - |U|^2 - |\bar{U}|^2),\\
&\re \rho_{10} = \frac{1}{\sqrt{2}}\re (UL^* + \bar{U}\bar{L}^*) , \\
&\re \rho_{0S}=\re(LS^* + \bar{L}\bar{S}^*), \\
&\re \rho_{1S}=\frac{1}{\sqrt{2}}\re(US^* + \bar{U}\bar{S}^*), 
\label{eq:rho1s} \\
&\rho^Y_{SS}+\rho^Y_{00}+2\rho^Y_{11} = \nonumber \\
&|S|^2-|\bar{S}|^2+|L|^2-|\bar{L}|^2+|U|^2-|\bar{U}|^2-|N|^2+|\bar{N}|^2 
= A , \\
&\rho^Y_{00}-\rho^Y_{11}=|L|^2 - |\bar{L}|^2 - \half (-|N|^2 + 
|\bar{N}|^2 + |U|^2 - |\bar{U}|^2) , \\
&\rho^Y_{1-1}=-\frac{1}{2} (|N|^2 - |\bar{N}|^2 + |U|^2 - |\bar{U}|^2),\\
&\re \rho^Y_{10} = \frac{1}{\sqrt{2}}\re (UL^* - \bar{U}\bar{L}^*) , \\
&\re \rho^Y_{0S}=\re(LS^* - \bar{L}\bar{S}^*), \\
&\re \rho^Y_{1S}= \frac{1}{\sqrt{2}}\re(US^* - \bar{U}\bar{S}^*), \\
&\im \rho^X_{1-1} = - \re (NU^* - \bar{N}\bar{U}^*), \\
\label{eq:rhox10}
&\im \rho^X_{10} = \frac{1}{\sqrt{2}}\re (NL^* - \bar{N}\bar{L}^*), \\
&\im \rho^X_{1S} = \frac{1}{\sqrt{2}}\re (NS^* - \bar{N}\bar{S}^*) \, .
\end{eqnarray}

It is also useful to introduce partial-wave intensities $I_A$
and partial-wave polarizations $P_A$:
\begin{equation}
I_A=|A|^2+|\bar{A}|^2 , \qquad 
P_A=|A|^2-|\bar{A}|^2 ,
\end{equation}
where $A=S,L,U,N$.

The absolute values of the transversity amplitudes could be 
reconstructed from the experiment on a polarized target. This is 
not so for the helicity amplitudes. 
The matrix elements measured allow to reconstruct the absolute values 
of all the amplitudes involved and all relative phases between them, 
except the relative phase between the two groups of amplitudes with 
different recoil nucleon transversity. 
The relations between the SDME and the amplitudes
could be put in the form of two independent similar
systems of equations \cite{Lutz,Svec},
one for amplitudes $S$, $L$, $U$ and $\bar{N}$ and 
the other for amplitudes with the opposite recoiled nucleon 
transversity.
Each of the systems could be reduced to a cubic equation in respect 
to $|L|^2$ or $|\bar{L}|^2$, correspondingly,
which has two positive solutions. This provides two-fold ambiguity
in the resulting amplitudes and four-fold ambiguity in the partial
wave intensities.

\section{Experimental layout}
The apparatus used in this work is the further development of the
experimental setup {\bf SPIN} \cite{Spin,RhoOld} designed for the
investigation of polarization effects in two- and three-particle
reactions with two charged particles in the final state. It is a
two-arm magnetic spectrometer with wire chambers and a transversely
polarized proton target. Spectrometric capabilities of the setup
are provided by the magnet of the polarized target equipped
with spark chambers placed in its field. The apparatus is located
at the ITEP accelerator in the beam of particles with maximum
momentum 2.1~GeV/c. The typical beam intensity is $5 \cdot 10^5$
pions per spill, the spill duration is about 1~s and the frequency
is 20--25 spills per minute. The beam angular divergences are
$\pm 6.5$ and $\pm 3.5$ mrad in the horizontal and vertical planes
respectively. Its dimensions on the target are 8--10~mm in both
projections. The momentum spread of the beam is $\pm 2$\%. The
central beam momentum was determined with the precision
better than 0.5\% by bending of the incident particles in the well
known magnetic field of the polarizing magnet and by the time-of-flight
difference between negative pions and anti-protons of the beam 
\cite{TOF}.

\begin{figure}
\epsfig{figure=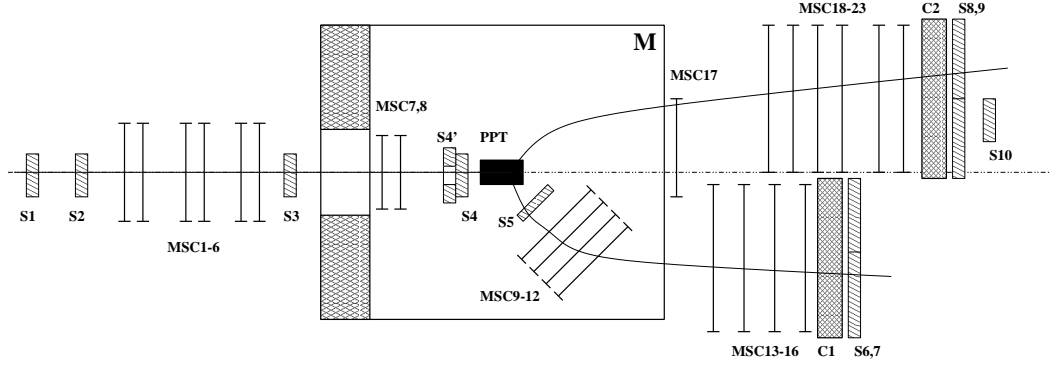,width=\hsize}
\caption{Experimental layout.}
\label{fig:SPIN}
\end{figure}

The setup {\bf SPIN} is shown in fig.~\ref{fig:SPIN}. The basic elements
are:
\begin{itemize}
\item Polarized proton target ({\bf PPT}), placed in the center of a
wide aperture magnet ({\bf M}). The magnet ({\bf M}) is C-shaped iron
magnet with approximately axially symmetrical horizontal field. The 
field
in the center is 2.55~T with non-uniformity in the region of the target
less than $2 \cdot 10^{-4}$. The bending power from the center is
0.45~T$\cdot$m. The protons of the target are polarized by the dynamic 
nuclear orientation method. The target material is propanediol 
C$_3$H$_8$O$_2$
doped by HBMA-Cr$^V$ complexes. Target dimensions are 
$21 \cdot 28 \cdot 60$~mm$^3$ (width$\cdot$height$\cdot$length). 
The working temperature of the target 0.5~K is achieved with a 
\nuc{3}{He}-evaporation type cryostat. The polarization is measured by 
the NMR method with the precision of 5\%. The average polarization 
during the data taking was $70 \pm 5$\%.
\item Liquid hydrogen target, which can be placed in the same cryostat
as the polarized target. This allows to make measurements on both 
types of targets
without additional readjustments of the setup --- just 
by replacing the
target. The liquid hydrogen is produced by the
cooling of the gaseous hydrogen
by the liquid helium at the atmospheric pressure. The helium stream in
the heat exchanger was automatically adjusted so as to keep the hydrogen
pressure
in the closed volume of the target constant. This allows to maintain
the volume of the liquid hydrogen with the precision better than 2\%.
\item Wire spark chambers with magnetostrictive readout 
({\bf MWSC1--23}),
which allow to measure trajectories of the incoming pion 
({\bf MWSC1--8})
and two outgoing charged particles: ({\bf MWSC9--16}) in the lower arm 
and
({\bf MWSC17--23}) in the upper arm. 
The chambers
({\bf MWSC9--12}) of the lower arm 
placed into the magnetic field 
provide the spectrometric capabilities 
of the apparatus.
\item System of scintillation and Cherenkov counters ({\bf S1--10} and 
{\bf \^C1,2}) for triggering. The threshold Cherenkov counters provide 
suppression
of protons and are used in the trigger, which was formed according 
to the equation:
\begin{eqnarray}
{\mathrm{Trig.}} = 
S1 \cdot S2 \cdot S3 \cdot S4 \cdot \overline{S4'} \cdot
\check C1 \cdot \check C2 \cdot \overline{S10} \cdot 
(S6+S7) \cdot (S8+S9)&&, \nonumber \\
\end{eqnarray}
where veto counter {\bf S10} is placed on the continuation of the 
beam line after the magnet.
\end{itemize}
The adjustment of the apparatus was performed in two subsidiary runs. In
one of them the special copper wire target was used to produce events
with known vertex coordinates and in the other the beam which momentum 
was controlled with
time-of-flight technique was bended by the magnet into different arms of
the setup. 

\begin{figure*}
\epsfig{figure=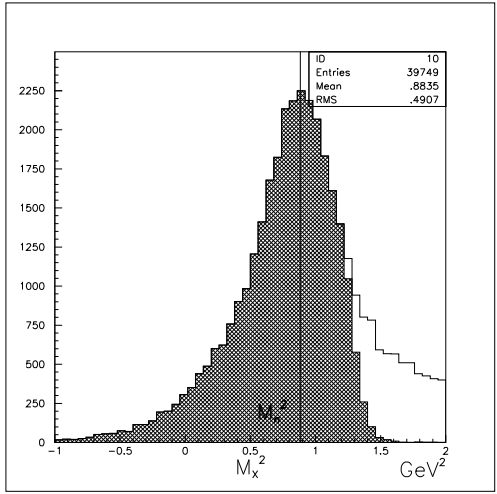,width=6.5cm}
\caption{Events distribution over missing mass squared.}
\label{fig:MMass}
\end{figure*}

\section{Data processing}
\begin{figure}
\begin{tabular}{ll}
\epsfig{figure=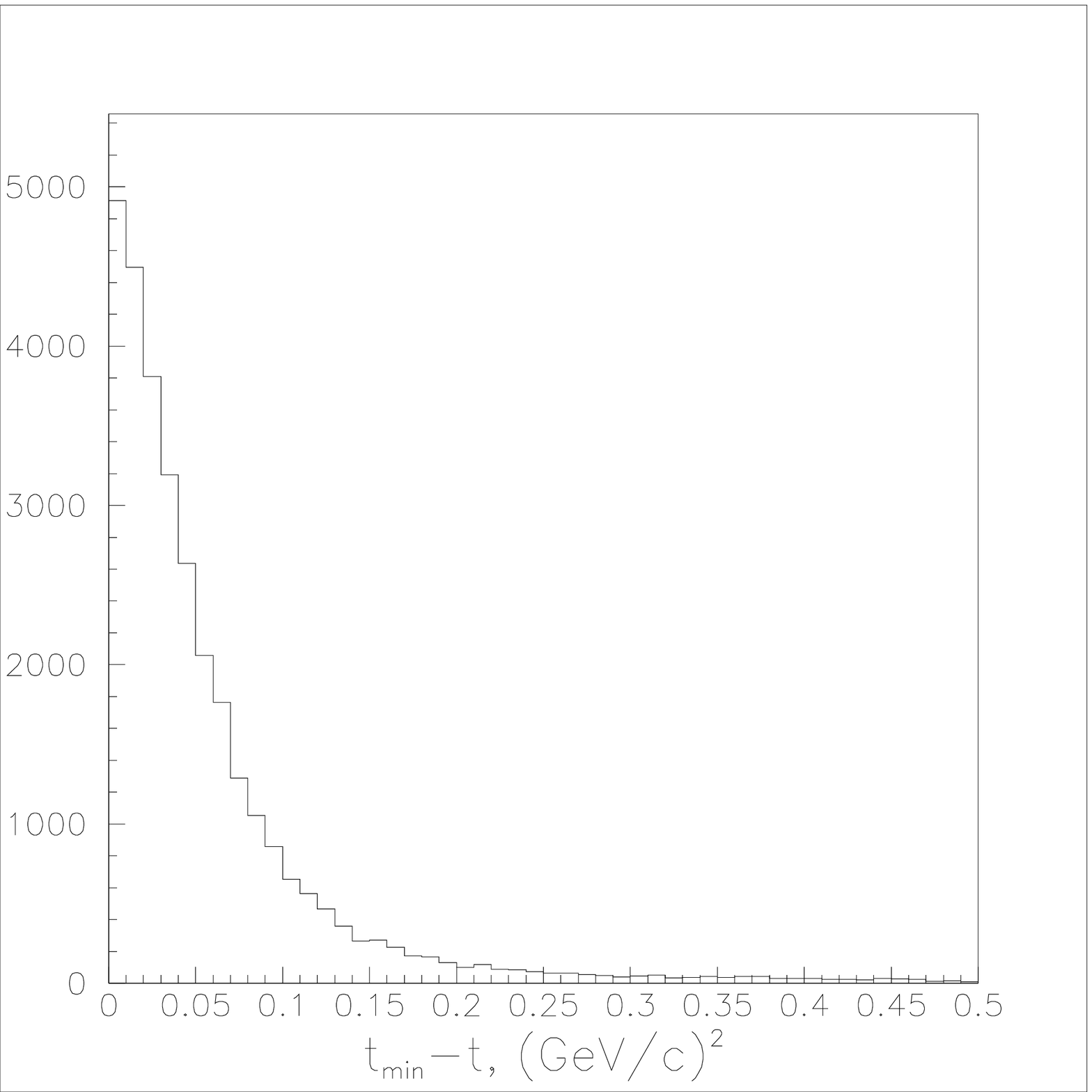,width=6.5cm}&
\epsfig{figure=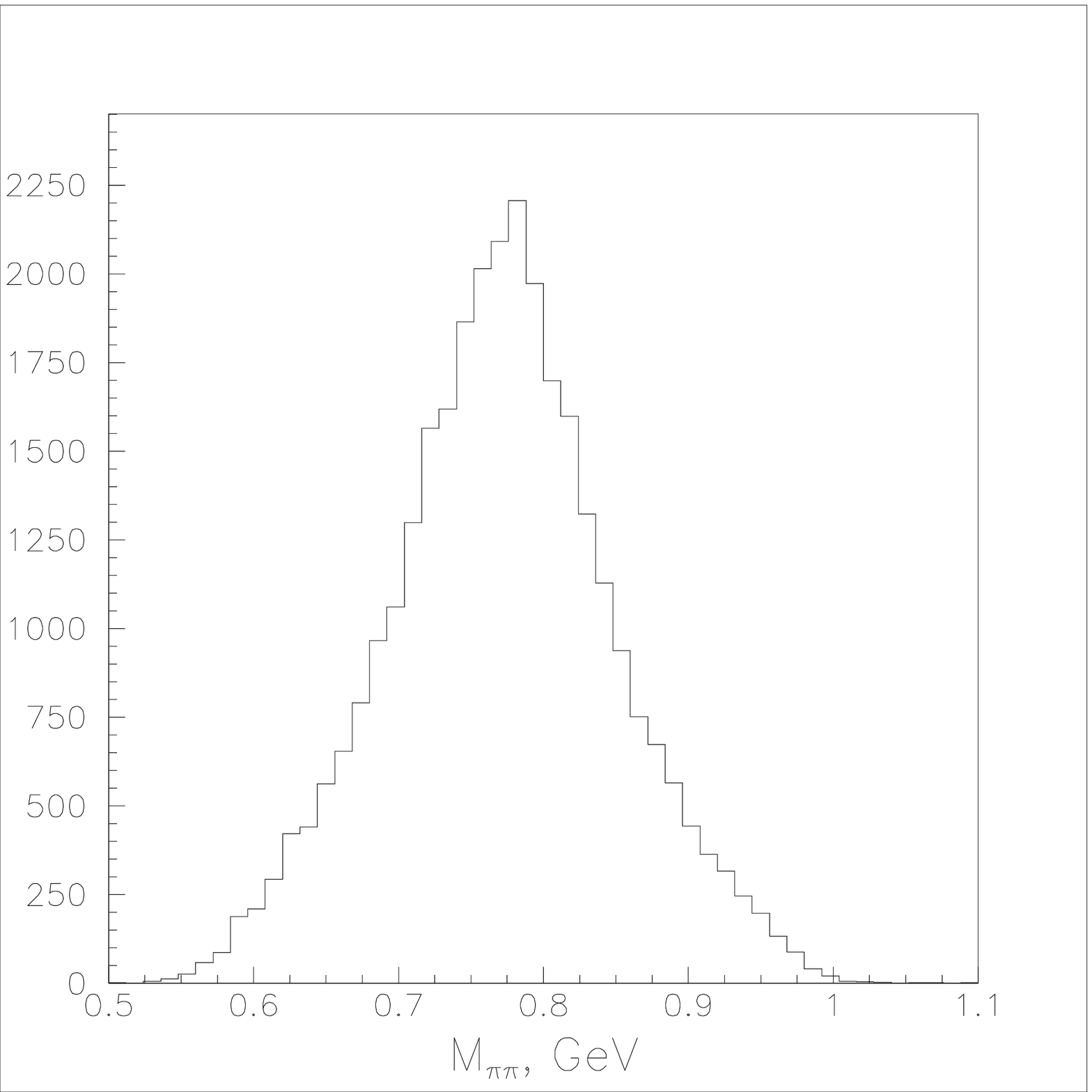,width=6.5cm}\\
a) & b)\\
\epsfig{figure=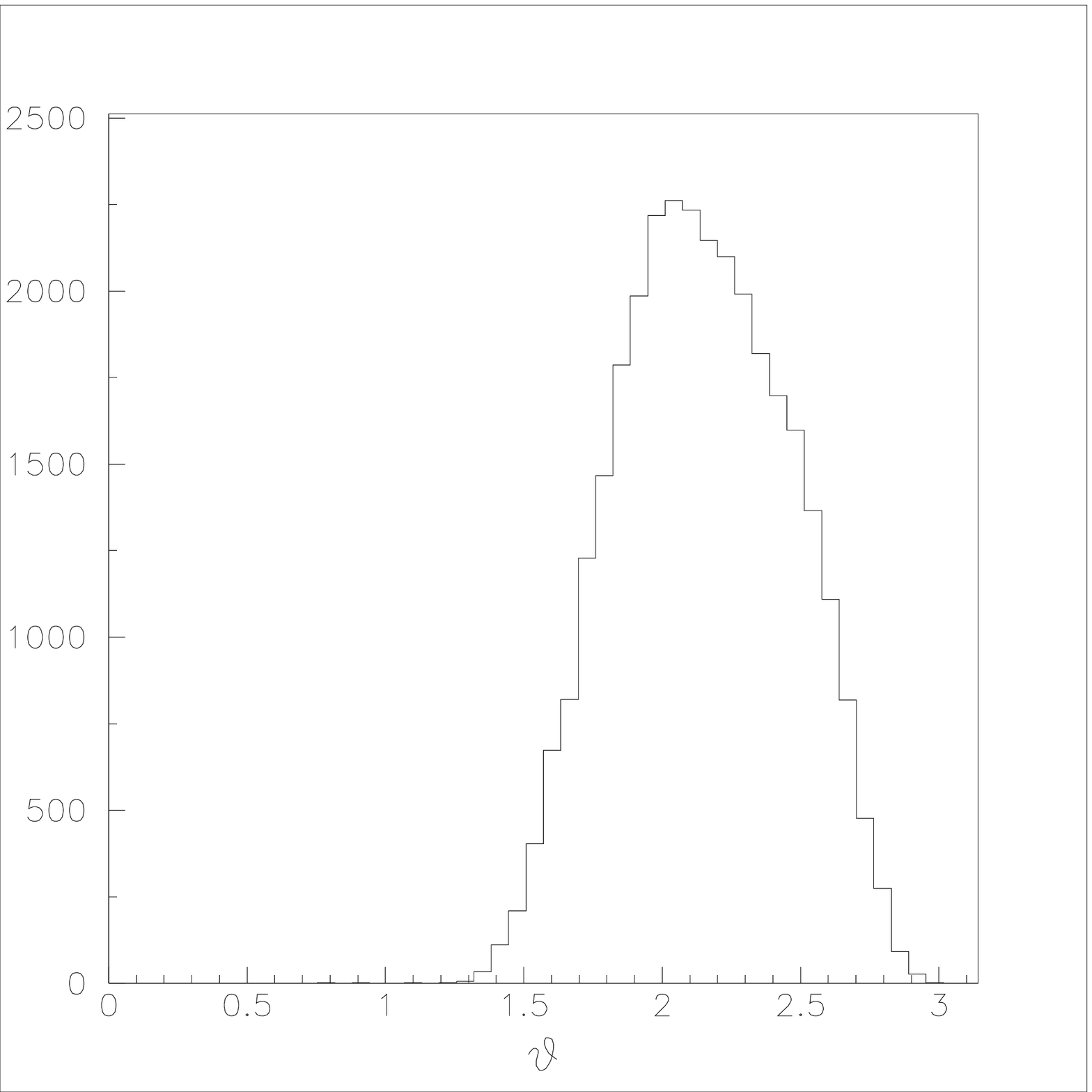,width=6.5cm}&
\epsfig{figure=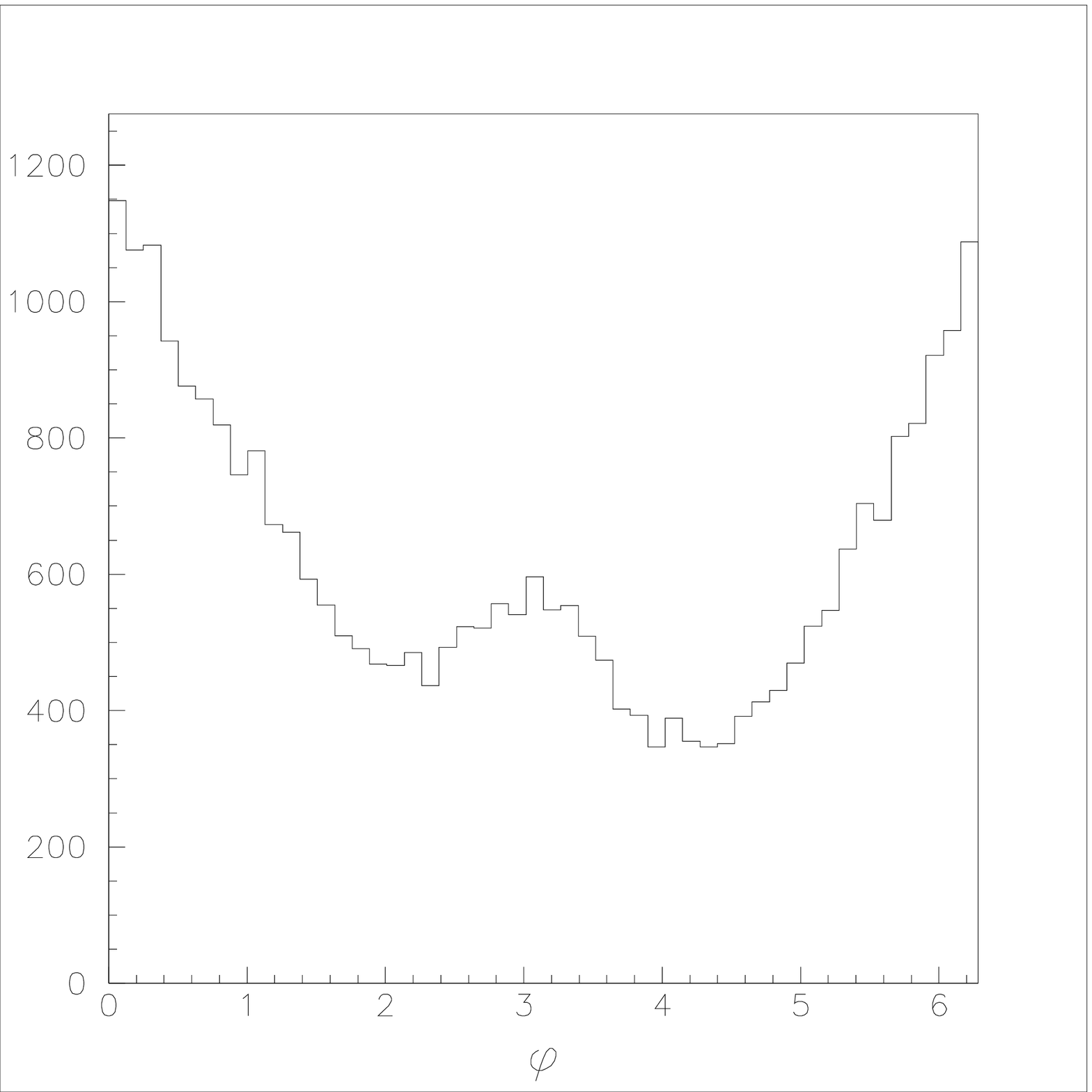,width=6.5cm}\\
c) & d)
\end{tabular}
\caption{Events distributions over momentum transferred 
$t_{\mathrm{min}} - t$ (a), dipion invariant mass $M_{\pi\pi}$ (b), 
Gotfried-Jackson angle $\theta$ (c) and Treiman-Yang angle $\phi$ (d).}
\label{fig:Events}
\end{figure}

Data processing was performed in two stages. At the first stage the
kinematic parameters
of the individual events were reconstructed. For every event 
the interaction point and particles momenta as well as their error
matrixes were reconstructed taking into account multiple scattering 
in the materials of the setup \cite{Eichinger}. The final kinematic
parameters were found using the hypothesis that the missing particle 
is a neutron.
The distribution of the events over missing mass squared is shown in
fig.~\ref{fig:MMass} (solid line). 
Further event selection was based on the $\chi^2$ of the 
hypothesis that the missing particle is a neutron. We assumed 
that in the background events one or more missed pions are 
produced and then the missing mass in such events should be greater 
the mass of a neutron. We made processing with different rejections 
over $\chi^2$ and missing mass and found no significant difference.
Finally we rejected events whose missing 
mass was above the mass of nucleon and $\chi^2 > 5$. 
In fig.~\ref{fig:MMass} ``Good'' events are shown by hatched 
area. 
The event distributions 
over momentum transferred $t_{\mathrm{min}} - t$,
dipion invariant mass $M_{\pi\pi}$, Gotfried-Jackson angle $\theta$ and
Treiman-Yang angle $\phi$ on the liquid hydrogen target
are shown in fig.~\ref{fig:Events}. These figures
describe the kinematic region covered by the setup.

At the second stage the SDME of the reaction were reconstructed using
the method of maximum likelihood \cite{Likelyhood} in bins over
dipion mass and momentum transferred. For a given angular distribution
of the events $I(\theta, \phi, \psi)$ the likelihood function could be 
expressed as a product over all the events observed :
\begin{equation}
L=\prod_i^N \lbrace I(\theta_i, \phi_i, \psi_i) \cdot 
  \eta(\theta_i, \phi_i, \psi_i) / C \rbrace ,
\end{equation}
where $N$ is the total number of events, 
$\eta(\theta_i, \phi_i, \psi_i)$
--- the probability of the apparatus to count the event with given 
kinematic parameters
and $C$ is normalizing constant which is defined as the 
integral:
\begin{equation}
C=\int I(\theta_i, \phi_i, \psi_i) \cdot \eta(\theta_i, \phi_i, \psi_i)
  \cdot \d \Omega ,
\end{equation}
which is taken over the whole phase space and can be found using
Monte-Carlo technique as a sum over all events hitting the modelled
setup:
\begin{equation}
C = \frac{1}{K} \sum_i^K I(\theta_i, \phi_i, \psi_i),
\label{eq:C}
\end{equation}
here $K$ is the number of Monte-Carlo events. We searched for a 
maximum of the function 
\begin{equation}
\ln{L}=\sum_i^N \ln{I(\theta_i, \phi_i, \psi_i)} +
  \sum_i^N \ln{\eta(\theta_i, \phi_i, \psi_i)} - N \cdot \ln{C} \, .
\end{equation}
The second term does not depend upon the SDME and can be omitted.
The Monte-Carlo simulation was performed with the aid of the
GEANT code \cite{GEANT}. The method of effective sample \cite{Sobol}
was used to reduce the computer time required. 
According to this method
the events were simulated with different probabilities over 
$M_{\pi\pi}$ and $t'$, so that distributions of simulated events
reminded the experimental distributions. 
Then the sum in (\ref{eq:C}) was calculated with necessary 
correction. The results of Monte-Carlo simulations were also used 
to ensure our understanding of apparatus acceptance and errors 
and to test program of event reconstruction: (i)  
we compared simulated distributions with real ones using 
reconstructed SDME; (ii) the Monte-Carlo generated events 
were processed by the program of event reconstruction and 
reconstructed parameters were compared to generated ones; (iii) 
missing mass squared distribution of Monte-Carlo events after 
reconstruction was built.

The uncertainty in the missing mass does not
allow to separate the reaction on the free protons from the 
background
reactions on the protons bound in the nuclei of the target material.
So we determined the spin-independent SDME in the experiment on liquid
hydrogen target only. The spin-dependent SDME were determined from the
data, obtained in the experiment on the polarized proton target,
in two-step processing. On the first step the spin-independent SDME 
on the target material mixture of nucleus were found. Then the
spin-dependent SDME on the polarized hydrogen of the target were
determined. To equalize the data samples with different target
polarization the weight function
\begin{equation}
\frac{w_-}{w_+}=\sqrt{\frac{N^+_\uparrow \cdot N^+_\downarrow}%
{N^-_\uparrow \cdot N^-_\downarrow}}
\end{equation}
was used \cite{Babou}. Here $w^\pm$ are the weights 
attributed to the events with corresponding
target polarization and $N^\pm_{\uparrow\downarrow}$ --- numbers of
the events with positive or negative target polarization 
and in which dipion goes up or down correspondingly.

The spin-dependent data obtained on polarized target is averaged by
the whole nuclei mixture of the target material. The correction 
for this effect was made by polarization dissolution factor,
given in the table below. The factor was calculated from
beam monitor counts, chamber efficiencies and spin-independent
SDME.

\begin{table*}
\label{tab:discoef}
\caption{Polarization dissolution factor}
\begin{tabular}{cc}
\hline
$M_{\pi\pi}$, GeV & $k$         \\
\hline
0.65--0.75    & $3.6\pm0.2$ \\
0.75--0.80    & $2.8\pm0.2$ \\
0.80--0.90    & $4.3\pm0.4$ \\
\hline
\end{tabular}
\end{table*}

\section{Results}
The statistics (after event reconstruction)
is $40 \cdot 10^3$ events on $LH_2$ target and 
about $320\cdot10^3$ events on the polarized target
in the kinematic range 
$0.005<t'<0.2$~(GeV/c)$^2$ and $0.6<M_{\pi\pi}<1.0$~GeV.

\subsection{Spin-independent SDME}
\begin{figure}
\begin{tabular}{ll}
\epsfig{figure=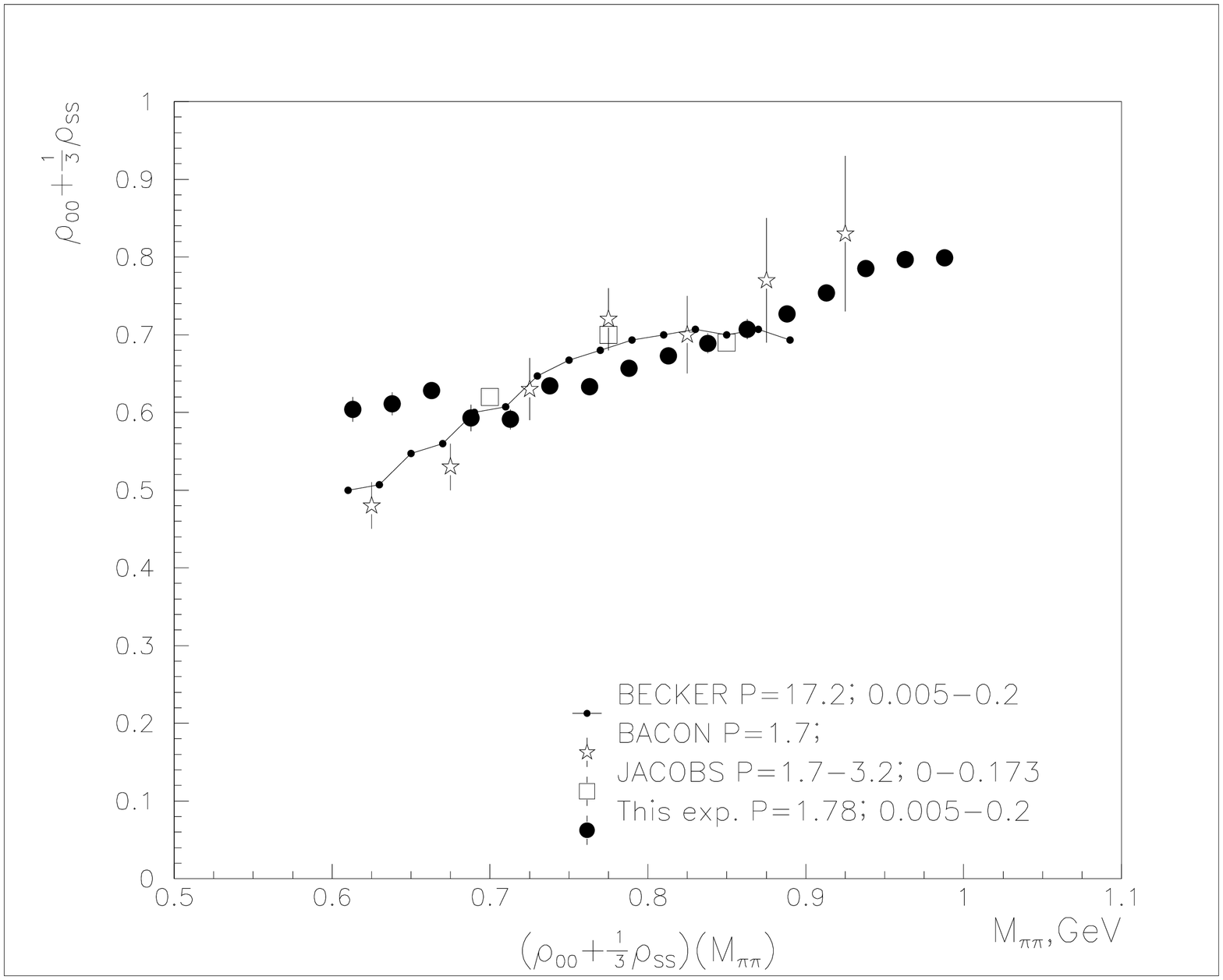,height=4cm,width=6.5cm}&
\epsfig{figure=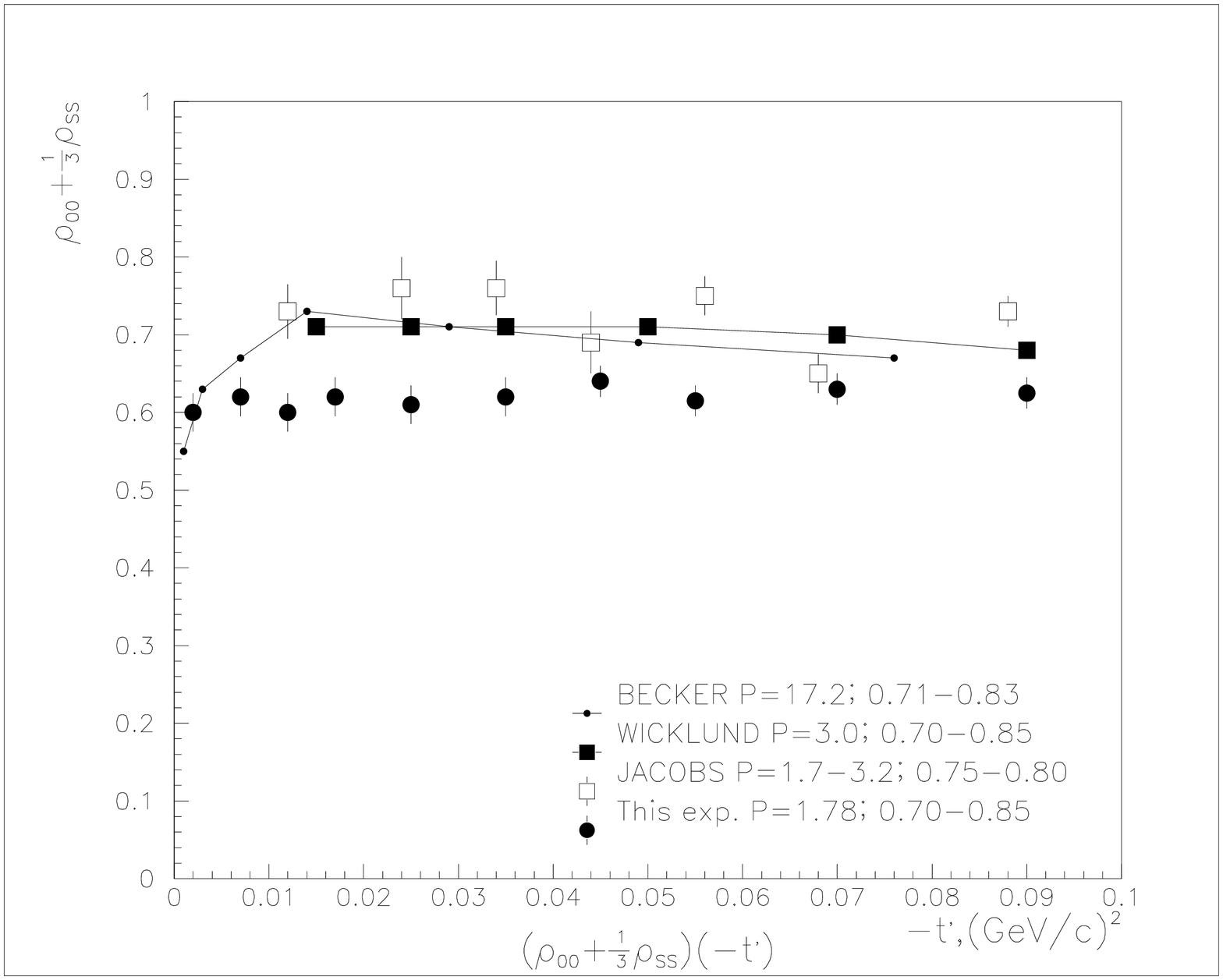,height=4cm,width=6.5cm}\\
\epsfig{figure=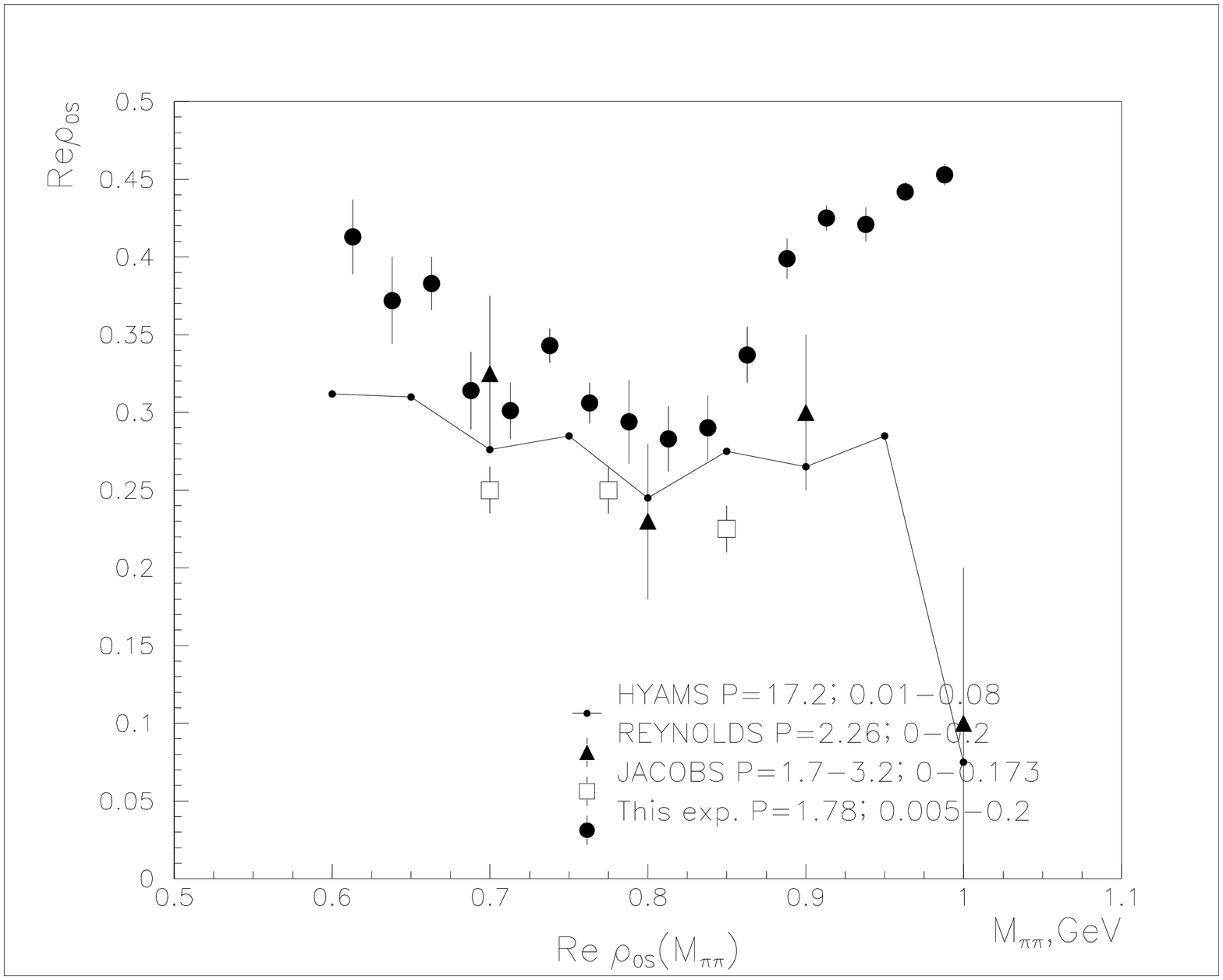,height=4cm,width=6.5cm}&
\epsfig{figure=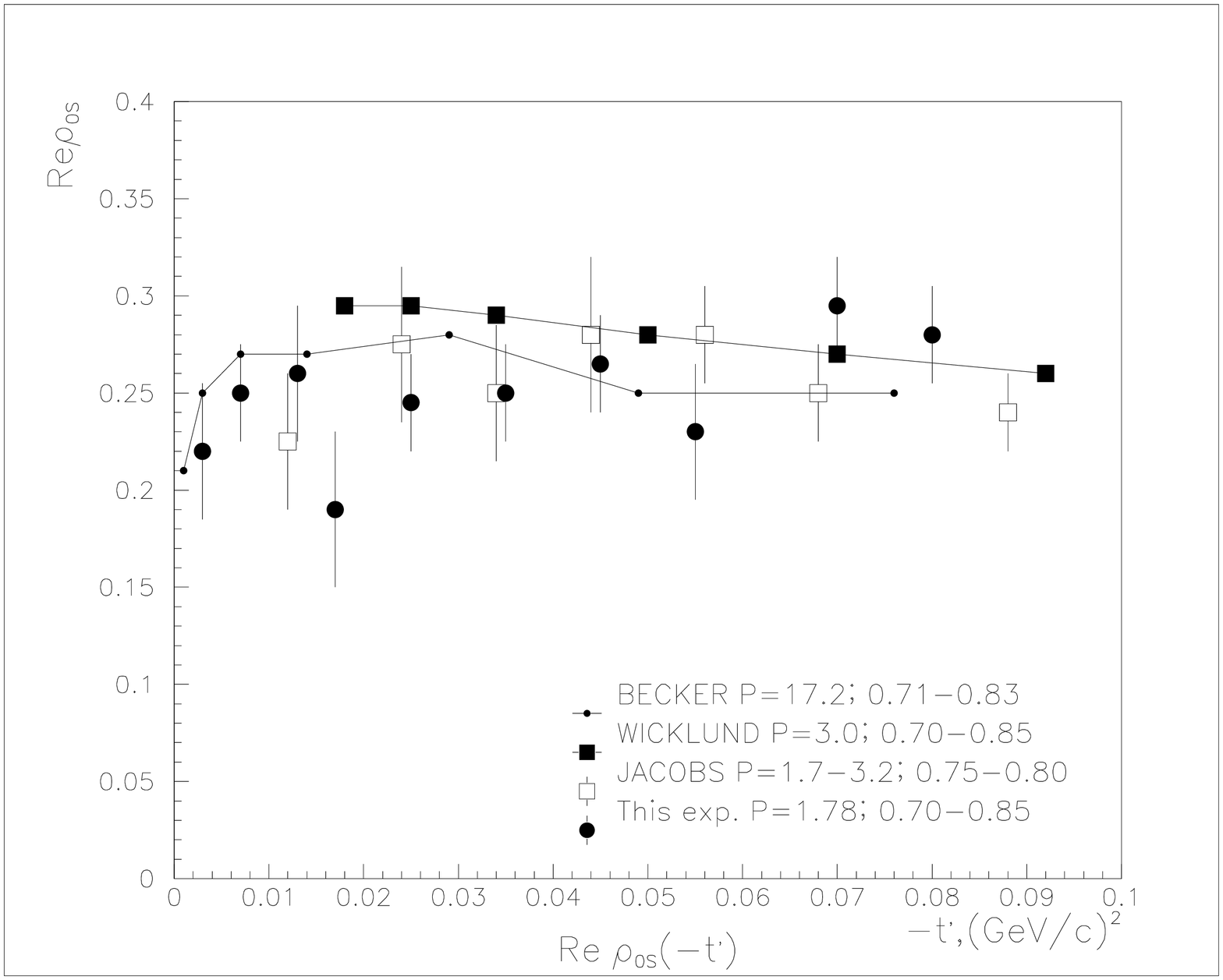,height=4cm,width=6.5cm}\\
\epsfig{figure=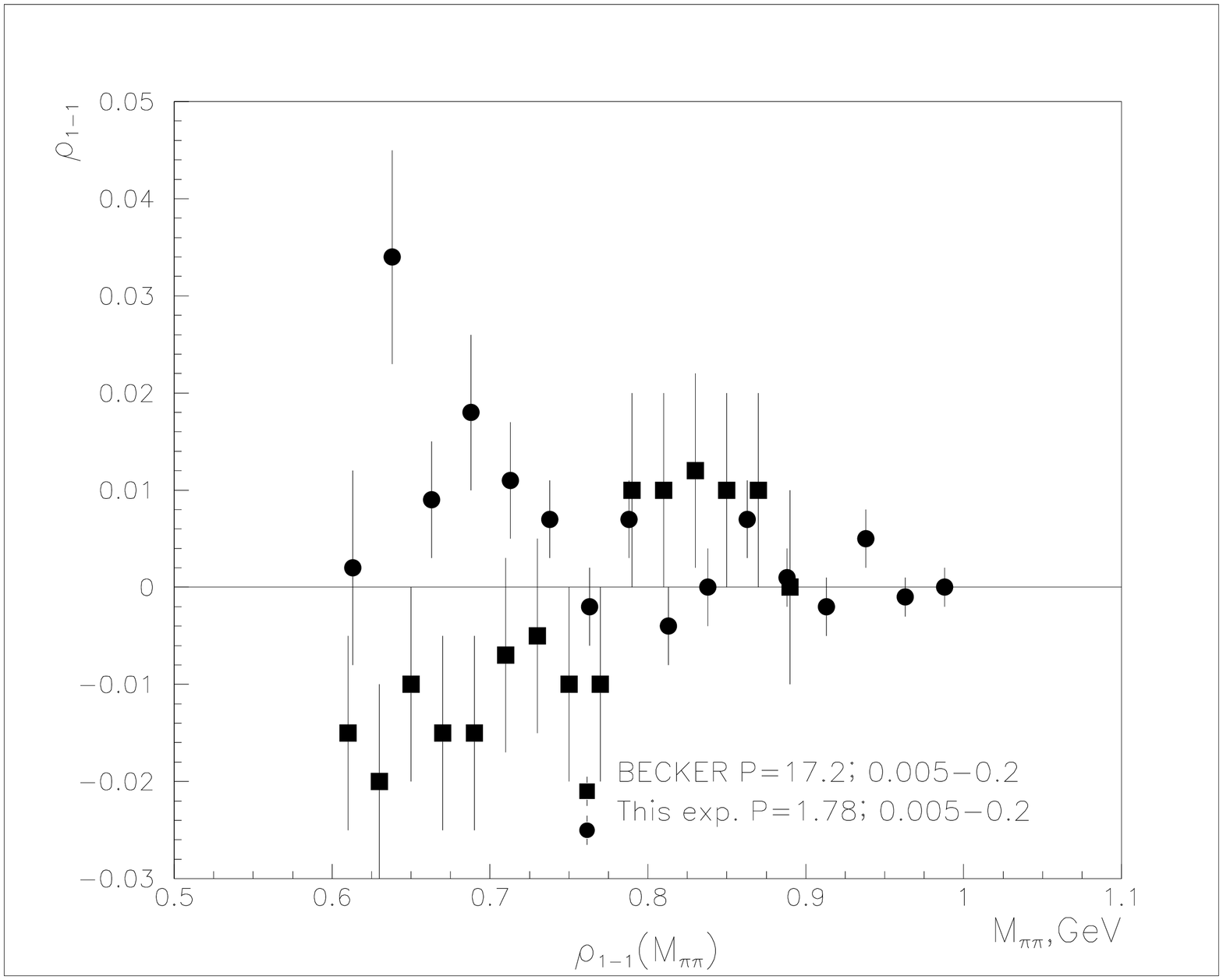,height=4cm,width=6.5cm}&
\epsfig{figure=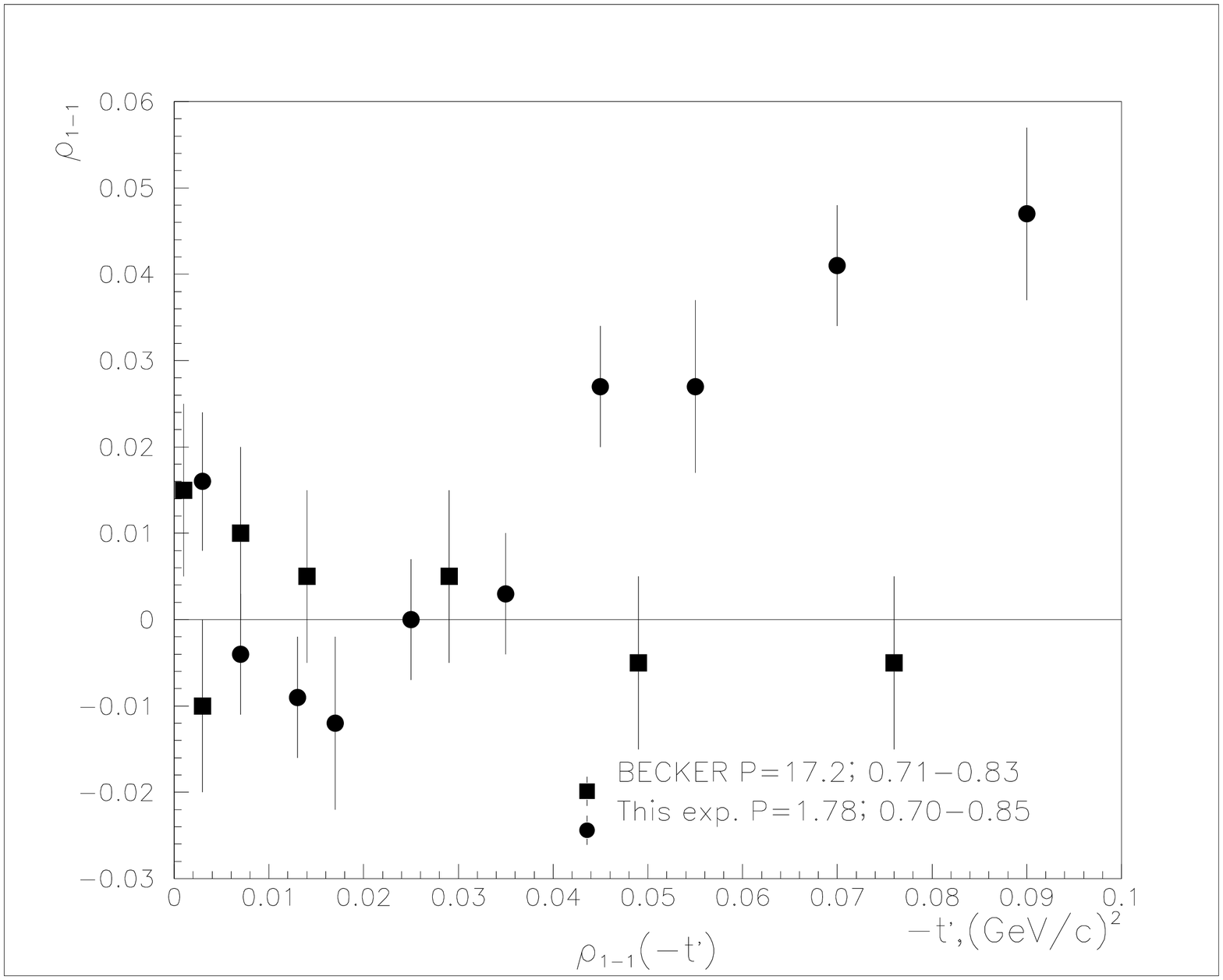,height=4cm,width=6.5cm}\\
\epsfig{figure=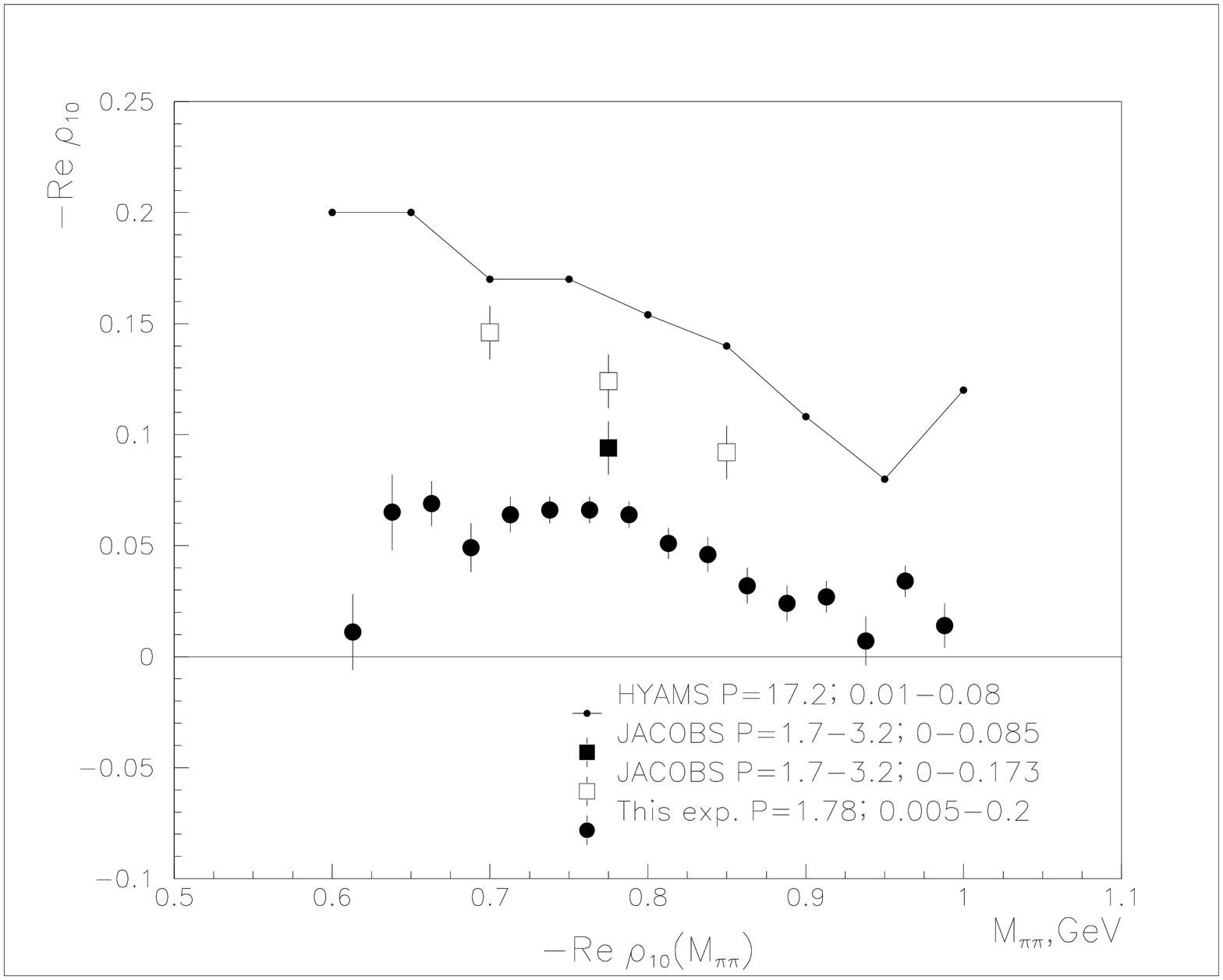,height=4cm,width=6.5cm}&
\epsfig{figure=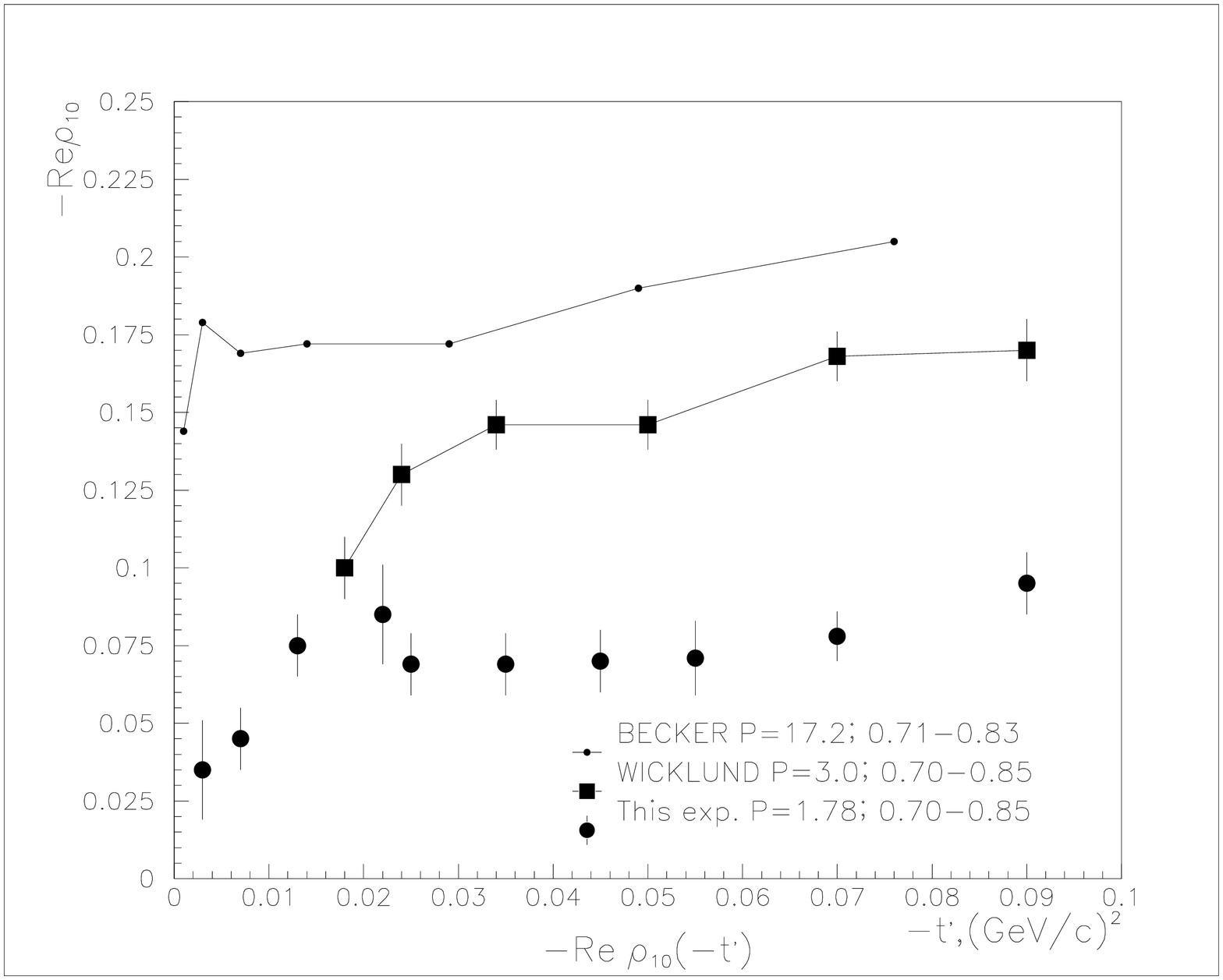,height=4cm,width=6.5cm}\\
\epsfig{figure=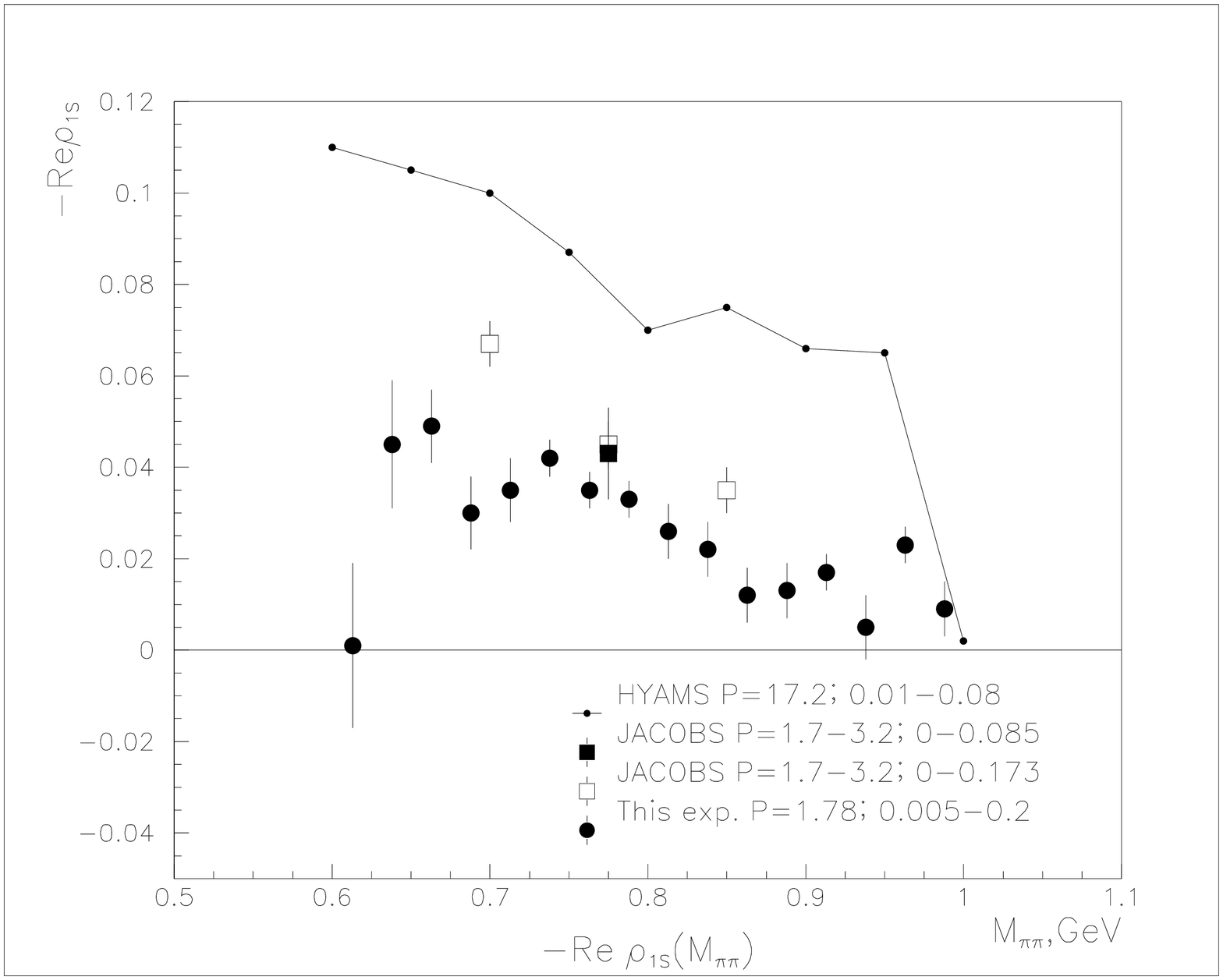,height=4cm,width=6.5cm}&
\epsfig{figure=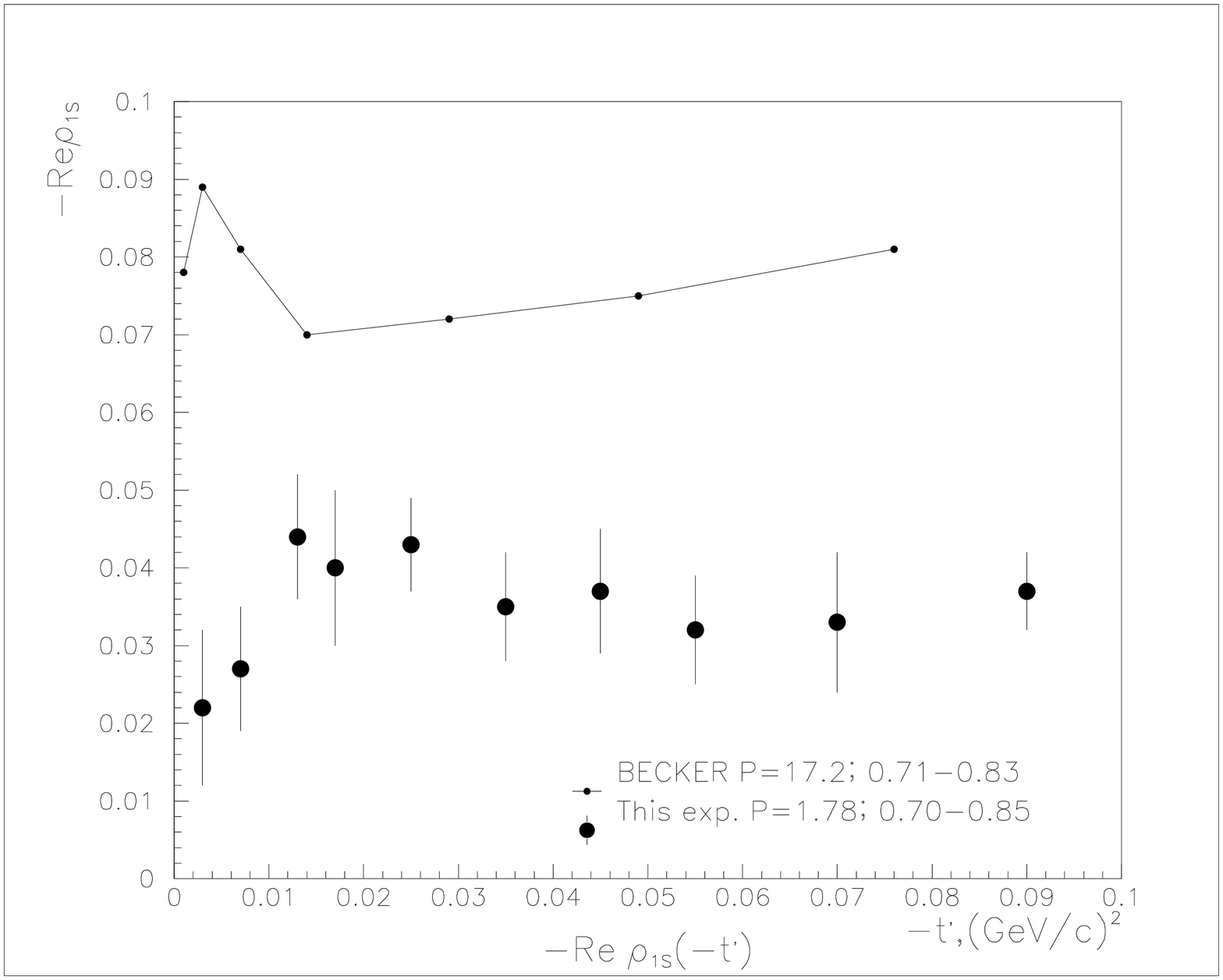,height=4cm,width=6.5cm}\\
a) & b)
\end{tabular}
\caption{Spin-independent density matrix elements as function 
of dipion invariant mass $M_{\pi\pi}$ (a) and momentum 
transfered $-t'$ (b).}
\label{fig:SDMEH}
\end{figure}

The normalized spin-independent SDME in the t-channel system 
(Jackson frame) are presented in fig.~\ref{fig:SDMEH} as
functions of dipion mass $M_{\pi\pi}$ and momentum transfered
$-t'$. Numeric data is given in the appendix.

The matrix element $\rho_{00}+\frac{1}{3}\rho_{SS}$
measured in all the experiments show approximately
the same behavior as a function of 
the dipion mass. 
The matrix element $\re \rho_{0S}$ in our experiment 
at $M_{\pi\pi} > 0.85$~GeV 
significantly differs from the data at the high energies. 
This probably could be attributed to 
the asymptotically vanishing amplitudes containing a term 
$\sqrt{\frac{t_{\mathrm{min}}}{t'}}$, which quickly increases
when $M_{\pi\pi}$ goes toward 1~GeV. The matrix
element $\rho_{1-1}$ is small, as it is at high energies. 
The matrix elements $\re \rho_{10}$ and $\re \rho_{1S}$ are decreasing
in the mass range (0.8--0.9)~GeV essentially quicker than at high
energies. These SDME also show obvious energy dependence. The
standard description of the dipion production in the frame of the
one pion exchange with absorption (OPEA) model \cite{Ochs}
gives the following mass dependence of the matrix element
ratios:
\begin{equation}
\frac{\re \rho_{10}}{\rho_{00} - \rho_{11}} =
\frac{\re \rho_{1S}}{\re \rho_{0S}} = 
- \frac{\mathrm{const.}}{M_{\pi\pi}} \, .
\end{equation}
The experimental data qualitatively follows the equation, but
demonstrates stronger dipion mass dependence in the range
(0.7--0.95)~GeV. 

Let us assume a small relative phase
between $S$ and $L$ waves and take the experimental data for the 
values of the SDME
$\rho_{00}+\frac{1}{3}\rho_{SS}$, $\re \rho_{0S}$ and $\rho_{1-1}$.
Then we can use equations (\ref{eq:norm2}-\ref{eq:rho1s})
connecting amplitudes to the SDME to show that
in these conditions intensity of the $U$-wave decreases 3 times in the
dipion mass region (0.8--0.92)~GeV, that agree with the strong
mass dependence of SDME $\re \rho_{10}$ and $\re \rho_{1S}$. We will
show below that the assumption about the phases corresponds to the only
physically justified solution of the amplitude analysis.

The behavior of the $t'$-dependencies is similar to 
those at high energies.
All SDME but $\rho_{1-1}$ show no significant dependence over $t'$
at $-t'>0.01$. At $t'$ near zero the expected kinematic suppression
of $\re \rho_{10}$ and $\re \rho_{1S}$ is observed. The matrix element
$\rho_{1-1}$ has a slow growth up to $-t'=0.1$ that agree with the 
predictions of the Regge model with moving branchings \cite{Kaydalov}.
The data does not exclude a slight energy dependence of
$\rho_{00}+\frac{1}{3}\rho_{SS}$.

\subsection{Spin dependence of the reaction}
\begin{figure*}
\epsfig{figure=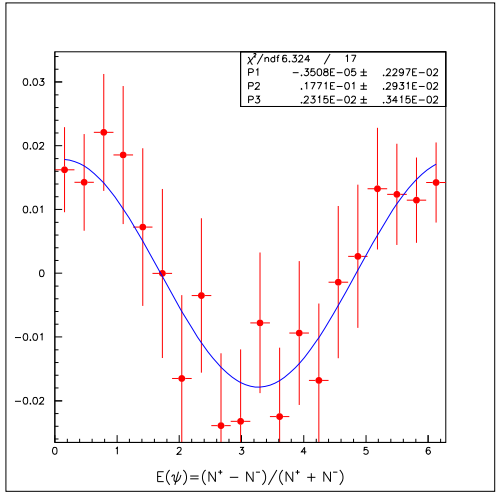,height=6.5cm}
\caption{Raw asymmetry (\ref{eq:rawass}) as function of 
the angle $\psi$.}
\label{fig:rawass}
\end{figure*}

The preliminary estimation of the asymmetry could be done using 
the function
\begin{equation}
E(\psi)=\frac{N^+(\psi)-N^-(\psi)}{N^+(\psi)+N^-(\psi)} ,
\label{eq:rawass}
\end{equation}
where $N^+,N^-$ are the normalized counts of the setup for positive and
negative target polarization. The fit with the cosine function of the
experimental values of $E(\psi)$ is shown in fig.~\ref{fig:rawass}.
The spin dependence of the reaction
is seen at the level of $6\cdot\sigma_{\mathrm{stat}}$.
Assuming only the asymmetry and no other terms in (\ref{eq:iY}) and 
making corrections for the mean target polarization and the dissolving
effect on unpolarized complex nuclei of the target one
can found that this fit corresponds to asymmetry about 9\%
with the same sign as at the high energies.

Only four SDME, namely $A=\rho^Y_{SS}+\rho^Y_{00}+2\rho^Y_{11}$,
$\rho^Y_{00}-\rho^Y_{11}$, $\re \rho^Y_{10}$ and
$\im \rho^X_{10}$ turned out to be nonzero within the experimental
errors. They are shown in fig.~\ref{fig:SDMEP} in comparison with the
data at 17.2~GeV/c \cite{Becker} and listed in the appendix. 
The signs of spin-dependent SDME coincide with those at high 
energies but the absolute values are significantly smaller.

\begin{figure}
\begin{tabular}{ll}
\epsfig{figure=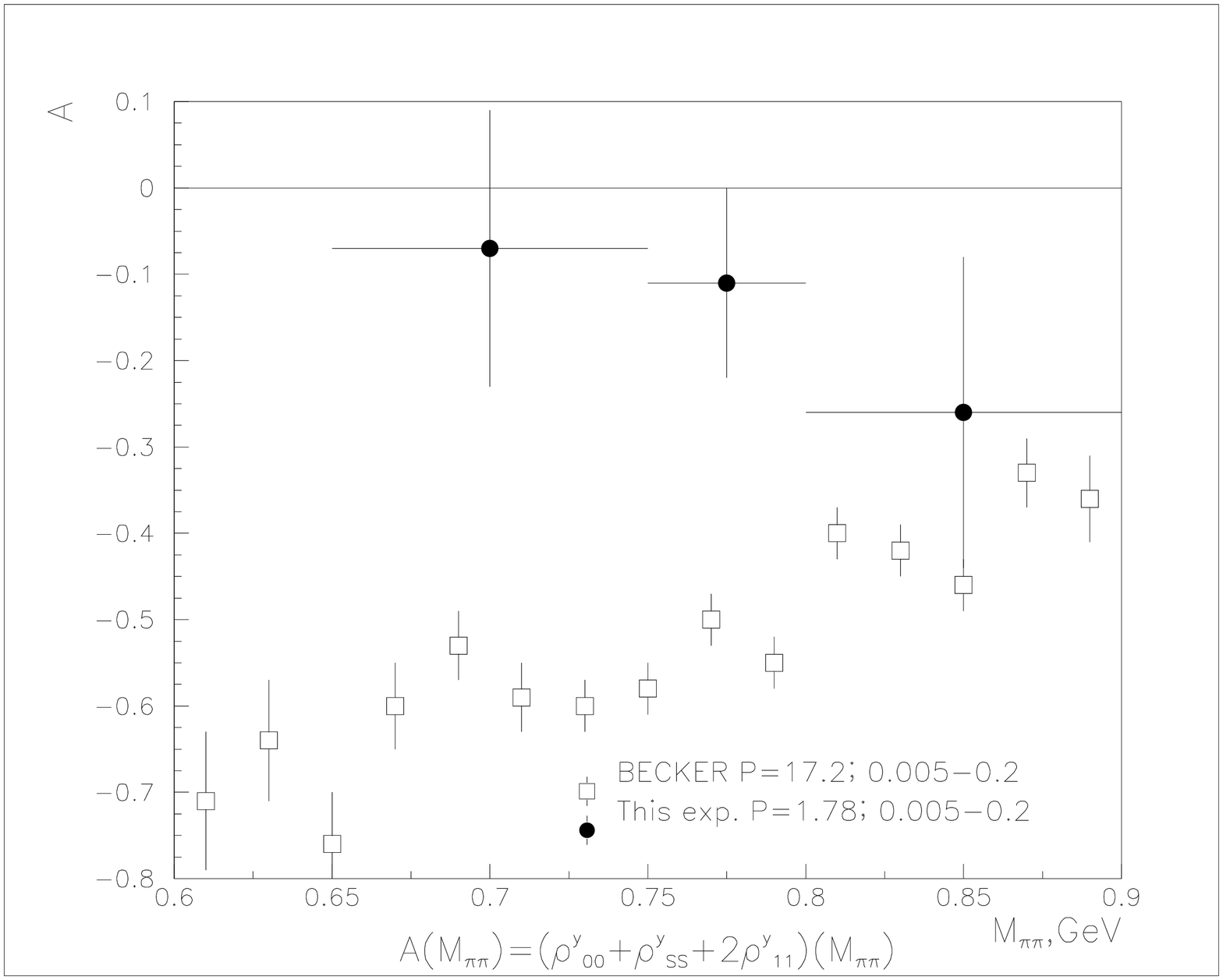,width=6.5cm}&
\epsfig{figure=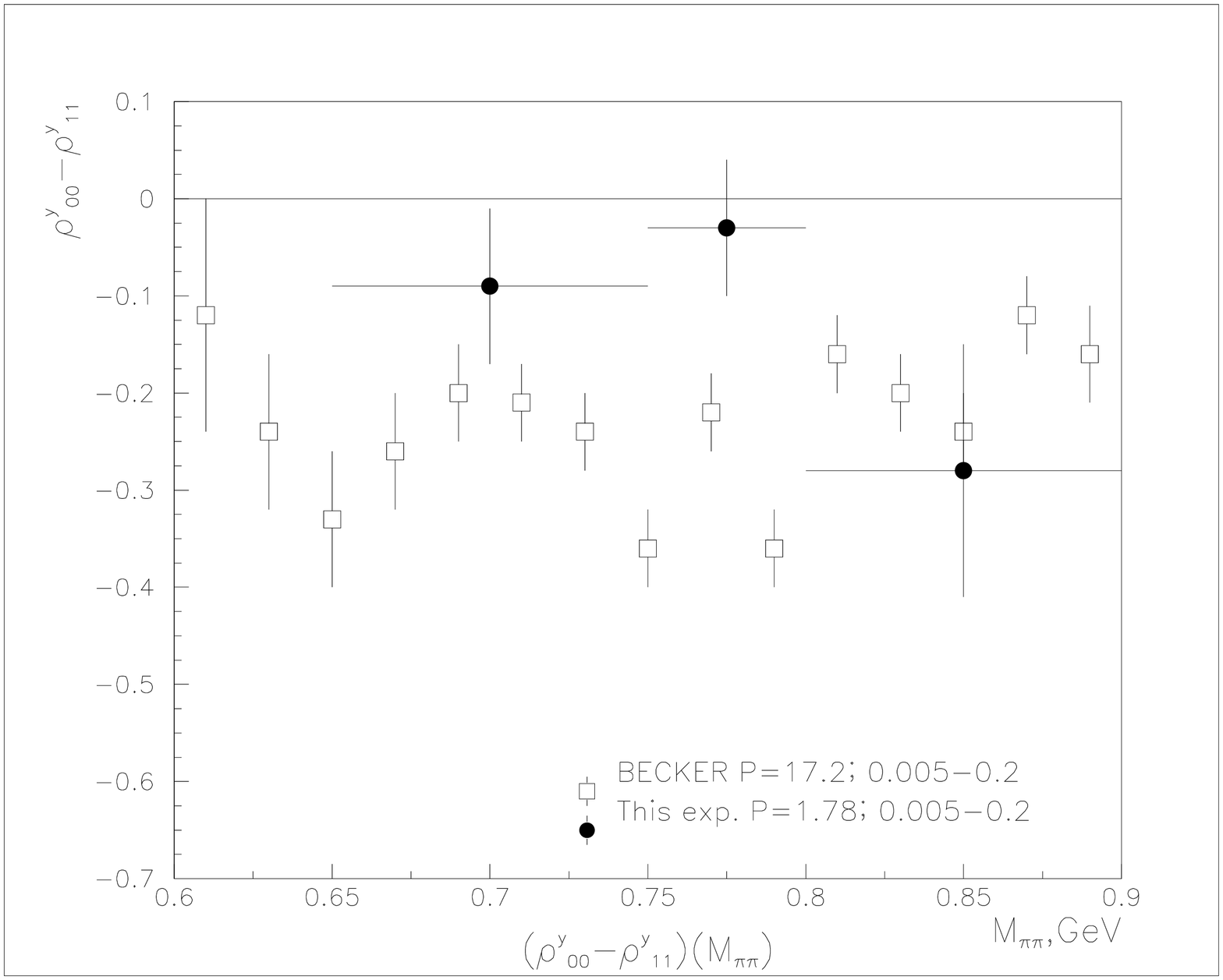,width=6.5cm}\\
\epsfig{figure=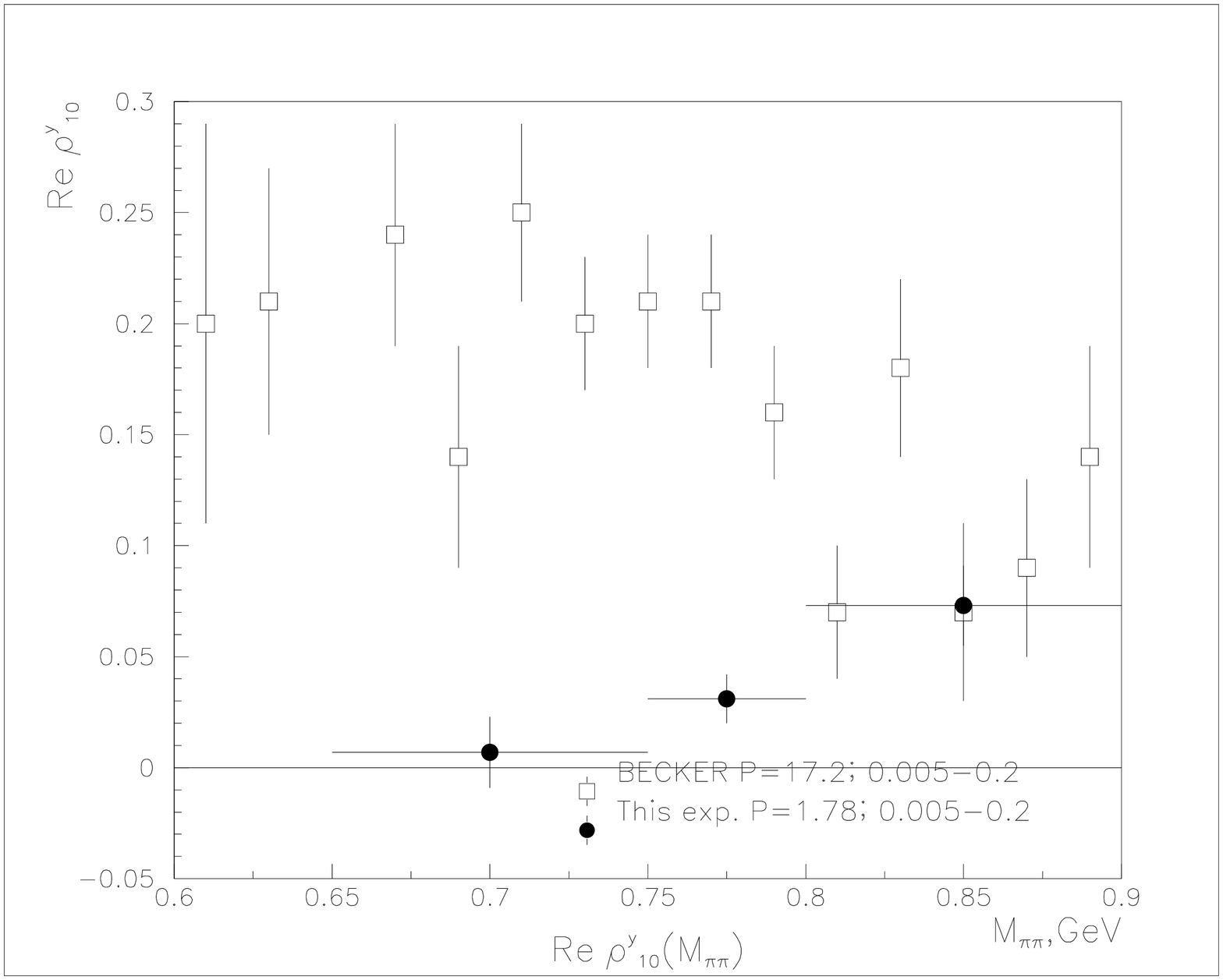,width=6.5cm}&
\epsfig{figure=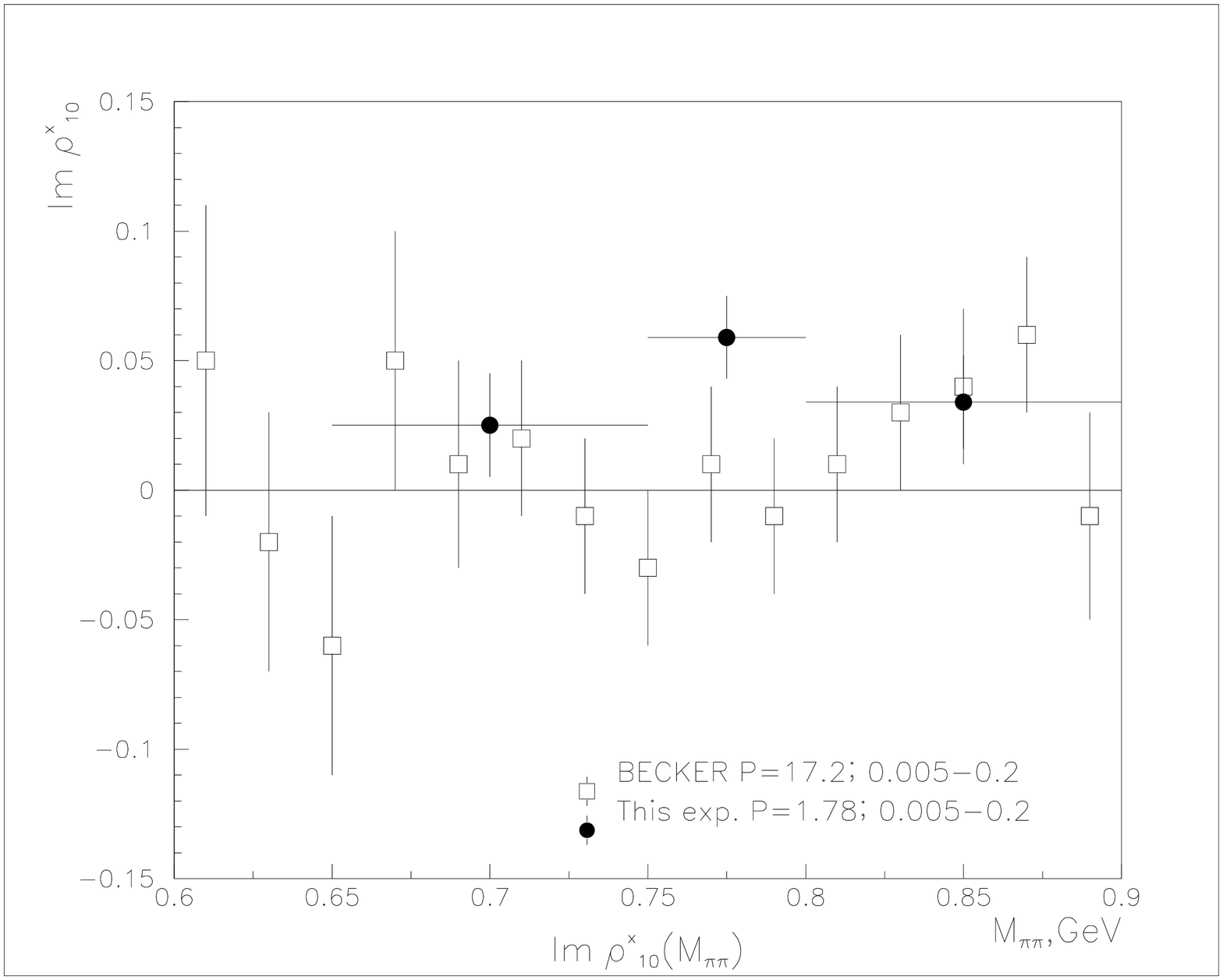,width=6.5cm}\\
\end{tabular}
\caption{Spin-dependent density matrix elements. Full dots --- this 
experiment, open squares --- Becker et al. \protect\cite{Becker}.} 
\label{fig:SDMEP}
\end{figure}

We would like to draw attention to the nonzero value of the SDME 
$\im \rho^X_{10}$, which describes interference between the waves with
different naturality (see equation (\ref{eq:rhox10})). This 
fact evidences
in favour of the contribution of amplitudes with natural parity and
the phase different from the phase of leading $(L,\bar{L})$ amplitudes.
In the frame of Regge model this corresponds to the contribution of
$a_2$-Regge trajectory ($a_2$-meson has $J^P=2^+$).

\subsection{Amplitude analysis}
\begin{figure}
\begin{tabular}{ll}
\epsfig{figure=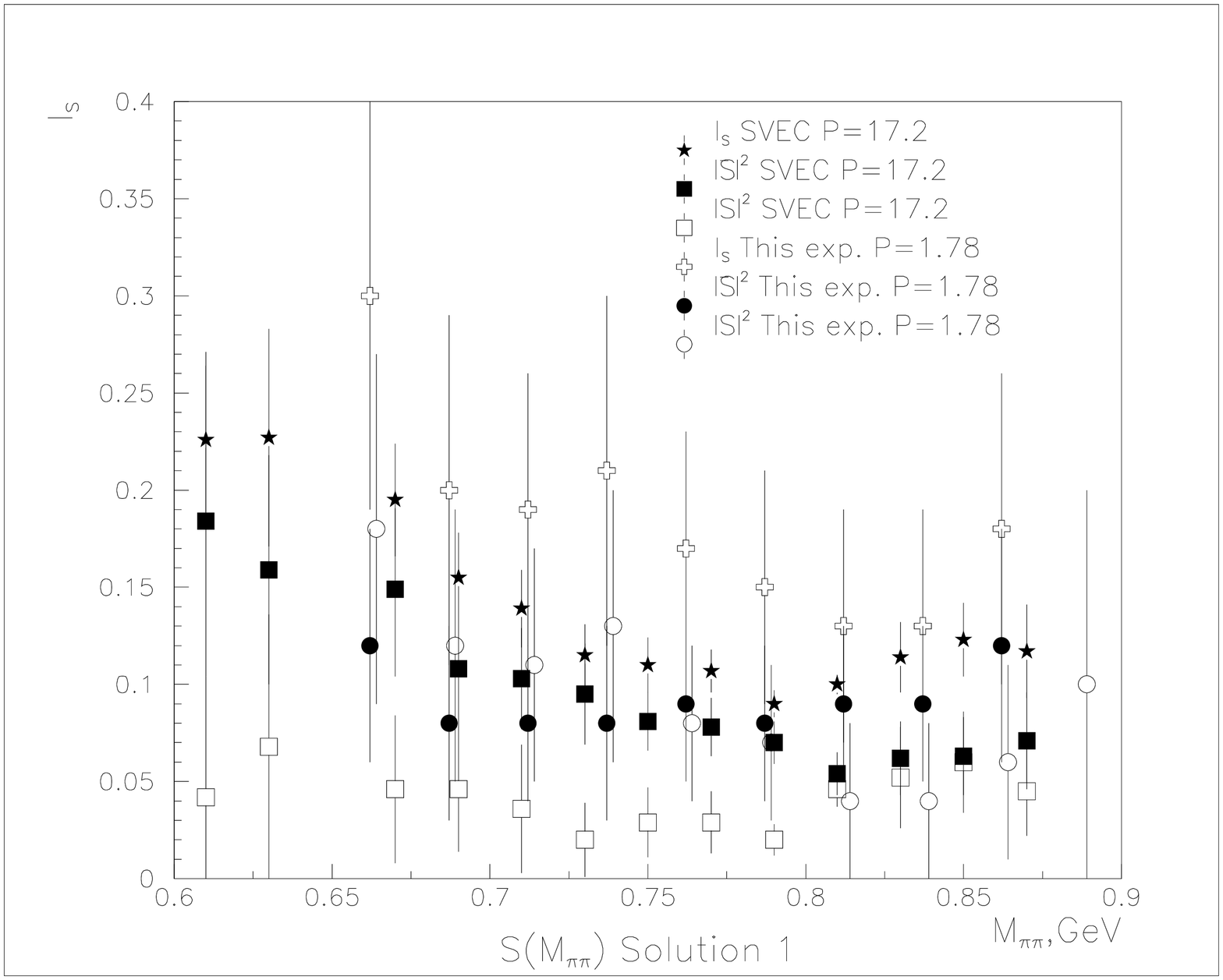,height=4cm,width=6.5cm}&
\epsfig{figure=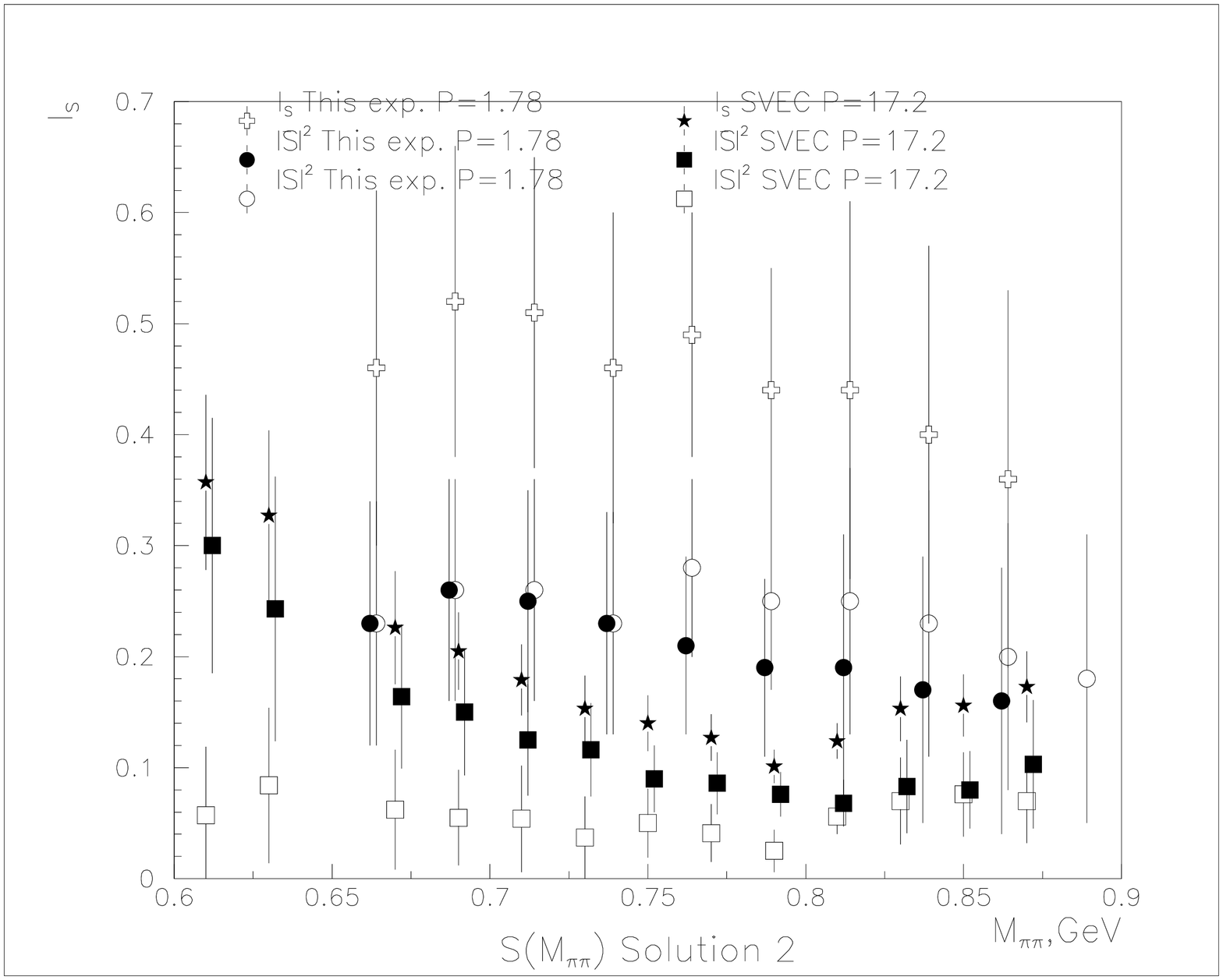,height=4cm,width=6.5cm}\\
\epsfig{figure=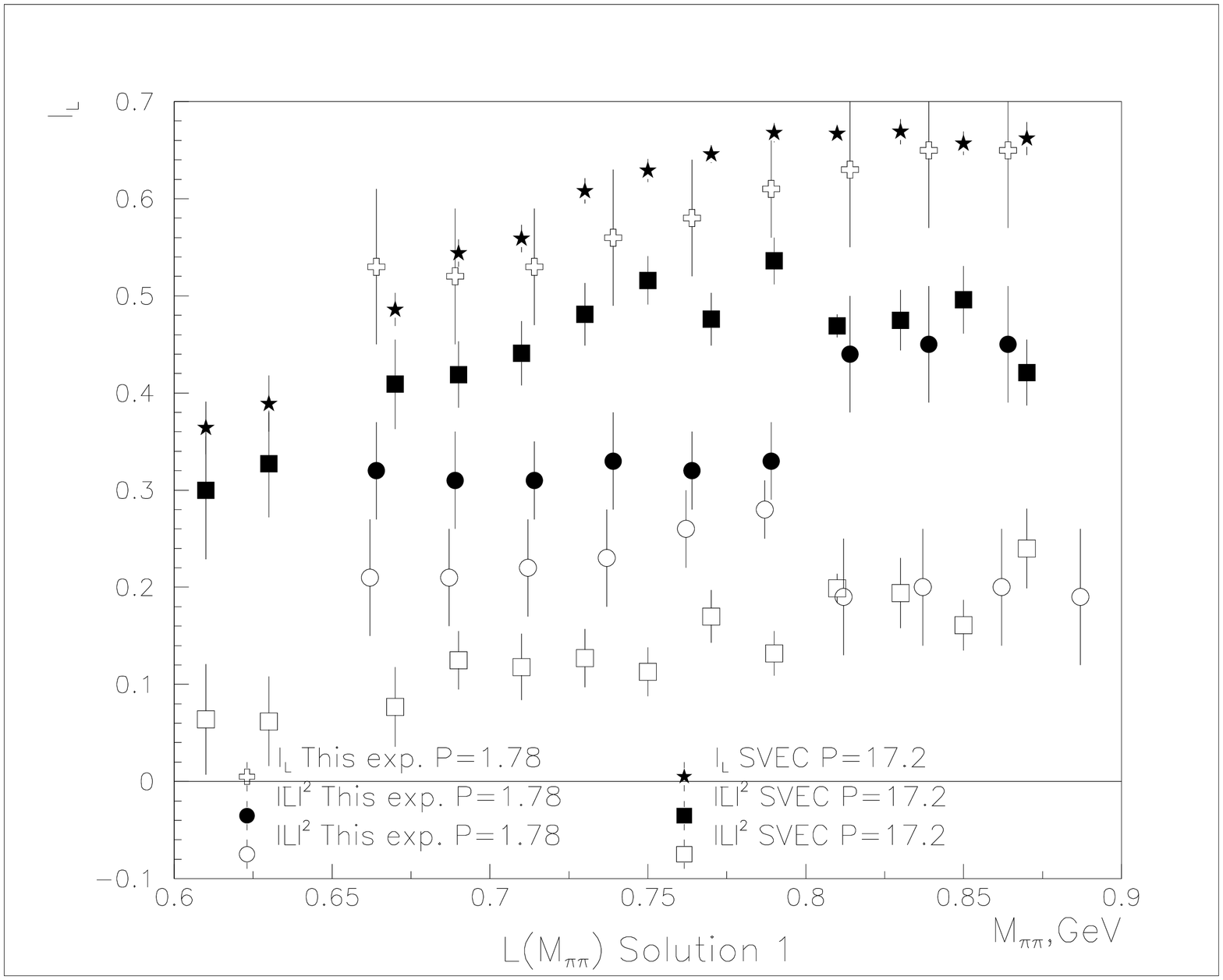,height=4cm,width=6.5cm}&
\epsfig{figure=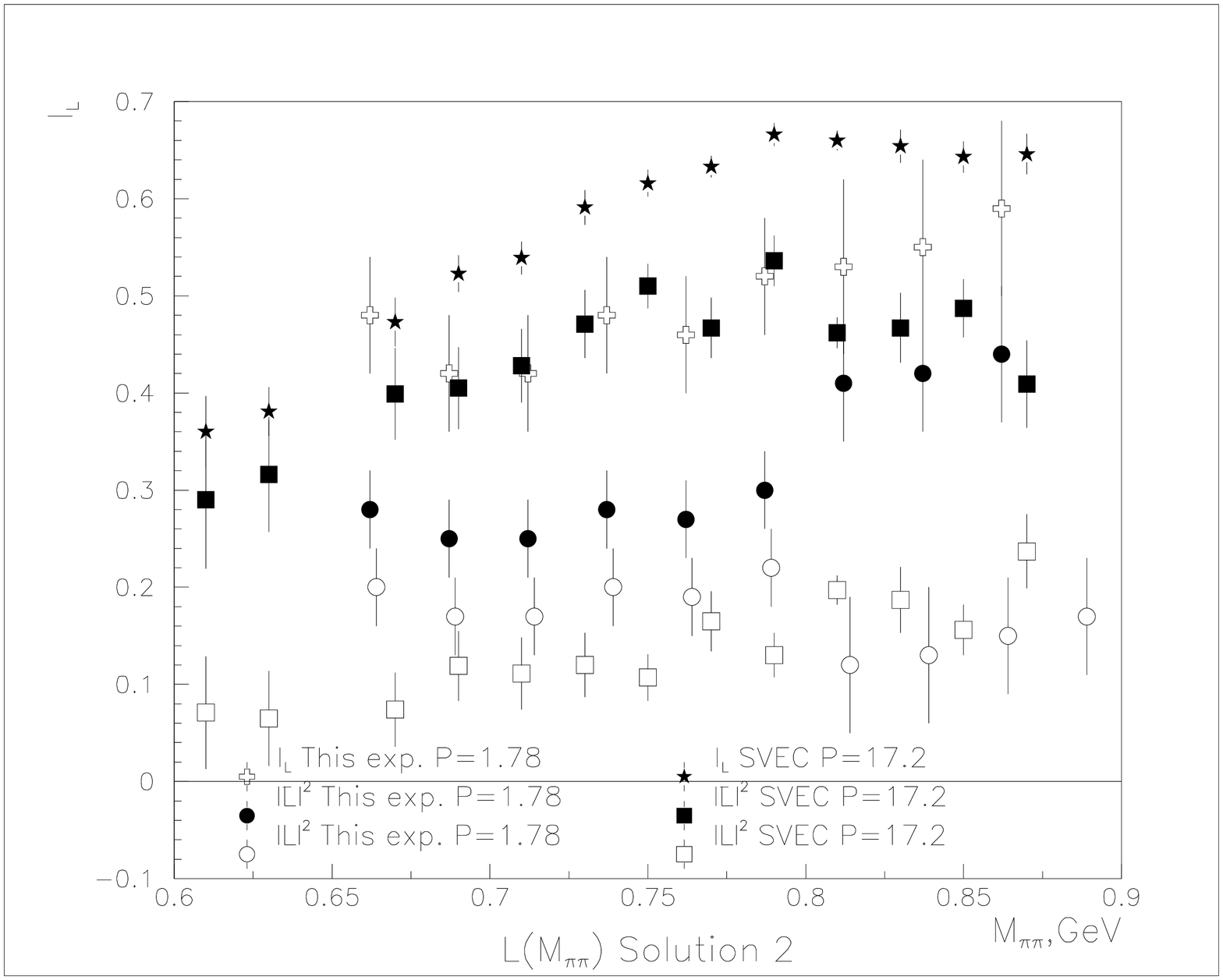,height=4cm,width=6.5cm}\\
\epsfig{figure=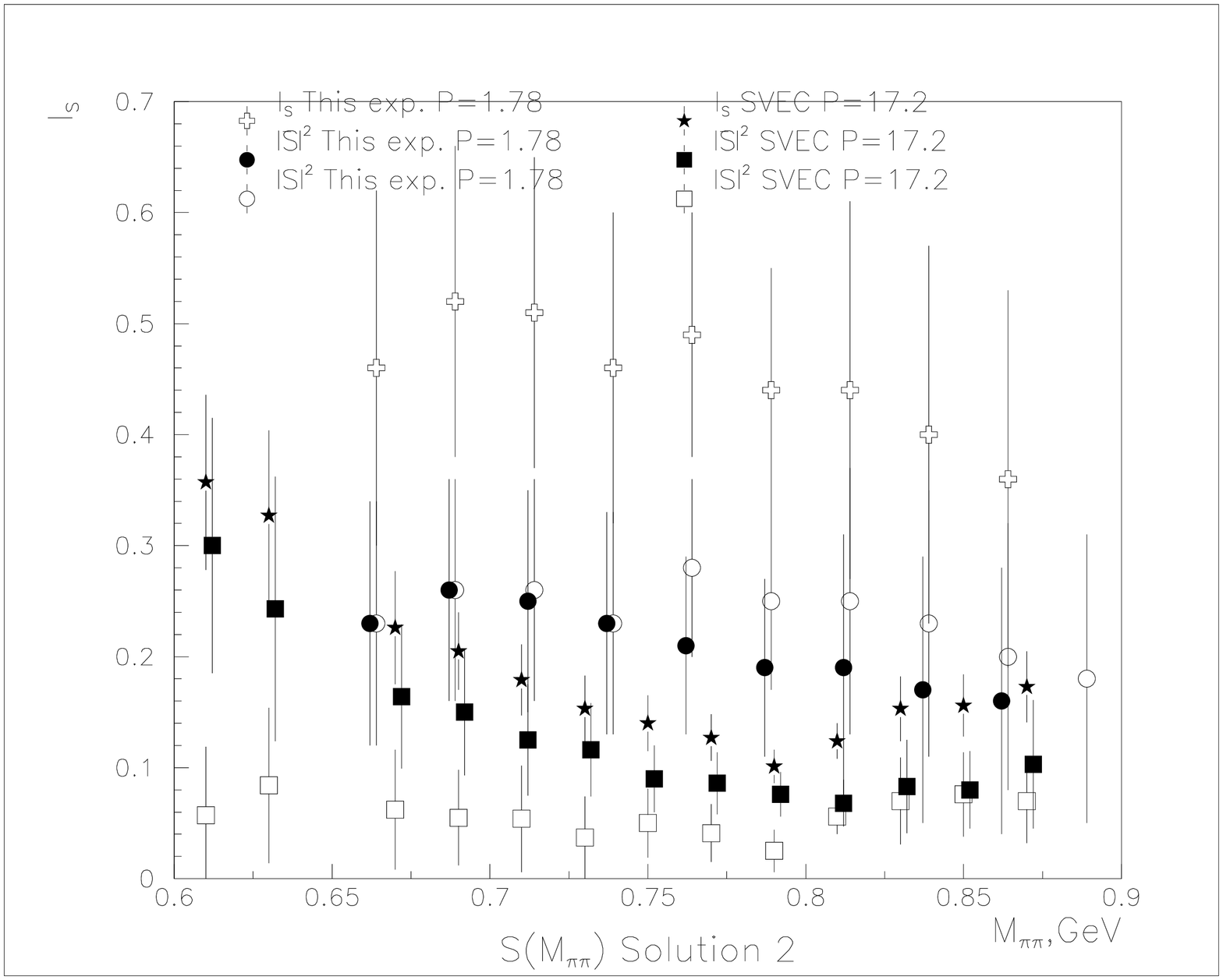,height=4cm,width=6.5cm}&
\epsfig{figure=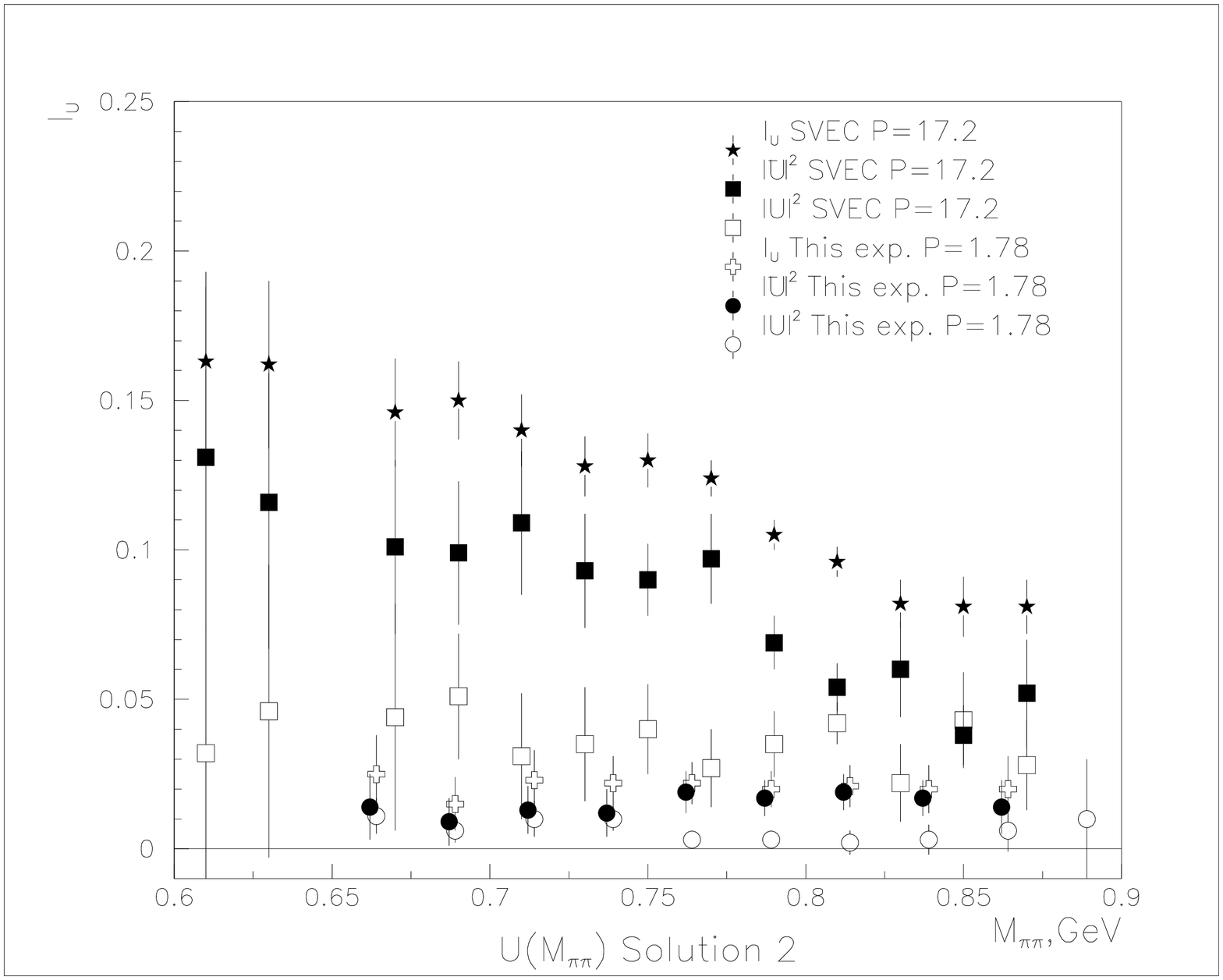,height=4cm,width=6.5cm}\\
\epsfig{figure=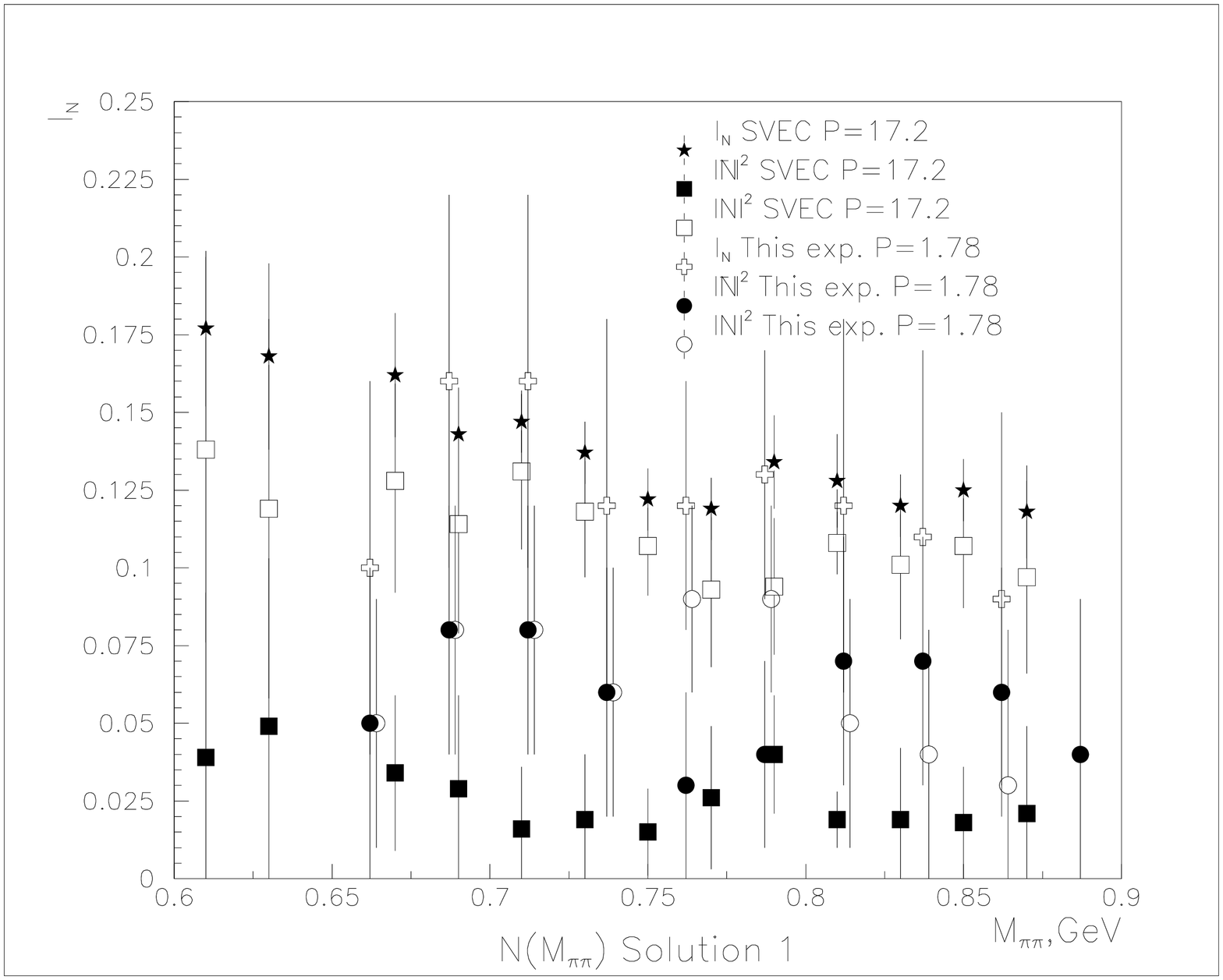,height=4cm,width=6.5cm}&
\epsfig{figure=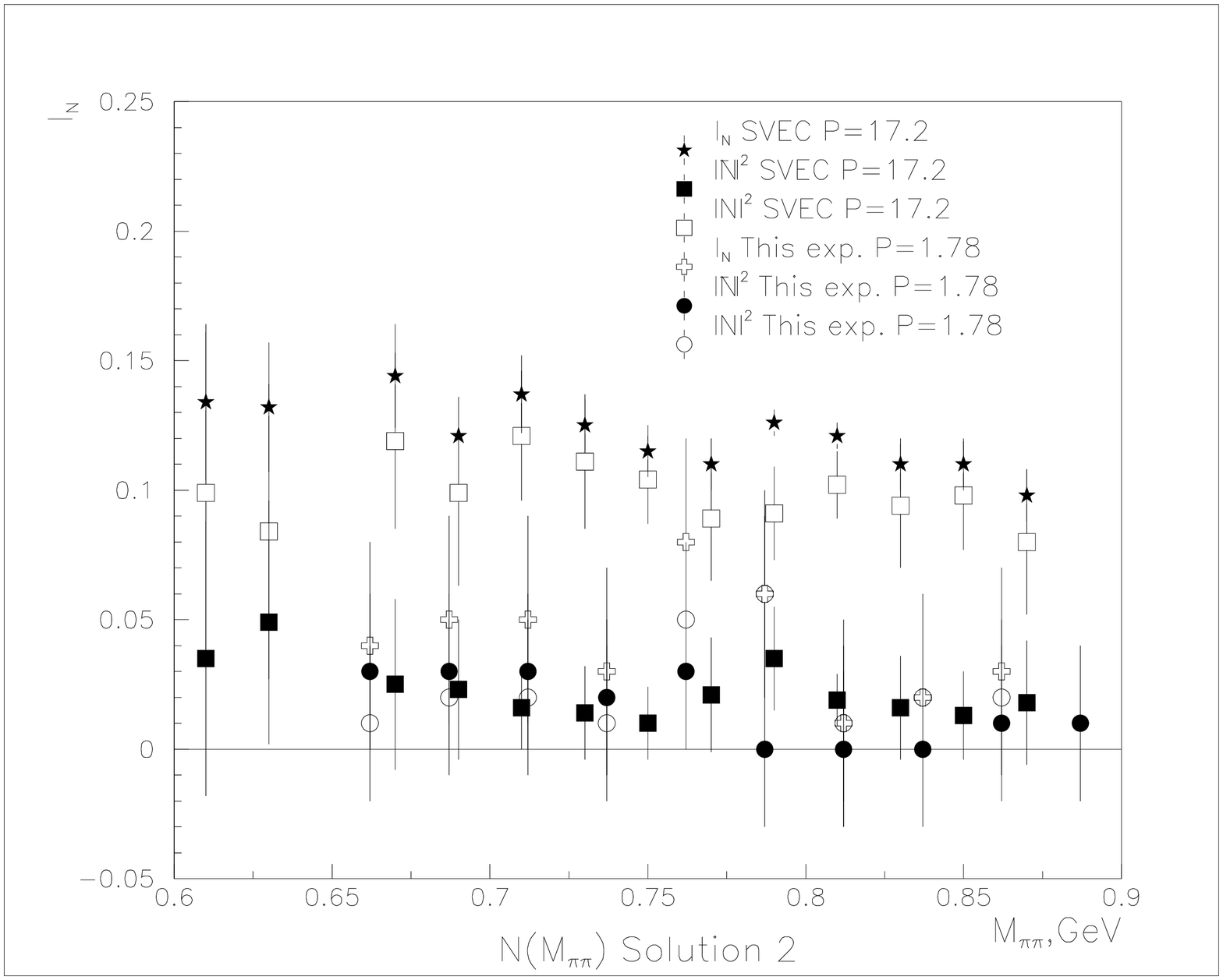,height=4cm,width=6.5cm}\\
a) & b)
\end{tabular}
\caption{Results of the model-independent amplitude
analysis. Solution 1 (a) and 2 (b). Notations: open/filled
circles/squares --- corresponding wave with recoiled nucleon 
transversity down/up (exept $N$-wave) this work/M.~Svec \cite{Svec};
open crosses/filled stars --- corresponding wave intensity this 
work/M.~Svec \cite{Svec}.}
\label{fig:AMPL}
\end{figure}

\begin{figure}
\begin{tabular}{ll}
\epsfig{figure=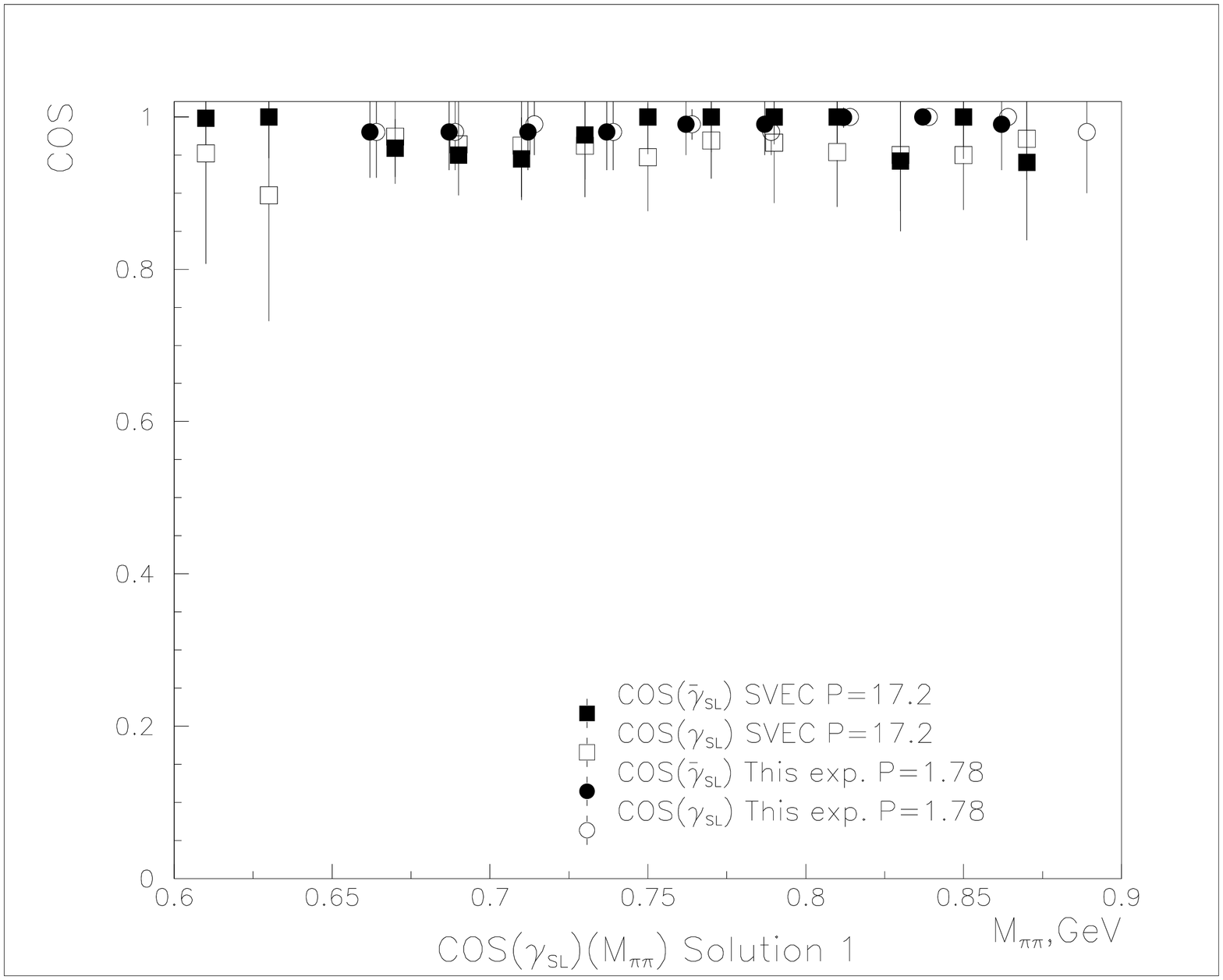,height=4cm,width=6.5cm}&
\epsfig{figure=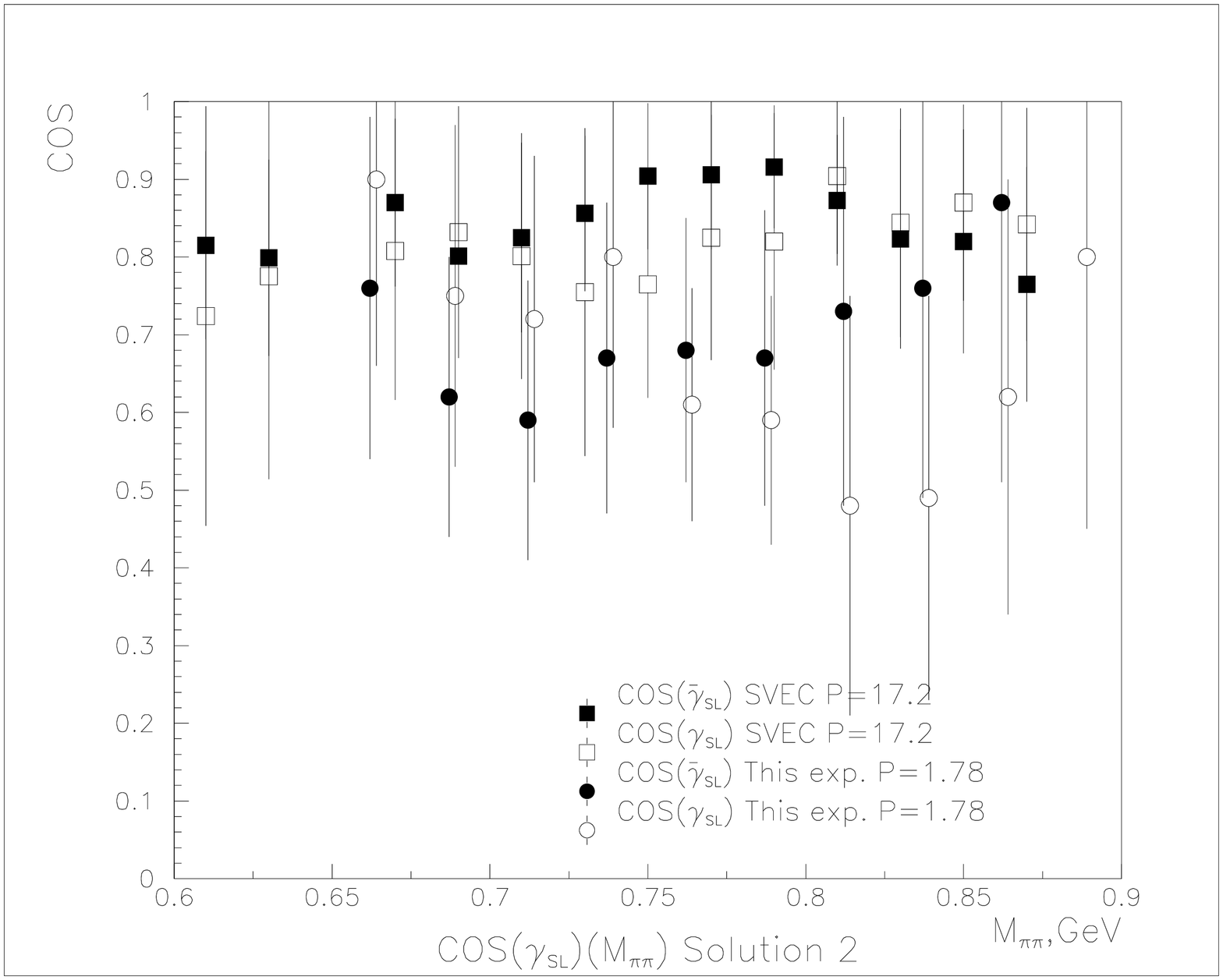,height=4cm,width=6.5cm}\\
\epsfig{figure=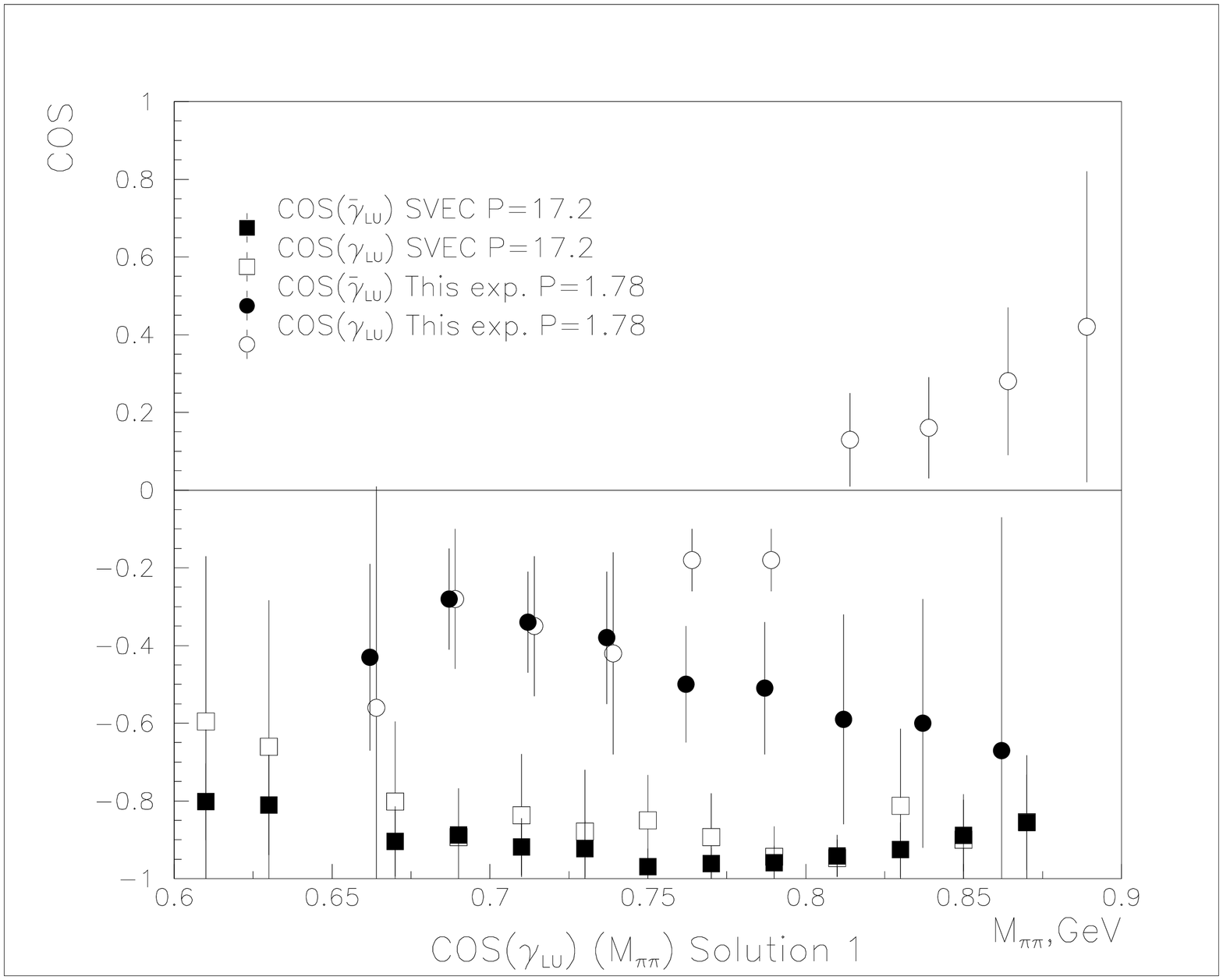,height=4cm,width=6.5cm}&
\epsfig{figure=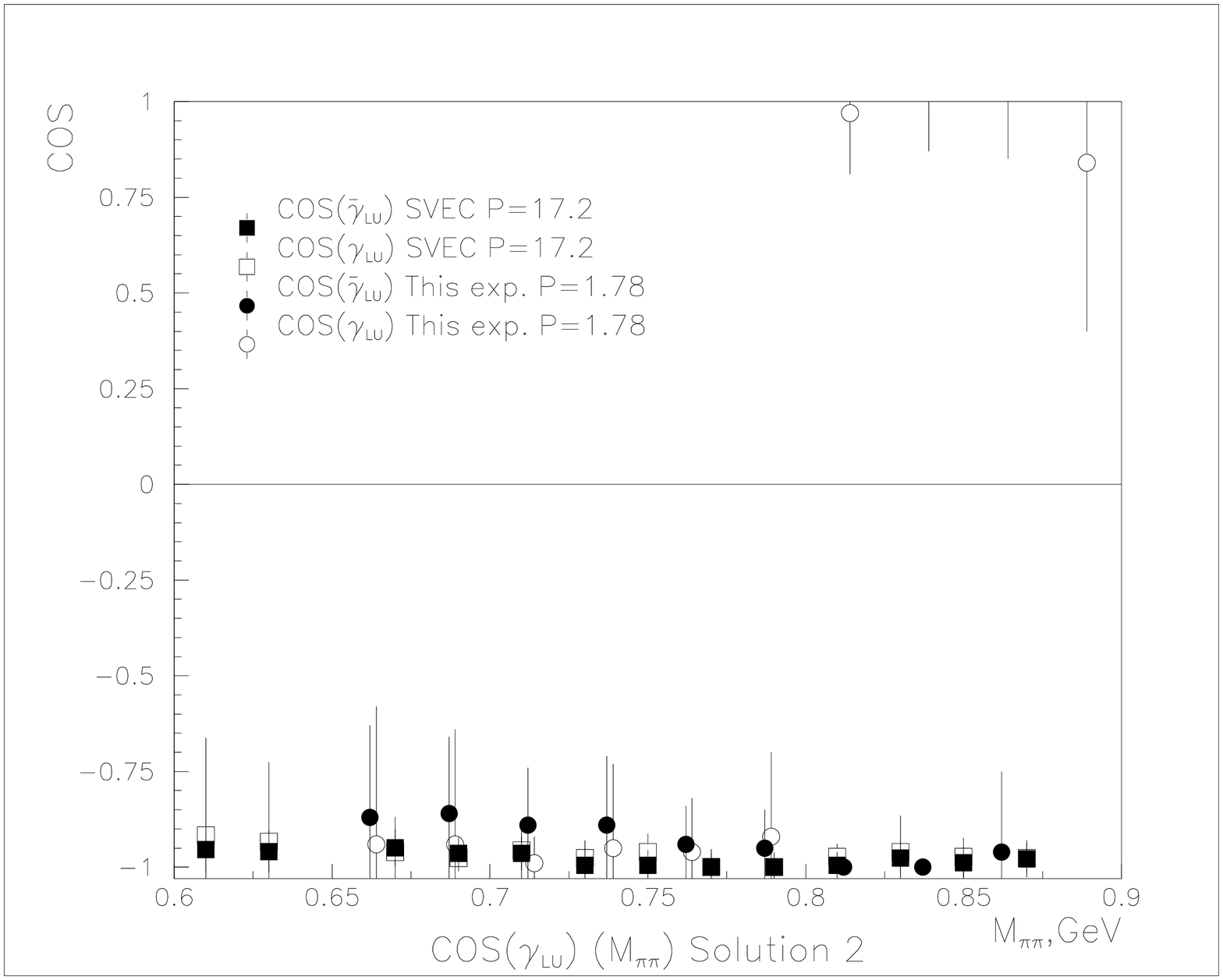,height=4cm,width=6.5cm}\\
\epsfig{figure=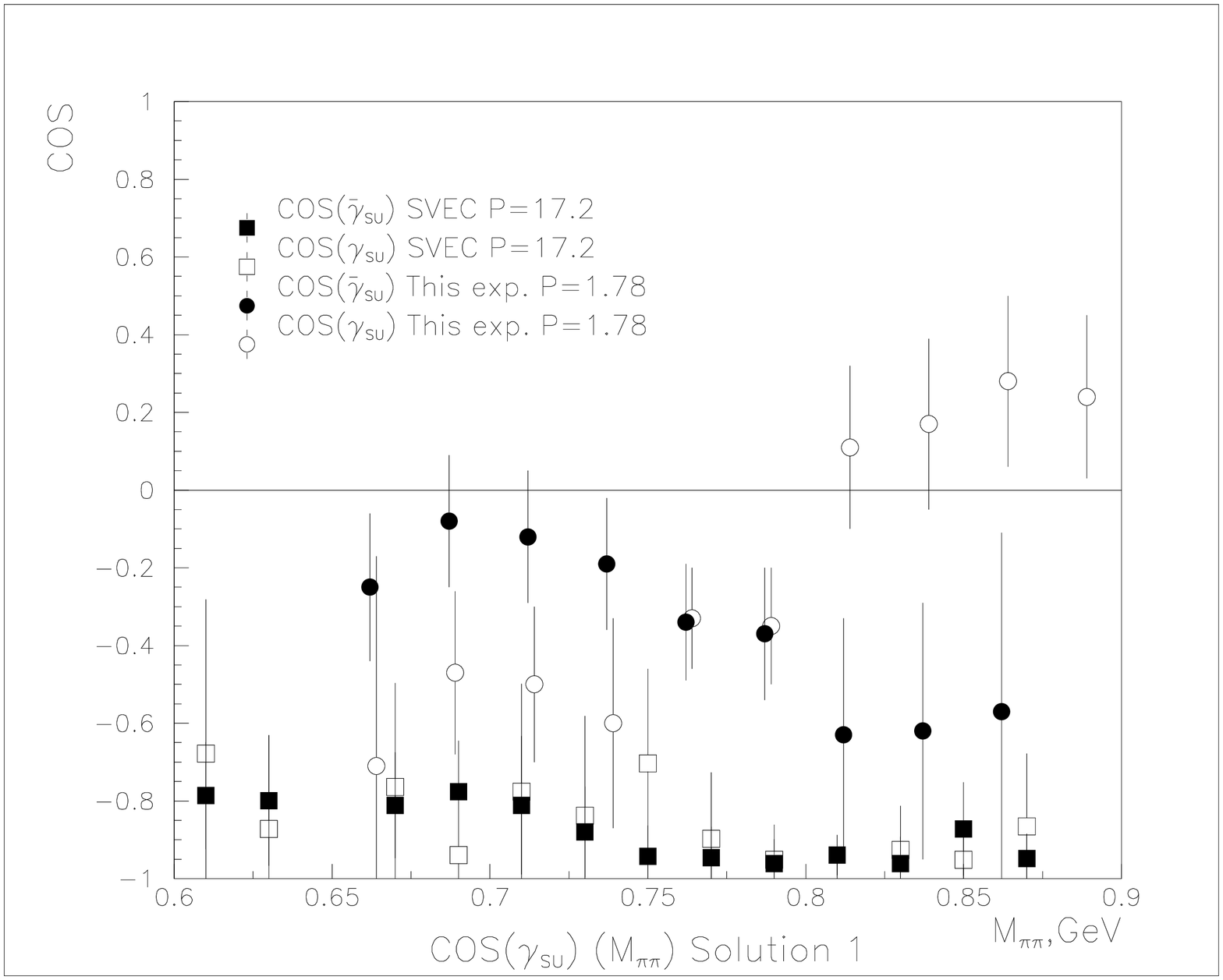,height=4cm,width=6.5cm}&
\epsfig{figure=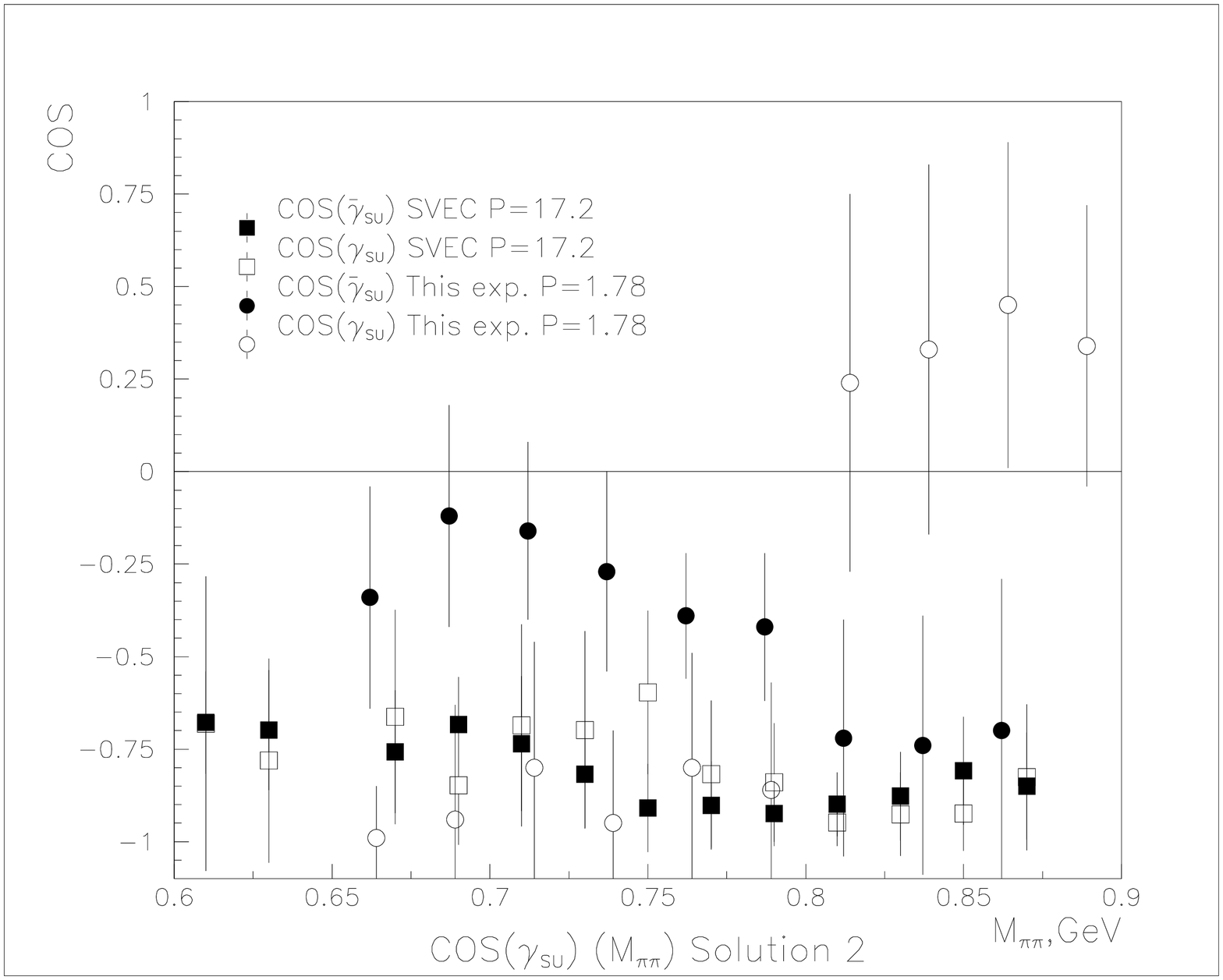,height=4cm,width=6.5cm}\\
a) & b)
\end{tabular}
\caption{Results of the model-independent amplitude
analysis. Solution 1 (a) and 2 (b). Continued. Notations are the same.}
\label{fig:AMPLC}
\end{figure}

The results of the model-independent amplitude analysis of the reaction
are shown in fig.~\ref{fig:AMPL},\ref{fig:AMPLC} in comparison with 
the results at 
17.2~GeV/c from the analysis \cite{Svec}. The analysis was performed in 
25~MeV bins over dipion mass. The polarization data was taken constant 
within wider bins in which it was available. 

Our model-independent analysis produced two significantly different
solutions 1 and 2. At all energies (1.78, 5.98, 11.85 and 17.2~GeV/c)
where model-independent analyses were performed they resulted in 
two-fold ambiguity solutions with the same clear signatures.
The solution 1 gives a very small angle between $S$ and $L$-waves and
the solution 2 gives an angle between $L$ and $U$-waves near $180^o$
independently of the recoil nucleon transversity. In addition the
solution 1 has essentially smaller amplitude of $S$-waves. These
signatures allow to trace the particular solution in the 
whole energy range.
On the other hand unlike the situation at
high energies we saw comparatively small
effect of the polarization. In terms of the amplitude analysis
it reveals in quite small difference between the barred and
unbarred amplitudes. In the limit of zero polarization effect
these amplitudes must coincide. So one and the same
solution should take place for
both sets of amplitudes. From this point of view only solutions
1.1 and 2.2 (in terms of M.~Svec \cite{Svec}) look reasonable.

In both solutions the visible asymmetry is most concentrated in the
$L$-wave. The partial polarization in this wave is 
$P_L=-0.10 \pm 0.02$. The effects of the spin are also observed in the
relative phases. For instance in the solution 1 the relative phase
between $U$ and $L$-waves ($\cos{\gamma_{LU}}$) differs noticeably from
the relative phase between $\bar{U}$ and $\bar{L}$-waves 
($\cos{\bar{\gamma}_{LU}}$) at $M_{\pi\pi} > 0.75$~GeV.

The difference between the barred and unbarred amplitudes in our data is not
large. This makes reasonable  to perform a model-dependent amplitude
analysis with assumption, that 
all spin-dependent matrix elements equal zero.
The model-dependent analysis allowed to spread the analysis to a
wider mass region and decrease errors. The resulting amplitudes
are shown in fig.~\ref{fig:AMPL1} and the numeric data is listed 
in the appendix. We tested that the intensities obtained in the
model-dependent analysis coincide within the errors with those
obtained in the model-independent one. The same was tested
for the mean relative phases.

\begin{figure}
\begin{tabular}{ll}
\epsfig{figure=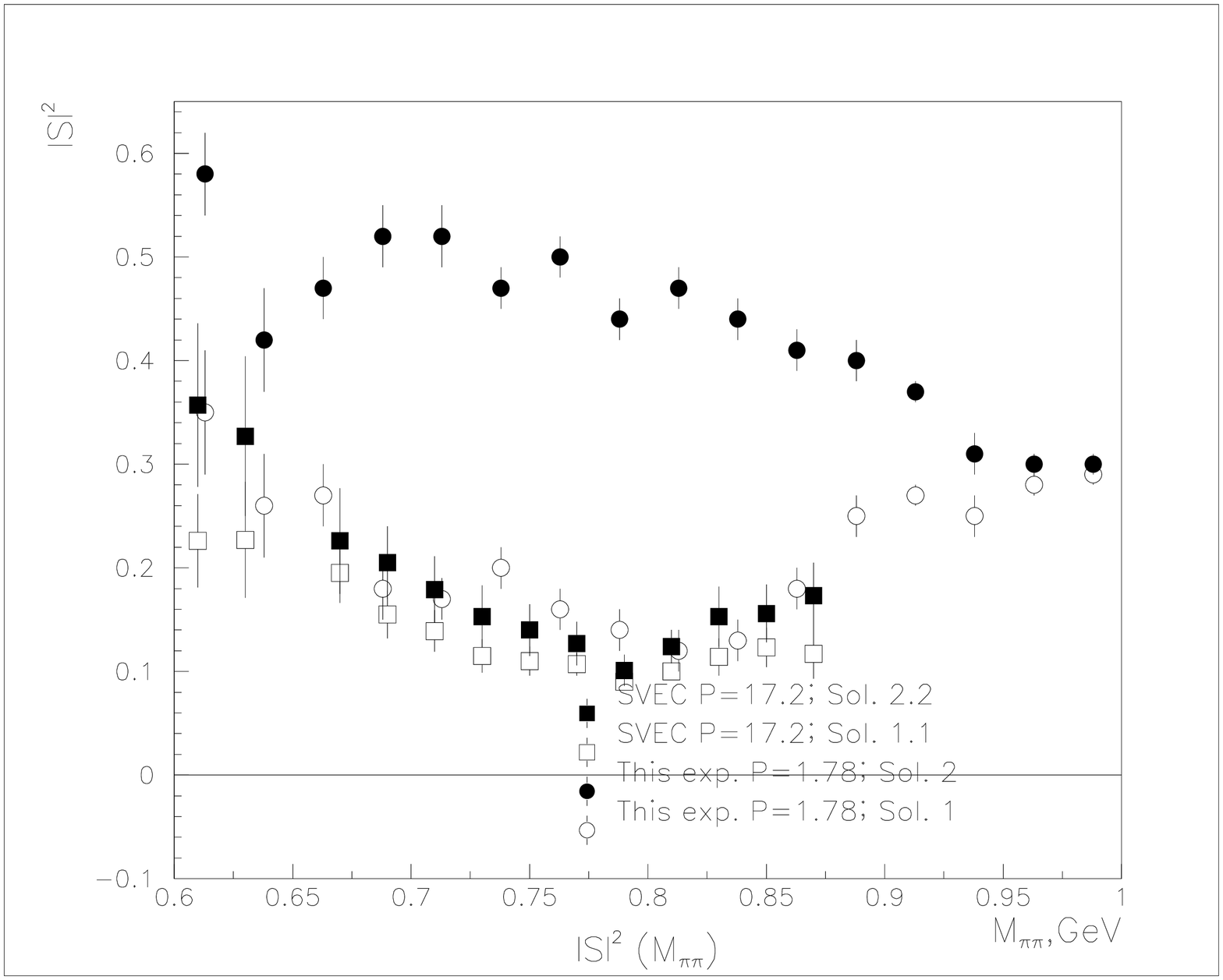,height=4cm,width=6.5cm}&
\epsfig{figure=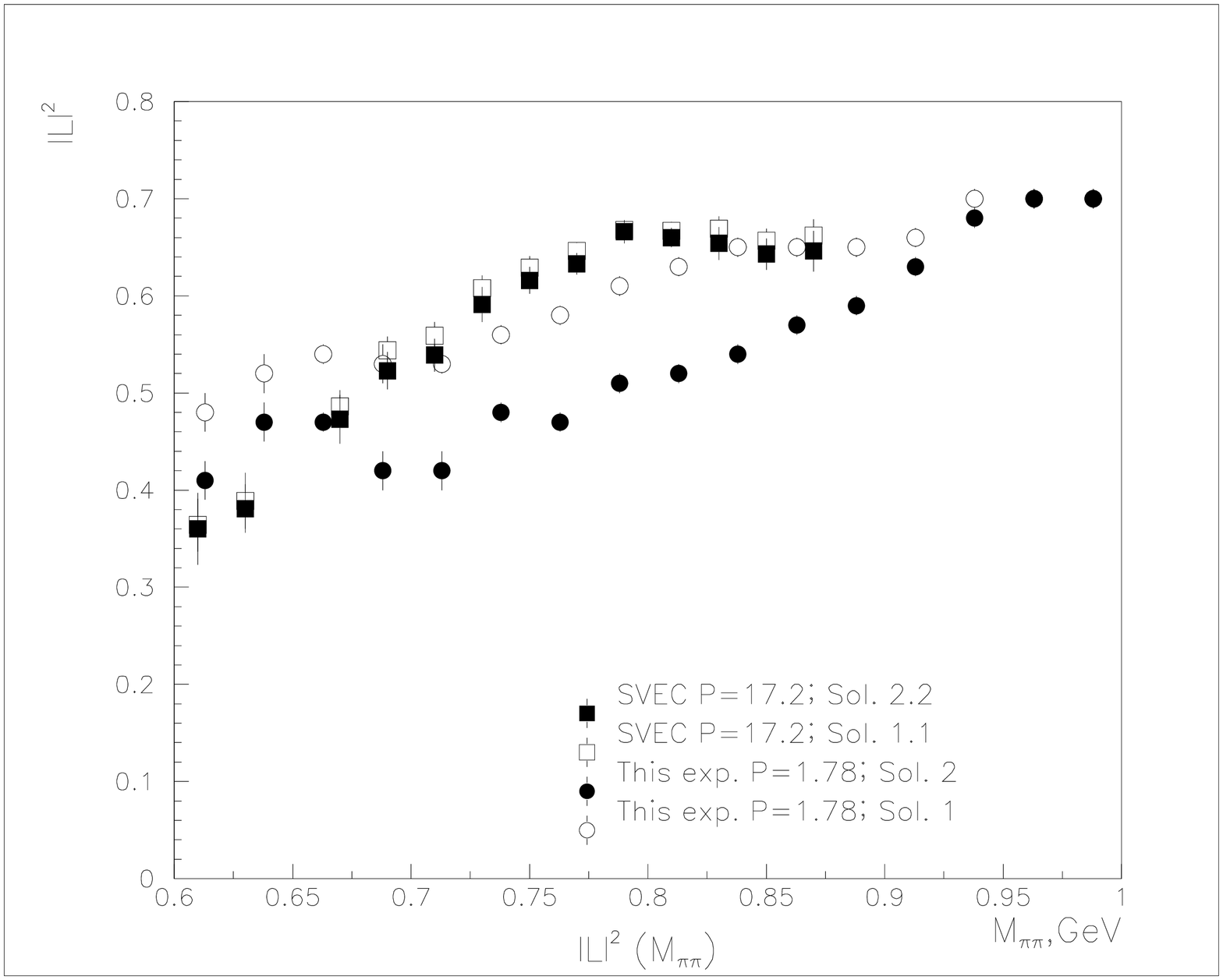,height=4cm,width=6.5cm}\\
\epsfig{figure=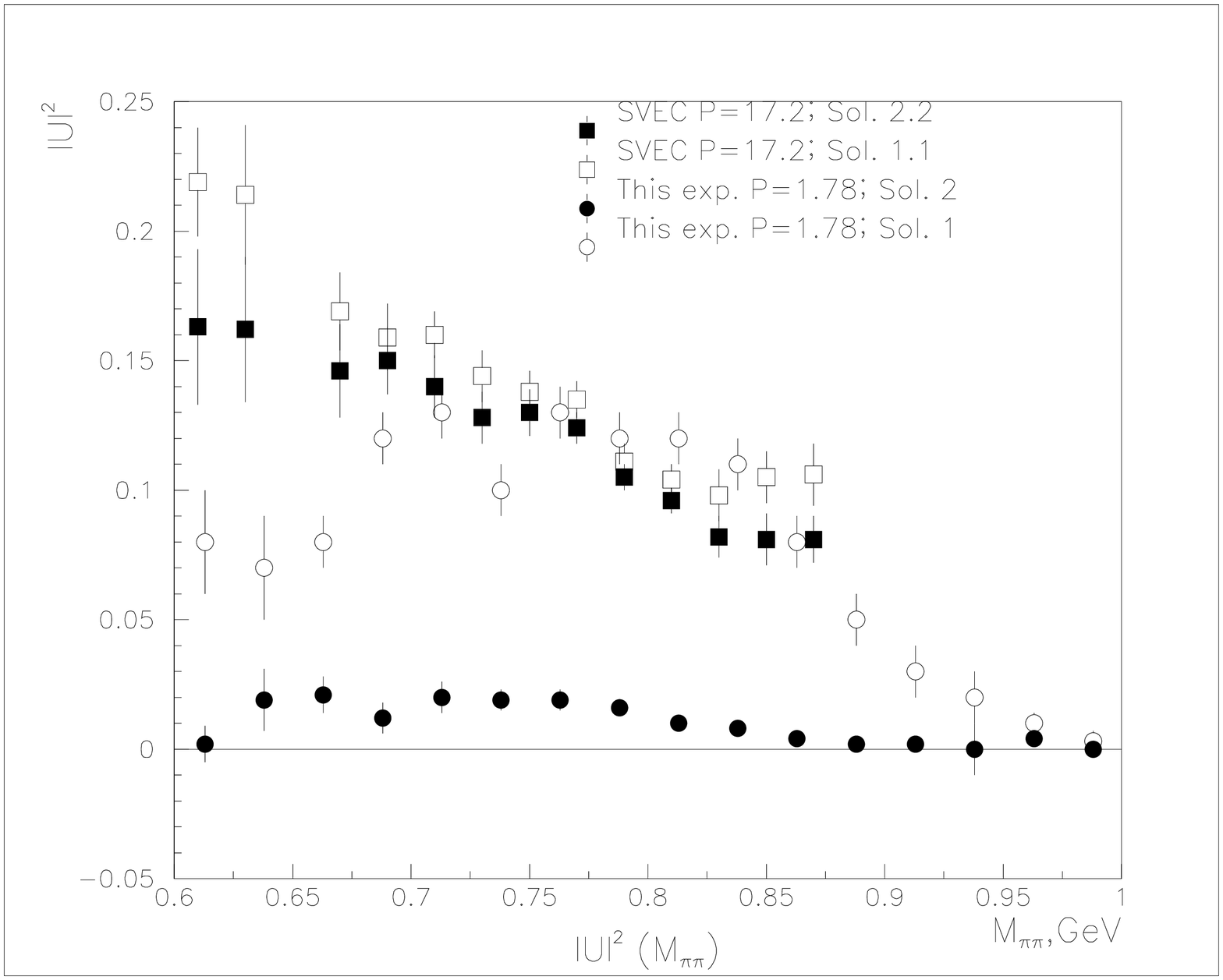,height=4cm,width=6.5cm}&
\epsfig{figure=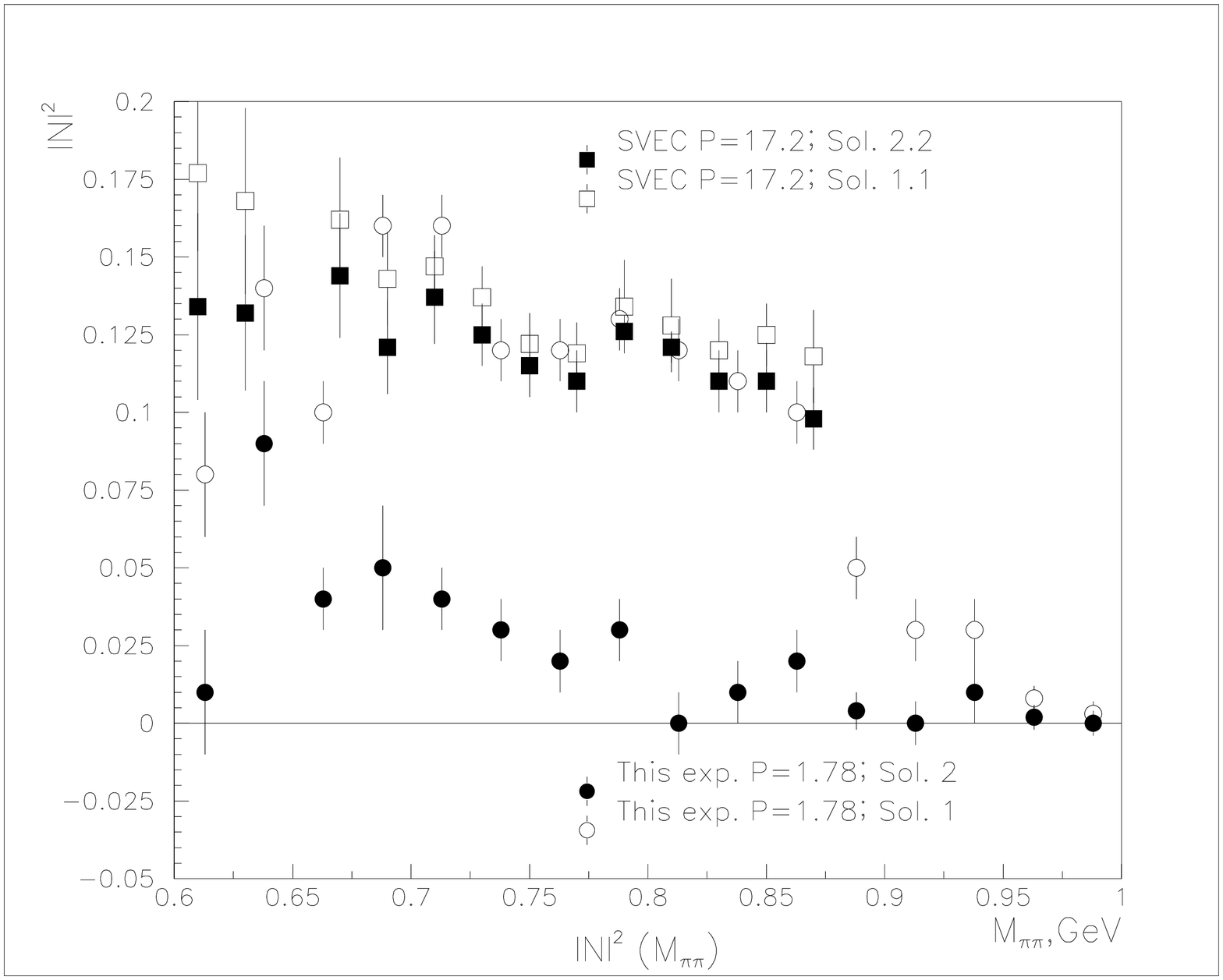,height=4cm,width=6.5cm}\\
\epsfig{figure=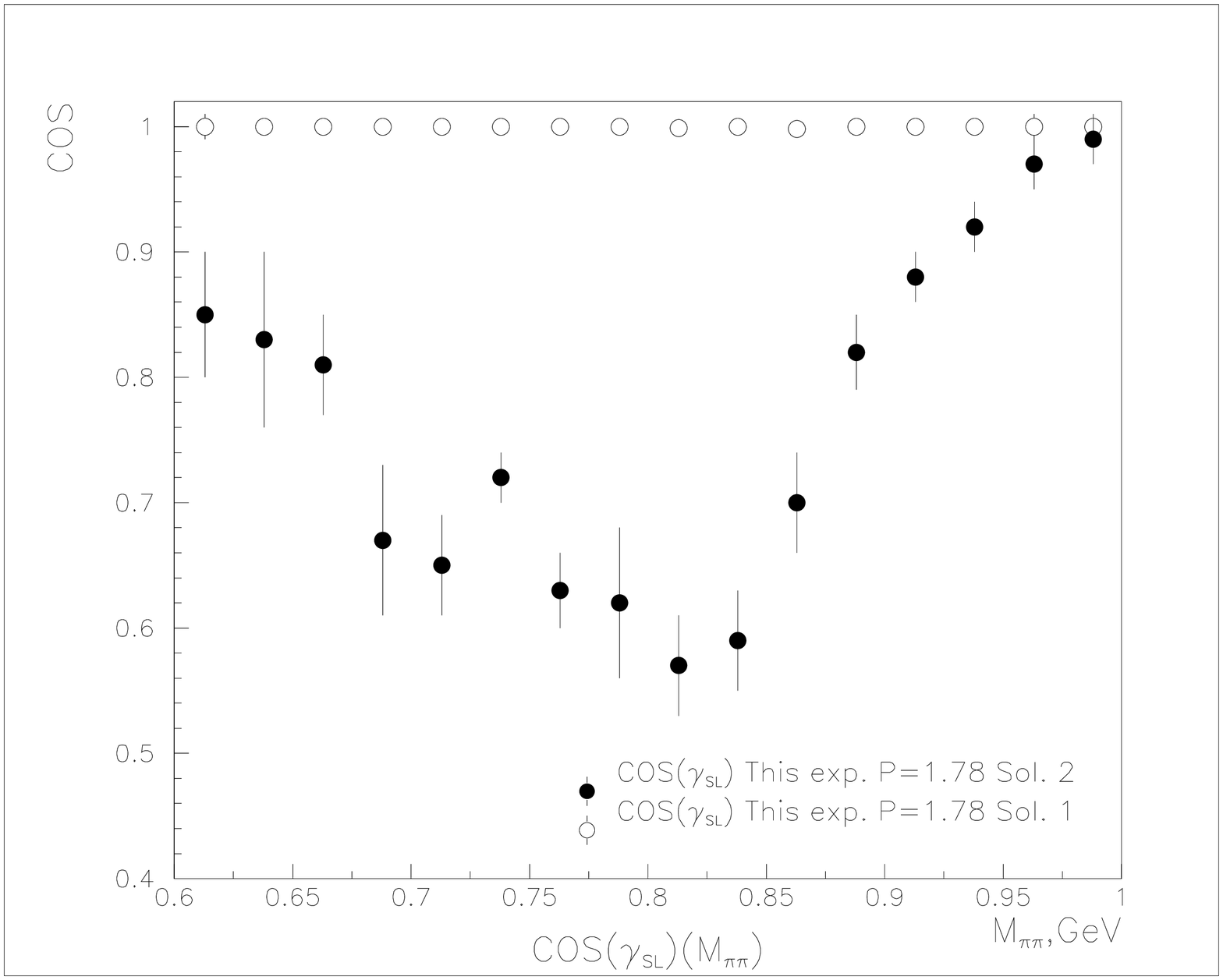,height=4cm,width=6.5cm}&
\epsfig{figure=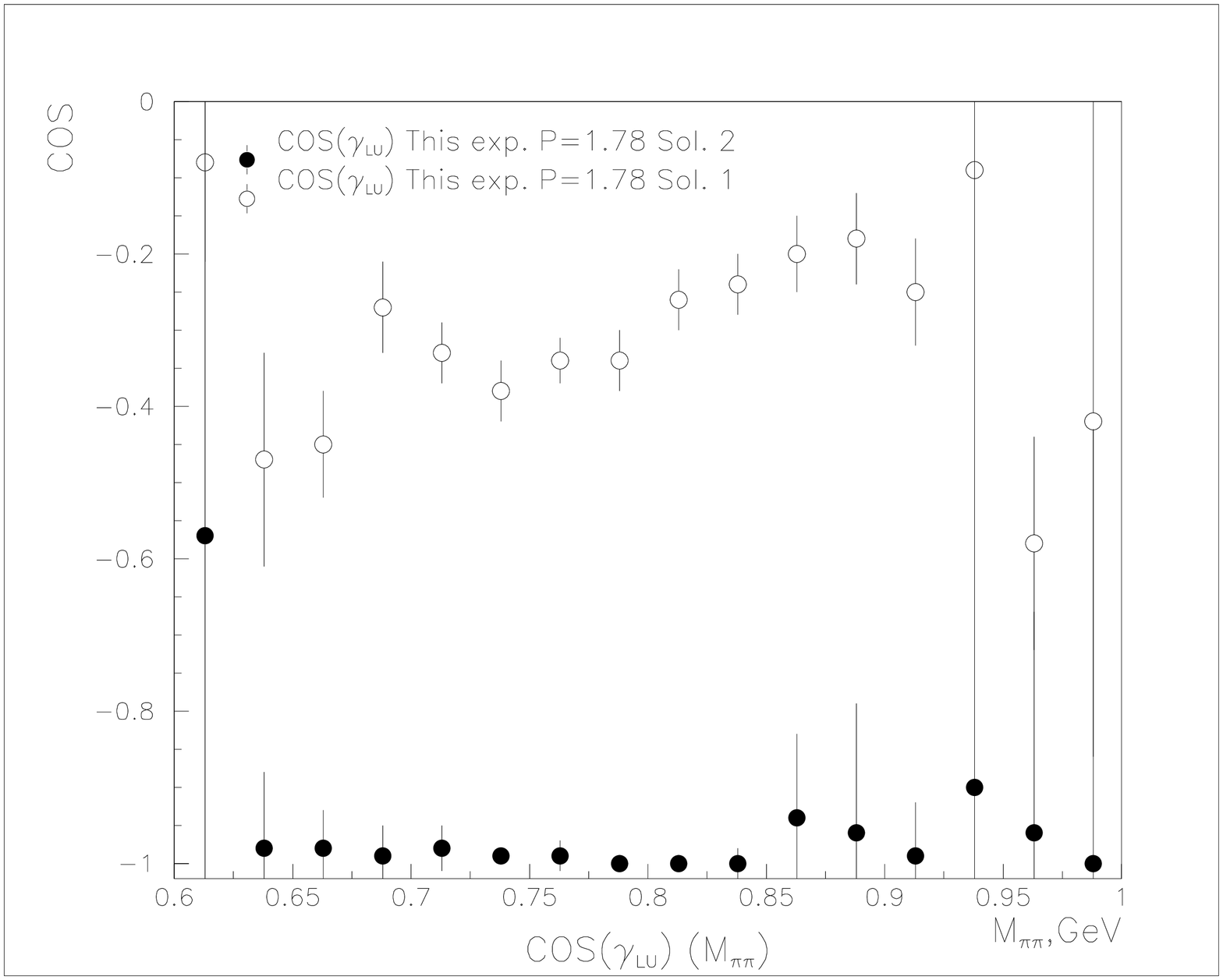,height=4cm,width=6.5cm}\\
\epsfig{figure=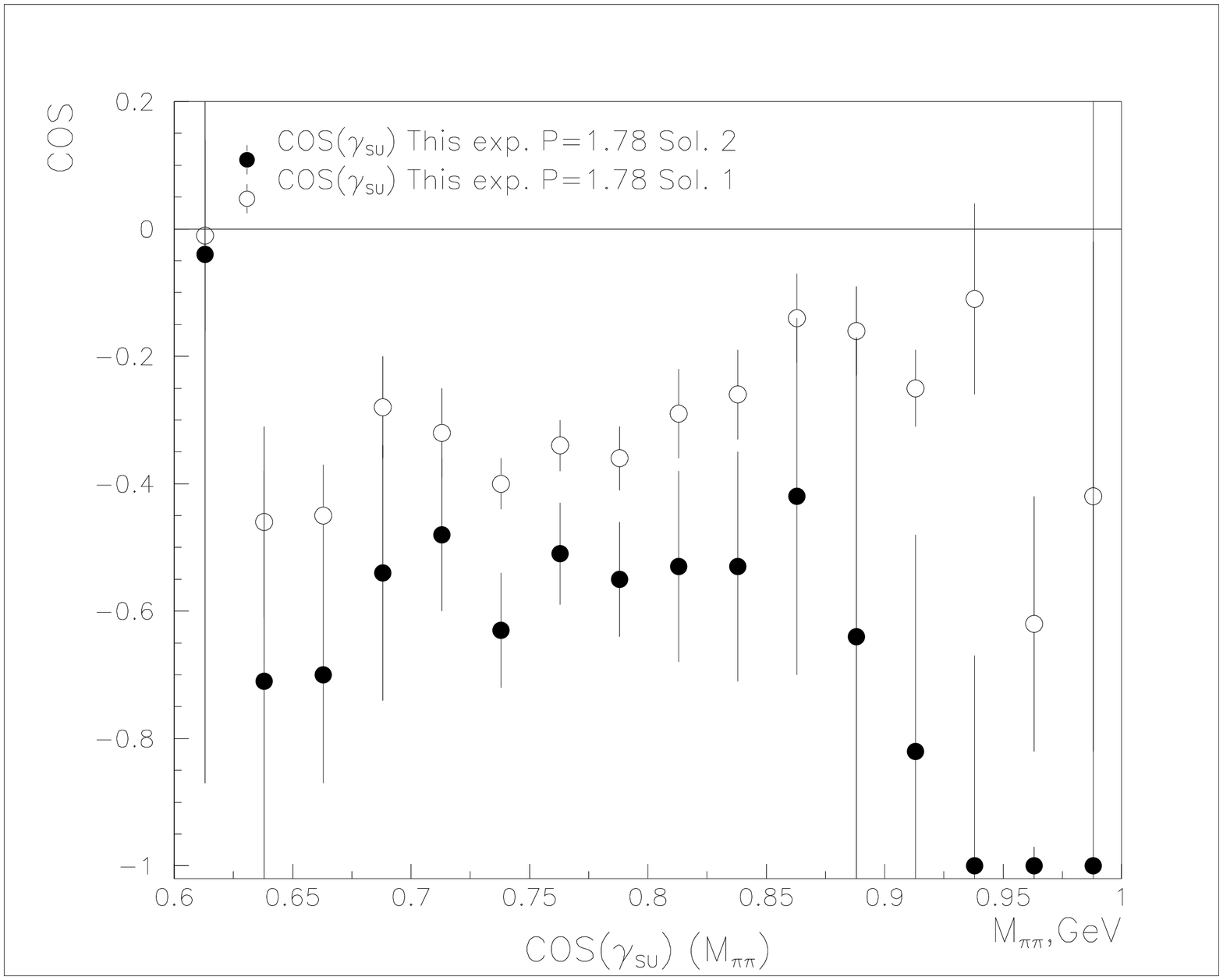,height=4cm,width=6.5cm}&\\
\end{tabular}
\caption{Results of the model-dependent amplitude analysis. Notations:
open/filled circles/squares --- solution 1/2 this work/M.~Svec 
\cite{Svec}.}
\label{fig:AMPL1}
\end{figure}

The intensity of the $S$-wave in the solution 2 is near the intensity
of $L$-wave but the intensities of $U$ and $N$-waves are abnormally
small. In this solution there is also significant energy dependence
of relative phase and intensities of $S$ and $L$-waves. The essential
energy dependence of 
the $S$-wave intensity in the solution of this type
was already noticed by the authors of the analysis \cite{Nilov},
performed on the base of the measurements at 4.0 and 4.5~GeV/c
\cite{Charlesworth,Nilov} with bubble chambers. 
It is difficult to explain all these features of the solution 2 
in the frames of the traditional ideas of the mechanism of the
peripheral pion production. 

Probably, the ambiguity could also be solved in a model-independent 
way if we complement our experiment with measurements on a 
longitudinally polarized target. Besides spin-independent SDME such
measurement provides three $\im \rho^z$ components of the 
spin density matrix:

\begin{eqnarray}
\im \rho^z_{1S} &=&\frac{1}{\sqrt{2}}\im (NS^* + \bar{N}\bar{S}^*),\\
\im \rho^z_{10} &=&\frac{1}{\sqrt{2}}\im (NL^* + \bar{N}\bar{L}^*),\\
\im \rho^z_{1-1}&=&-\im (NU^* + \bar{N}\bar{U}^*) ,
\end{eqnarray}

which do not vanish when barred and unbarred amplitudes are 
equal. The other interesting information, which could be 
obtained from these measurements is relative phases of 
$N$-waves, which are difficult to get from measurements on a 
transversally polarized target, because of large experimental 
errors.

In the solution 1 only the relative phase between $U$ and $L$ waves 
manifests significant energy dependence which should be explained by
theory.

\section{Evaluation of the parameters of $\sigma (750)$}
\begin{figure}
\begin{tabular}{ll}
\epsfig{figure=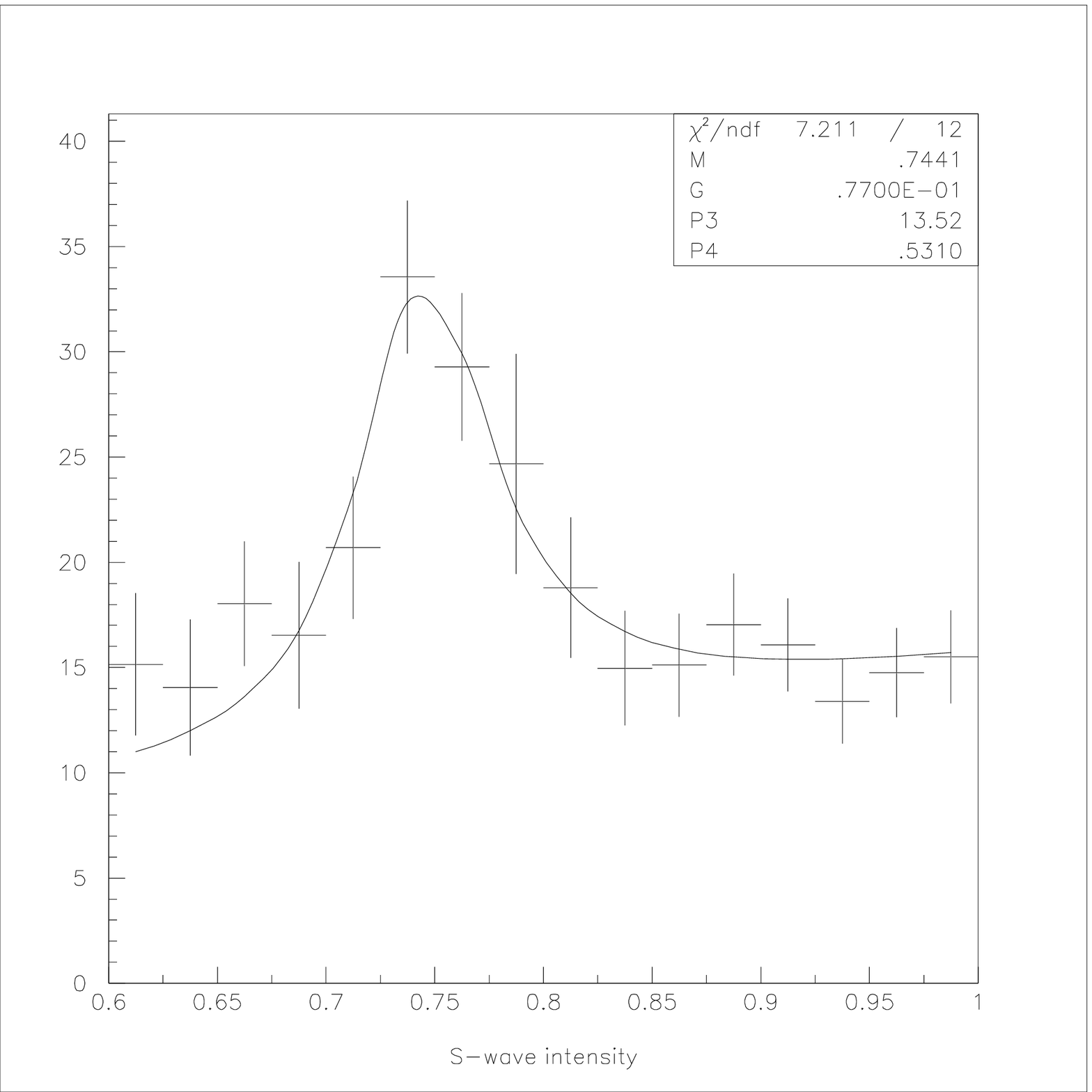,height=6.5cm}&
\epsfig{figure=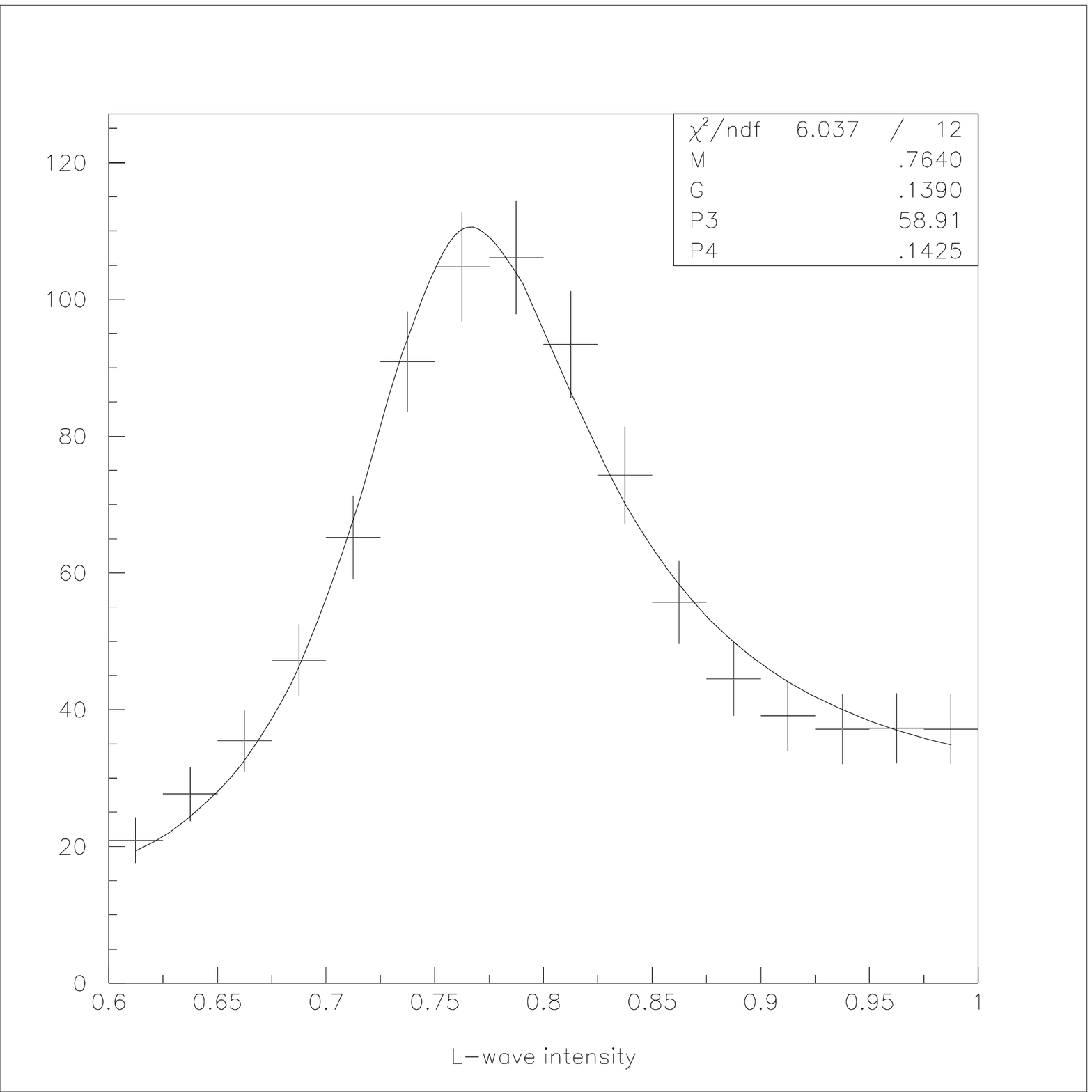,height=6.5cm}\\
a) & b) \\
\end{tabular}
\caption{Breit-Wigner fits for $S$ (a) and $L$-waves intensities.}
\label{fig:fits}
\end{figure}

Both solutions of the amplitude analysis have nearly constant relative
$S-L$-phase and do not have a dip in the ratio $I_S/I_L$. Yet, in the
$L$-wave a strong $\rho$-resonance is present. As a consequence
there should be a scalar-isoscalar\footnote{The state is symmetrical
so it could be only isospin 0 or 2. But the wave with isospin
2 is believed to be small and such a state will have an open exotic,
because a meson with isospin 2 can not be combined from quark-anti-quark
pair} resonance in $S$-wave with similar mass and width.
In order to estimate its parameters we plotted 
the unnormalised intensities
of $S$ and $L$-waves using the normalized amplitudes from our 
model-dependent analysis and the mass dependence of the cross section
obtained at 2.26~GeV/c in $4\pi$-geometry experiment \cite{Reynolds}.
The unnormalised amplitudes were fitted by relativistic Breit-Wigner
formula with constant incoherent background:
\begin{equation}
I(m)=qF(m)N(|BW(m)|^2+B) \, .
\end{equation}
Here $m=M_{\pi\pi}$ is the invariant mass of dipion, $N$ is a normalizing
constant, $B$ is the constant background, $q=\half\sqrt{m^2 - 4m_\pi^2}$ 
is the momentum of pions in the dipion rest frame,
$BW(m) = \frac{m_R \Gamma}{m_R^2 - m^2 - \mathrm{i} m_R \Gamma}$ is 
relativistic Breit-Wigner amplitude \cite{PDG}, 
$F(m)=(2J+1)(\frac{m}{q})^2$ is Pi\v{s}\'ut-Roos resonance shape
formula \cite{Pisut}. The mass dependence of the width is given by the
equation:
$\Gamma=\Gamma_R (\frac{q}{q_R})^{2J+1} \frac{D_J(q_R r)}{D_J(qr)}$.
$ D_J(qr)=\left\{ \begin{array}{ccc}
                          1     & , & J=0 \\
                     1 + (qr)^2 & , & J=1 \\
                  \end{array} \right.
$ is centrifugal barrier functions of Blatt and Weishopf \cite{Blatt}.
And at last $m_R$ and $\Gamma_R$ are the mass and width of the resonance
and $q_R$ is the $q$ at the point $m=m_R$. The fits for $S$ and $L$-wave
intensities in the solution 1 are shown in fig.~\ref{fig:fits}. The
description of $L$-wave gives $M_\rho=764 \pm 3$~MeV and 
$\Gamma_\rho = 139 \pm 14$~MeV which is in good agreement with world
data over $\rho$-meson ($M_\rho=768.5 \pm 0.6$~MeV and 
$\Gamma_\rho = 150.7 \pm 1.2$~MeV \cite{PDG}). The results of our fitting
to $S$-wave are shown in tab.~\ref{tab:sigma} together with the data
at other energies for the same solution.

\begin{table}
\caption{Parameters of $\sigma(750)$}
\label{tab:sigma}
\begin{tabular}{lcclll}
\hline
$p_{\mathrm{beam}}$ & Mass & Width & Reaction & Experiment & Analysis \\
GeV/c               & MeV  & MeV   &          &            &          \\
\hline
1.78  & $744\pm 5$  & $77\pm 22$  &$\pi^- p_\uparrow \to \pi^- \pi^+ n$&
     \multicolumn{2}{c}{this work} \\
5.98  & $746\pm 16$ & $145\pm 69$ &$\pi^+ n_\uparrow \to \pi^- \pi^+ p$&
     \cite{Lesquen} & \cite{Svec}  \\
11.85 & $782\pm 16$ & $117\pm 26$ &$\pi^+ n_\uparrow \to \pi^- \pi^+ p$&
     \cite{Lesquen} & \cite{Svec}  \\
17.2  & $771\pm 13$ & $161\pm 22$ &$\pi^- p_\uparrow \to \pi^- \pi^+ n$&
     \cite{Becker}  & \cite{Svec1} \\
\hline
\end{tabular}
\end{table}

The average data over listed 4 estimates of $\sigma(750)$ parameters is:

\begin{tabular}{ll}
$M_\sigma      = 750 \pm 4$ MeV  & $\chi^2/ndf=8.11/3$ ;\\
$\Gamma_\sigma = 119 \pm 13$ MeV & $\chi^2/ndf=7.44/3$ .\\
\end{tabular}

The available information about the $\sigma(750)$-meson is quite
contradictory. In the last edition of the {\it Review of Particle 
Physics} \cite{PDG} this meson is present as $f_0(400-1200)$ with rather
unfixed parameters (mass 400--1200~MeV and width 600--1000~MeV).
It is seen in a number of works but the situation remains rather
unclear. On one hand amplitude analysis of the experimental
data obtained on polarized targets supports narrow (70--200~MeV) 
resonance decaying into two charged pions. On the other hand
partial wave analyses of the data from unpolarized targets
resulted in a wide weak resonance and its parameters provided
by different analyses are much different. The possible reason lays
in the approximation of spin-independence used in partial wave analyses.
Particularly these analyses disregarded the contribution of 
the helicity non flip amplitudes. From the experiments performed
on polarized targets \cite{Becker,Lesquen} it is known that this
approximation is rather strongly violated especially at high 
energies, where asymmetry reaches 50\%. 
But it is high energy region where most recent and
precise measurements were made. New interesting data 
concerning the problem
of $\sigma$-meson could be obtained in the experimental study of
the reaction $\pi^- p \to \pi^0 \pi^0 n$, in which $P$-wave is 
forbidden, on a polarized target \cite{Svec2}. 
This would be a natural continuation of the
experiments performed on unpolarized targets by E852-collaboration
at 18~GeV/c \cite{E852} and by GAMS-collaboration
at 38~GeV/c \cite{GAMS}. On the other hand the study of the same
reaction at intermediate energies even on unpolarized target 
is also interesting. Small polarization effects observed in this
work at 1.78~GeV/c make believe that correct amplitude analysis
could be done without information about spin-dependent SDME.

\section{Phenomenological analysis}
In this section we want to present some
estimates of energy dependence of 
the amplitudes in the frame of Regge model. This
allows to compare our results to those at high energies. It is well
known that Regge phenomenology successfully describes energy dependence
of amplitudes. Reggitized one pion exchange was successfully used
for interpretation of peripheral pion production \cite{Nilov}.
We will use the model by Kimel and Owens \cite{Kimel}, which takes into
account the asymptotically dying amplitudes, that are essential in the 
range of this experiment, as well as $a_1$-exchange amplitudes,
which describe the spin dependence of the dynamics of the pion 
generation.
The model was parameterized by the data at 17.2~GeV/c 
\cite{Becker,Hyams}.
The authors worked with helicity $s$-channel amplitudes. We used
the crossing equations to transfer to the Jackson system used in this
work:
\begin{eqnarray}
&& S_{+\pm}=S^S_{+\pm}, \qquad N_{+\pm}=N^S_{+\pm}, \\
&& L_{+\pm}=\cos \chi \cdot L^S_{+\pm} + \sin \chi \cdot U^S_{+\pm}, \\
&& U_{+\pm}=-\sin \chi \cdot L^S_{+\pm} + \cos \chi \cdot U^S_{+\pm}, 
\end{eqnarray}
where $\chi$ is the crossing angle of the vector meson and upper script 
$^S$ denotes $s$-channel helicity amplitudes. The model parameterizes
the pion pole terms as:
\begin{eqnarray}
\label{eq:pipole1}
&& L^S_{+-}=\sqrt{-t'} \cdot \beta_\pi \frac{m_\rho}{2} 
  \e^{C^0_\pi(t-m^2_\pi)} \xi_\pi , \qquad
L^S_{++}=-rL^S_{+-}, \\
&& S^S_{+\pm}=L^S_{+\pm} \Gamma \e^{\mathrm{i}\Delta} ,\\
&& U^{SP}_{+-}=(-t') \cdot \beta_\pi \frac{m_\rho}{2} 
  \e^{C^1_\pi(t-m^2_\pi)} \xi_\pi , \\
&& U^S_{++}=-rU^{SP}_{+-} , 
\label{eq:pipole2}
\end{eqnarray}
where $\xi_\pi=\Gamma (-\alpha_\pi)(1+\e^{-\mathrm{i}\pi\alpha_\pi})
(\frac{s}{s_0})^{\alpha_\pi}$, $\alpha_\pi = 0.7(t-m^2_\pi)$,
$ r=\sqrt{\frac{t_{\mathrm{min}}}{t'}} $, $\Gamma=0.4$,
$\Delta=0.4$, $s_0=1.0$~GeV$^2$. The pion pole residues correspond
to those used in the standard pion exchange model. All the amplitudes
written above have common Regge phase independent on the energy.
The contribution of the Regge cut is described as:
\begin{equation}
U^{SC}_{+-}=N^{SC}_{+-}=\beta_C \e^{C_c t} s^{\alpha_C} 
\e^{\mathrm{i}\pi\alpha_C / 2} ,
\end{equation}
where $\alpha_C=0.0 + 0.4 t$. At low $|t|$ this provides the relative
phase of contributions to $U^S$- amplitude from cut and pole terms
near $180^o$. The constants $\beta_\pi$, $\beta_C$, $C^0_\pi$,
$C^1_\pi$ and $C_C$ are parameters of the model. 

From equations (\ref{eq:pipole1}-\ref{eq:pipole2}) one can 
expect neither the 
energy dependence in the $S$ to $L$ intensity ratio nor in the relative
phase for $s$-channel transversity amplitudes, though their
phases change with decreasing energy, because of increasing contribution
of amplitudes without helicity flip $S^S_{++}$ and $L^S_{++}$,
which contain term $r$. The amplitude $U^S_{+-}$ contains 
contributions from the pion pole and the cut with opposite signs. 
At 17.2~GeV/c it changes its sign at $t'=-0.02$~(GeV/c)$^2$. The
pole amplitude $U^{SP}_{+-}$ is decreasing with the decrease of
energy because of the growth of minimum momentum transferred.
So the point where amplitude $U^{S}_{+-}$ becomes zero moves to
$t'=-0.06$~(GeV/c)$^2$ at 1.78~GeV/c for $M_{\pi\pi}=0.77$~GeV.
From the crossing relations one can see that contributions of the 
amplitudes
$L^S_{++}$ and $U^S_{++}$, which are taken as a reference point 
for phases, to the real part 
$r(-\sin \chi \cdot L^S_{++} + \cos \chi \cdot U^S_{++})$ of 
transversity $t$-channel $U$-amplitude compensate each other, while
there is no compensation of the imaginary part of
amplitude, because of the zero
in $U^S_{+-}$. That's why $U$-amplitude is imaginary at the point
$U^S_{+-}=0$ and the phase of $L$-amplitude is determined by 
the parameter
$r$, which is a function of the initial energy. At $r=1$ and
$t'=-0.06$~(GeV/c)$^2$ the estimation of relative phase between
$U$ and $L$-amplitudes basing on the model parameters \cite{Kimel}
gives 130$^o$, which is in agreement with the amplitude analysis
solution 1 and in contradiction to the solution 2.

Spin dependence of the production process in model \cite{Kimel} is
determined by the exchange by the axial-vector meson $a_1$ $(J^P=1^+)$,
which is parameterized as contributions of the Regge pole and cut:
\begin{equation}
L^P_{++}=\beta^0_{a_1} \e^{C^0_{a_1}t} \xi_{a_1}, \qquad
L^C_{++}=- \mathrm{i} \beta^{a_1}_C \e^{C^{a_1}_C t} 
(\frac{s}{s_0})^{\alpha^{a_1}_C} 
\e^{-\mathrm{i}\pi\frac{\alpha^{a_1}_C}{2}},
\label{eq:a1exchange}
\end{equation}
where $\xi_{a_1}=\Gamma (1-\alpha_{a_1})
(1-\e^{-\mathrm{i}\pi \alpha_{a_1}}) (\frac{s}{s_0})^{\alpha^{a_1}_C}$. 
The Regge trajectories are given as: $\alpha_{a_1} = -0.3+0.9t$ and
$\alpha^C_{a_1}=-0.4+0.45t$. At small momenta transferred amplitudes
$L^P_{++}$ and $L^C_{++}$ have opposite phases and 
values for $\beta^0_{a_1}$ and $\beta^{a_1}_C$ fitted at 17.2~GeV/c
satisfy the relation
$\beta^0_{a_1} \approx - \beta^{a_1}_C$. The partial polarization of
$L$-wave is defined as:
\begin{equation}
P_L=\frac{2 \im L^S_{++} \cdot L^S_{+-}}{\sigma} ,
\end{equation}
where $\sigma$ is the sum of 
modules squared of the all $s$-channel amplitudes.
Using (\ref{eq:a1exchange}) we can write the energy dependence of the
partial polarization in this model:
\begin{equation}
P_L \sim \frac{(\frac{s}{s_0})^{\alpha_{a_1}} -
(\frac{s}{s_0})^{\alpha^{a_1}_C}}{1+r^2} , 
\end{equation}
because the part of amplitude $L^S_{++}$, determined by the
contribution of pion pole has the same phase as $L^S_{+-}$ and does
not contribute to the polarization. Then taking intercept of Regge 
cut of $a_1$ from Kimel and Owens we see that 
the decrease of partial 
polarization observed $\frac{P_L(17.2)}{P_L(1.78)}=3.0 \pm 0.7$
correspond to the intercept of Regge trajectory of $a_1$-meson
$-0.1 \pm 0.2$ and does not contradict to the one, estimated by
Kimel and Owens $-0.3 \pm 0.1$ \cite{Kimel}.

\section{Conclusions}
In the experiment presented here all 14 SDME of the reaction 
(\ref{eq:pip}) were measured for the first time in the resonance
region. This allowed us to perform both model-dependent and 
model-independent amplitude analyses and phenomenological analysis
of the data obtained. These analyses lead us to the conclusion that the
only physically justified solution of the amplitude analysis is 
solution 1, which corresponds to the solution "UP" of 
the $\pi\pi$ partial
wave analysis. The clear signature of the solution allows to state
that at high energies only solution with minimum $S$-wave intensity
under the peak of $\rho$-meson is true (solution 1.1 from M.~Svec's
analysis \cite{Svec}). We should mention that it is solution 1.1
from all four M.~Svec's solutions for the intensity, where
resonant behaviour of $S$-wave is especially clear. This could be an
additional argument in favour of the existing of a narrow $\sigma$-meson.
The mass dependence of $S$-wave intensity obtained in this work
is similar to the one observed at high energies and correspond to
existing of narrow ($\Gamma \sim 100$~MeV) S-wave
scalar-isoscalar resonance, originally proposed in \cite{Lesquen}.

The constituent structure of $\sigma(750)$ is still an open question.
It is very doubtful that this is a quarkonium state. Hybrid 
quark-gluonium or pure gluonium nature of the $\sigma$-meson looks
much more probable.
M.~Svec, using results of works \cite{Ellis,Shifman}, which connected
mass and width of gluonium in the frames of low-energy theorems of
broken chiral symmetry, proposed that this state could be a 
low-energy gluonium. These theorems predict width about 100~MeV for
a gluonium with mass 750~MeV, which does not contradict to the 
experiment.
On the other hand quantum chromodynamic calculations on lattices
\cite{Bali,Butler} set the lowest limits for scalar gluonium mass as
$1550 \pm 50$ and $1740 \pm 70$~MeV, correspondingly. But earlier
works gave the value $740 \pm 40$~MeV for the mass of the basic state 
of gluonium \cite{Ishikawa}. Here we also want to mention
an interpretation of light scalar resonances as ``new hadrons''
or ``vacuum scalars'' with small width of the decay into two pions.
This model beyond standard QCD was recently proposed by V.N.~Gribov 
et al. \cite{Gribov}.

The experimental data show the spin dependence of the dipion production
dynamics at the level of 6 standard deviations at intermediate energies.
The sign of the asymmetry coincides with the one at high energies but the
value is (3--4) times smaller. The largest effects are observed in
$\rho$-meson production. Our analysis of the energy dependence of
the polarization in the frame of Regge model by Kimel and Owens
resulted in a value of $a_1$ Regge trajectory intercept 
$-0.1 \pm 0.2$.

\section*{Acknowledges}
We are grateful to the ITEP accelerator division for the pion
beam production
allowing us to conduct these measurements. Authors want
to thank V.V.~Vladimirskii, A.B.~Kaidalov and Yu.S.~Kalashnikova
for the fruitful discussions of our results. We want to thank
M.~Svec and R.~Kaminski for their comments on their works on
scalar mesons and amplitude analysis.

The work was partially supported by the Russian Fund for Basic 
Research and Russian State Program "Fundamental Nuclear Physics".

\appendix

\begin{table}
\caption{Spin-independent SDME as function of the dipion mass 
integrated over momentum transferred in the region
$0.005 < t_{\mathrm{min}} - t < 0.2$~(GeV/c)$^2$.}
\begin{tabular}{cccccc}
\hline
$M_{\pi\pi}$,& $\rho_{00}+\frac{1}{3}\rho_{SS}$ &
$\rho_{1-1}$ & $\re \rho_{10}$ & $\re \rho_{0S}$ & $\re \rho_{1S}$ \\
MeV &&&&&\\
\hline
613 & $.604\pm .016$ & $.002\pm .010 $ & $-.011\pm .017$ &
$.413\pm .024$ & $.001\pm .018$\\
638 & $.611\pm .015$ & $.034\pm .011 $ & $-.065\pm .017$ &
$.372\pm .028$ & $.045\pm .014$\\
663 & $.628\pm .010$ & $.009\pm .006 $ & $-.069\pm .010$ &
$.383\pm .017$ & $.049\pm .008$\\
688 & $.593\pm .017$ & $.018\pm .008 $ & $-.049\pm .011$ &
$.314\pm .025$ & $.030\pm .008$\\
713 & $.591\pm .013$ & $.011\pm .006 $ & $-.064\pm .008$ &
$.301\pm .018$ & $.035\pm .007$\\
738 & $.634\pm .008$ & $.007\pm .004 $ & $-.066\pm .006$ &
$.343\pm .011$ & $.042\pm .004$\\
763 & $.633\pm .009$ & $-.002\pm .004$ & $-.066\pm .006$ &
$.306\pm .013$ & $.035\pm .004$\\
788 & $.657\pm .008$ & $.007\pm .004 $ & $-.064\pm .006$ &
$.294\pm .027$ & $.033\pm .004$\\
813 & $.673\pm .011$ & $-.004\pm .004$ & $-.051\pm .007$ &
$.283\pm .021$ & $.026\pm .006$\\
838 & $.689\pm .013$ & $.000\pm .004 $ & $-.046\pm .008$ &
$.290\pm .021$ & $.022\pm .006$\\
863 & $.707\pm .013$ & $.007\pm .004 $ & $-.032\pm .008$ &
$.337\pm .018$ & $.012\pm .006$\\
888 & $.727\pm .010$ & $.001\pm .003 $ & $-.024\pm .008$ &
$.399\pm .013$ & $.013\pm .006$\\
913 & $.754\pm .008$ & $-.002\pm .003$ & $-.027\pm .007$ &
$.425\pm .008$ & $.017\pm .004$\\
938 & $.785\pm .009$ & $.005\pm .003 $ & $-.007\pm .011$ &
$.421\pm .011$ & $.005\pm .007$\\
963 & $.797\pm .007$ & $-.001\pm .002$ & $-.034\pm .007$ &
$.442\pm .006$ & $.023\pm .004$\\
988 & $.799\pm .009$ & $.000\pm .002 $ & $-.014\pm .010$ &
$.453\pm .007$ & $.009\pm .006$\\
\hline
\end{tabular}
\end{table}

\newsavebox{\tableone}
\begin{table}
\savebox{\tableone}[\textheight][l]{\vbox{
\hsize=\textheight
\caption{Spin-independent SDME as function of the momentum transferred
$-t'=t - t_{\mathrm{min}}$ integrated over dipion mass in the region
$700 < M_{\pi\pi} < 850$~MeV.}
\begin{tabular}{cccccc}
\hline
$-t'$, (GeV/c)$^2$ & $\rho_{00}+\frac{1}{3}\rho_{SS}$ &
$\rho_{1-1}$ & $\re \rho_{10}$ & $\re \rho_{0S}$ & $\re \rho_{1S}$ \\
\hline
.002 & $.600\pm .025$ & $.016\pm .008 $ & $ -.035\pm .016$ &
$ .220\pm .035$ & $ .022\pm .018$\\
.007 & $.620\pm .025$ & $-.004\pm .007$ & $ -.045\pm .010$ &
$ .250\pm .025$ & $ .027\pm .014$\\
.012 & $.600\pm .025$ & $-.009\pm .007$ & $ -.075\pm .010$ &
$ .260\pm .035$ & $ .044\pm .008$\\
.017 & $.620\pm .025$ & $-.012\pm .010$ & $ -.085\pm .016$ &
$ .190\pm .040$ & $ .040\pm .008$\\
.025 & $.610\pm .025$ & $.000\pm .007 $ & $ -.069\pm .010$ &
$ .245\pm .025$ & $ .043\pm .007$\\
.035 & $.620\pm .025$ & $.003\pm .007 $ & $ -.069\pm .010$ &
$ .250\pm .025$ & $ .035\pm .004$\\
.045 & $.640\pm .020$ & $.027\pm .007 $ & $ -.070\pm .010$ &
$ .265\pm .025$ & $ .037\pm .004$\\
.055 & $.615\pm .020$ & $.027\pm .010 $ & $ -.071\pm .012$ &
$ .230\pm .035$ & $ .032\pm .004$\\
.070 & $.630\pm .020$ & $.041\pm .007 $ & $ -.078\pm .008$ &
$ .295\pm .025$ & $ .033\pm .006$\\
.090 & $.625\pm .020$ & $.047\pm .010 $ & $ -.095\pm .010$ &
$ .280\pm .025$ & $ .037\pm .006$\\
\hline
\end{tabular}
\\
%\end{table}
%\begin{table}
\caption{
Spin-dependent SDME integrated over momentum transferred
in the region $0.005 < t_{\mathrm{min}} - t < 0.2$~(GeV/c)$^2$.}
\begin{tabular}{cccccccccc}
\hline
$M_{\pi\pi}$, & $A=\rho^Y_{SS}+$ &
$\rho^Y_{00}-\rho^Y_{11}$ & $\rho^Y_{1-1}$ & $\re \rho^Y_{10}$ &
$\re \rho^Y_{0S} $ & $\re \rho^Y_{1S}$ & $\im \rho^X_{0S}$ &
$\im \rho^X_{10}$ & $\im \rho^X_{1S}$\\
GeV & $\rho^Y_{00}+2\rho^Y_{11}$ &&&&&&&&\\
\hline
.65--.75 & $-.07\pm .16$ & $-.09\pm .08$ & $ .008\pm .025$ & 
$.007 \pm .016$ & $.00 \pm .07$ & $-.022\pm .016$ & 
$.018 \pm .025$ & $.025\pm .020 $ & $.022 \pm .016$ \\
.75--.80 & $-.11\pm .11$ & $-.03\pm .07$ & $-.034\pm .028$ & 
$.031 \pm .011$ & $-.02\pm .06$ & $.000 \pm .011$ & 
$-.017\pm .028$ & $.059\pm .016 $ & $.028 \pm .011$ \\
.80--.90 & $-.26\pm .18$ & $-.28\pm .13$ & $ .00 \pm .03$ & 
$.073 \pm .018$ & $-.12\pm .08$ & $.034 \pm .013$ & 
$.04 \pm .03$ & $.034\pm .018 $ & $.017 \pm .013$    \\
\hline
\end{tabular}}}
\rotl\tableone
\end{table}

\newsavebox{\tabletwo}
\begin{table}
\savebox{\tabletwo}[\textheight][l]{\vbox{
\hsize=\textheight
\caption{Model-independent analysis. $t$-channel transversity
amplitudes as function of the dipion mass. Solution 1.}
\begin{tabular}{cccccccc}
\hline
$M_{\pi\pi}$, MeV & $|S|^2$ & $|L|^2$ & $|\bar{N}|^2$ & $|U|^2$ &
$\cos\gamma_{LS}$ & $\cos\gamma_{LU}$ & $\cos\gamma_{SU}$ \\
\hline
663 & $.18\pm .09$ & $.21\pm .06$ & $.05\pm .05$ &
$.03\pm .05$ & $.98\pm .06$ & $-.56\pm .57$ & $ -.71\pm .54$ \\
688 & $.12\pm .07$ & $.21\pm .05$ & $.08\pm .04$ &
$.05\pm .04$ & $.98\pm .05$ & $-.28\pm .18$ & $ -.47\pm .21$ \\
713 & $.11\pm .06$ & $.22\pm .05$ & $.08\pm .04$ &
$.06\pm .04$ & $.99\pm .04$ & $-.35\pm .18$ & $ -.50\pm .20$ \\
738 & $.13\pm .07$ & $.23\pm .05$ & $.06\pm .04$ &
$.04\pm .04$ & $.98\pm .05$ & $-.42\pm .26$ & $ -.60\pm .27$ \\
763 & $.08\pm .04$ & $.26\pm .04$ & $.03\pm .03$ &
$.07\pm .03$ & $.99\pm .02$ & $-.18\pm .08$ & $ -.33\pm .13$ \\
788 & $.07\pm .04$ & $.28\pm .03$ & $.04\pm .03$ &
$.06\pm .03$ & $.98\pm .03$ & $-.18\pm .08$ & $ -.35\pm .15$ \\
813 & $.04\pm .04$ & $.19\pm .06$ & $.07\pm .04$ &
$.08\pm .04$ & $1.00\pm .005$ & $.13\pm .12$ & $ .11\pm .21 $ \\
838 & $.04\pm .04$ & $.20\pm .06$ & $.07\pm .04$ &
$.07\pm .04$ & $1.00\pm .001$ & $.16\pm .13$ & $ .17\pm .22 $ \\
863 & $.06\pm .05$ & $.20\pm .06$ & $.06\pm .04$ &
$.05\pm .04$ & $1.00\pm .002$ & $.28\pm .19$ & $ .28\pm .22 $ \\
888 & $.10\pm .10$ & $.19\pm .07$ & $.04\pm .05$ &
$.04\pm .05$ & $.98\pm .08 $ & $.42\pm .40$ & $ .24\pm .21 $ \\
\hline
\end{tabular}}}
\rotl\tabletwo
\end{table}

\newsavebox{\tabletwoi}
\begin{table}
\savebox{\tabletwoi}[\textheight][l]{\vbox{
\hsize=\textheight
\caption{Model-independent analysis. $t$-channel transversity
amplitudes as function of the dipion mass. Solution 2.}
\begin{tabular}{cccccccc}
\hline
$M_{\pi\pi}$, MeV & $|S|^2$ & $|L|^2$ & $|\bar{N}|^2$ & $|U|^2$ &
$\cos\gamma_{LS}$ & $\cos\gamma_{LU}$ & $\cos\gamma_{SU}$ \\
\hline
663 & $.23\pm .11$ & $.20\pm .04$ & $.03\pm .03$ &
$.011\pm .006$ & $.90\pm .24$ & $-.94\pm .36$ & $ -.99\pm .14$ \\
688 & $.26\pm .10$ & $.17\pm .04$ & $.03\pm .03$ &
$.006\pm .004$ & $.75\pm .22$ & $-.94\pm .30$ & $ -.94\pm .31$ \\
713 & $.26\pm .10$ & $.17\pm .04$ & $.03\pm .03$ &
$.010\pm .006$ & $.72\pm .21$ & $-.99\pm .07$ & $ -.80\pm .34$ \\
738 & $.23\pm .10$ & $.20\pm .04$ & $.02\pm .03$ &
$.010\pm .004$ & $.80\pm .22$ & $-.95\pm .22$ & $ -.95\pm .25$ \\
763 & $.28\pm .08$ & $.19\pm .04$ & $.03\pm .03$ &
$.003\pm .002$ & $.61\pm .15$ & $-.96\pm .14$ & $ -.80\pm .31$ \\
788 & $.25\pm .08$ & $.22\pm .04$ & $.00\pm .03$ &
$.003\pm .002$ & $.59\pm .16$ & $-.92\pm .22$ & $ -.86\pm .29$ \\
813 & $.25\pm .12$ & $.12\pm .07$ & $.00\pm .03$ &
$.002\pm .004$ & $.48\pm .27$ & $ .97\pm .16$ & $  .24\pm .51$ \\
838 & $.23\pm .12$ & $.13\pm .07$ & $.00\pm .03$ &
$.003\pm .005$ & $.49\pm .26$ & $ .98\pm .11$ & $  .33\pm .50$ \\
863 & $.20\pm .12$ & $.15\pm .06$ & $.01\pm .03$ &
$.006\pm .007$ & $.62\pm .28$ & $ .98\pm .13$ & $  .45\pm .44$ \\
888 & $.18\pm .13$ & $.17\pm .06$ & $.01\pm .03$ &
$.010\pm .020$ & $.80\pm .35$ & $ .84\pm .44$ & $  .34\pm .38$ \\
\hline
\end{tabular}}}
\rotl\tabletwoi
\end{table}

\newsavebox{\tablethree}
\begin{table}
\savebox{\tablethree}[\textheight][l]{\vbox{
\hsize=\textheight
\caption{Model-independent analysis. $t$-channel transversity
amplitudes as function of the dipion mass. Solution 1.}
\begin{tabular}{cccccccc}
\hline
$M_{\pi\pi}$, MeV & $|\bar{S}|^2$ & $|\bar{L}|^2$ & $|N|^2$ &
$|\bar{U}|^2$ & $\cos\bar{\gamma}_{LS}$ &
$\cos\bar{\gamma}_{LU}$ & $\cos\bar{\gamma}_{SU}$ \\
\hline
663 & $.12\pm .06$ & $.32\pm .05$ & $.05\pm .04$ &
$.05\pm .04$ & $.98\pm .06 $ & $-.43\pm .24$ & $  -.25\pm .19 $ \\
688 & $.08\pm .05$ & $.31\pm .05$ & $.08\pm .04$ &
$.07\pm .04$ & $.98\pm .05 $ & $-.28\pm .13$ & $  -.08\pm .17 $ \\
713 & $.08\pm .04$ & $.31\pm .04$ & $.08\pm .04$ &
$.07\pm .04$ & $.98\pm .05 $ & $-.34\pm .13$ & $  -.12\pm .17 $ \\
738 & $.08\pm .05$ & $.33\pm .05$ & $.06\pm .04$ &
$.06\pm .04$ & $.98\pm .05 $ & $-.38\pm .17$ & $  -.19\pm .17 $ \\
763 & $.09\pm .04$ & $.32\pm .04$ & $.09\pm .03$ &
$.06\pm .03$ & $.99\pm .04 $ & $-.50\pm .15$ & $  -.34\pm .15 $ \\
788 & $.08\pm .04$ & $.33\pm .04$ & $.09\pm .03$ &
$.05\pm .03$ & $.99\pm .04 $ & $-.51\pm .17$ & $  -.37\pm .17 $ \\
813 & $.09\pm .04$ & $.44\pm .06$ & $.05\pm .04$ &
$.05\pm .04$ & $.999\pm .013$ & $-.59\pm .27$ & $  -.63\pm .30$ \\
838 & $.09\pm .04$ & $.45\pm .06$ & $.04\pm .04$ &
$.04\pm .04$ & $1.000\pm .008$ & $-.60\pm .32$ & $  -.62\pm .33$ \\
863 & $.12\pm .06$ & $.45\pm .06$ & $.03\pm .05$ &
$.03\pm .05$ & $.99\pm .06 $ & $-.67\pm .60$ & $  -.57\pm .46 $ \\
\hline
\end{tabular}}}
\rotl\tablethree
\end{table}

\newsavebox{\tablethreei}
\begin{table}
\savebox{\tablethreei}[\textheight][l]{\vbox{
\hsize=\textheight
\caption{Model-independent analysis. $t$-channel transversity
amplitudes as function of the dipion mass. Solution 2.}
\begin{tabular}{cccccccc}
\hline
$M_{\pi\pi}$, MeV & $|\bar{S}|^2$ & $|\bar{L}|^2$ & $|N|^2$ &
$|\bar{U}|^2$ & $\cos\bar{\gamma}_{LS}$ &
$\cos\bar{\gamma}_{LU}$ & $\cos\bar{\gamma}_{SU}$ \\
\hline
663 & $.23\pm .11$ & $.28\pm .04$ & $.01\pm .03$ &
$.014\pm .011$ & $.76\pm .22$ & $ -.87\pm .24$ & $  -.34\pm .30$ \\
688 & $.26\pm .10$ & $.25\pm .04$ & $.02\pm .03$ &
$.009\pm .008$ & $.62\pm .18$ & $ -.86\pm .20$ & $  -.12\pm .30$ \\
713 & $.25\pm .10$ & $.25\pm .04$ & $.02\pm .03$ &
$.013\pm .008$ & $.59\pm .18$ & $ -.89\pm .15$ & $  -.16\pm .24$ \\
738 & $.23\pm .10$ & $.28\pm .04$ & $.01\pm .03$ &
$.012\pm .008$ & $.67\pm .20$ & $ -.89\pm .18$ & $  -.27\pm .27$ \\
763 & $.21\pm .08$ & $.27\pm .04$ & $.05\pm .03$ &
$.019\pm .007$ & $.68\pm .17$ & $ -.94\pm .10$ & $  -.39\pm .17$ \\
788 & $.19\pm .08$ & $.30\pm .04$ & $.06\pm .03$ &
$.017\pm .006$ & $.67\pm .19$ & $ -.95\pm .10$ & $  -.42\pm .20$ \\
813 & $.19\pm .12$ & $.41\pm .06$ & $.01\pm .03$ &
$.019\pm .006$ & $.73\pm .25$ & $ -1.00\pm .005$ & $ -.72\pm .32$ \\
838 & $.17\pm .12$ & $.42\pm .06$ & $.02\pm .03$ &
$.017\pm .006$ & $.76\pm .27$ & $ -1.00\pm .01$ & $  -.74\pm .35$ \\
863 & $.16\pm .12$ & $.44\pm .07$ & $.02\pm .03$ &
$.014\pm .009$ & $.87\pm .36$ & $ -.96\pm .21$ & $  -.70\pm .41$ \\
\hline
\end{tabular}}}
\rotl\tablethreei
\end{table}

\newsavebox{\tablefour}
\begin{table}
\savebox{\tablefour}[\textheight][l]{\vbox{
\hsize=\textheight
\caption{Model-dependent analysis. $t$-channel transversity
amplitudes as function of the dipion mass. Solution 1.}
\begin{tabular}{cccccccc}
\hline
$M_{\pi\pi}$, MeV & $I_S$ & $I_L$ & $I_N$ & $I_U$ &
$\cos\gamma_{LS}$ & $\cos\gamma_{LU}$ & $\cos\gamma_{SU}$ \\
\hline
613 & $.35\pm .06$ & $.48\pm .02$ & $.08\pm .02$ &
$.08\pm .02$ & $  1.00\pm .01  $ & $-.08\pm .13$ & $-.01\pm .15$ \\
638 & $.26\pm .05$ & $.52\pm .02$ & $.14\pm .02$ &
$.07\pm .02$ & $  1.000\pm .003$ & $-.47\pm .14$ & $-.46\pm .15$ \\
663 & $.27\pm .03$ & $.54\pm .01$ & $.10\pm .01$ &
$.08\pm .01$ & $  1.000\pm .001$ & $-.45\pm .07$ & $-.45\pm .08$ \\
688 & $.18\pm .03$ & $.53\pm .02$ & $.16\pm .01$ &
$.12\pm .01$ & $  1.000\pm .001$ & $-.27\pm .06$ & $-.28\pm .08$ \\
713 & $.17\pm .02$ & $.53\pm .01$ & $.16\pm .01$ &
$.13\pm .01$ & $  1.000\pm .001$ & $-.33\pm .04$ & $-.32\pm .07$ \\
738 & $.20\pm .02$ & $.56\pm .01$ & $.12\pm .01$ &
$.10\pm .01$ & $  1.000\pm .001$ & $-.38\pm .04$ & $-.40\pm .04$ \\
763 & $.16\pm .02$ & $.58\pm .01$ & $.12\pm .01$ &
$.13\pm .01$ & $  1.000\pm .001$ & $-.34\pm .03$ & $-.34\pm .04$ \\
788 & $.14\pm .02$ & $.61\pm .01$ & $.13\pm .01$ &
$.12\pm .01$ & $  1.000\pm .002$ & $-.34\pm .04$ & $-.36\pm .05$ \\
813 & $.12\pm .02$ & $.63\pm .01$ & $.12\pm .01$ &
$.12\pm .01$ & $  .999\pm .003$ & $-.26\pm .04$ & $-.29\pm .07$ \\
838 & $.13\pm .02$ & $.65\pm .01$ & $.11\pm .01$ &
$.11\pm .01$ & $  1.000\pm .001$ & $-.24\pm .04$ & $-.26\pm .07$ \\
863 & $.18\pm .02$ & $.65\pm .01$ & $.10\pm .01$ &
$.08\pm .01$ & $  .998\pm .005$ & $-.20\pm .05$ & $-.14\pm .07$ \\
888 & $.25\pm .02$ & $.65\pm .01$ & $.05\pm .01$ &
$.05\pm .01$ & $  1.000\pm .002$ & $-.18\pm .06$ & $-.16\pm .07$ \\
913 & $.27\pm .01$ & $.66\pm .01$ & $.03\pm .01$ &
$.03\pm .01$ & $  1.000\pm .001$ & $-.25\pm .07$ & $-.25\pm .06$ \\
938 & $.25\pm .02$ & $.70\pm .01$ & $.03\pm .01$ &
$.02\pm .01$ & $  1.000\pm .003$ & $-.09\pm .14$ & $-.11\pm .15$ \\
963 & $.28\pm .01$ & $.70\pm .01$ & $.008\pm .004$ &
$.010\pm .004$ & $1.00\pm .01	$ & $-.58\pm .14$ & $-.62\pm .20$ \\
988 & $.29\pm .01$ & $.70\pm .01$ & $.003\pm .004$ &
$.003\pm .004$ & $1.000\pm .002$ & $-.42\pm .44$ & $-.42\pm .40$ \\
\hline
\end{tabular}}}
\rotl\tablefour
\end{table}

\newsavebox{\tablefive}
\begin{table}
\savebox{\tablefive}[\textheight][l]{\vbox{
\hsize=\textheight
\caption{Model-dependent analysis. $t$-channel transversity
amplitudes as function of the dipion mass. Solution 2.}
\begin{tabular}{cccccccc}
\hline
$M_{\pi\pi}$, MeV& $I_S$ & $I_L$ & $I_N$ & $I_U$ &
$\cos\gamma_{LS}$ & $\cos\gamma_{LU}$ & $\cos\gamma_{SU}$ \\
\hline
613 & $.58\pm .04$ & $.41\pm .02$ & $.01\pm .02$ &
$.002\pm .007$ & $ .85\pm .05$ & $-.57\pm .70$ & $-.04\pm .83$ \\
638 & $.42\pm .05$ & $.47\pm .02$ & $.09\pm .02$ &
$.019\pm .012$ & $ .83\pm .07$ & $-.98\pm .10$ & $-.71\pm .33$ \\
663 & $.47\pm .03$ & $.47\pm .01$ & $.04\pm .01$ &
$.021\pm .007$ & $ .81\pm .04$ & $-.98\pm .05$ & $-.70\pm .17$ \\
688 & $.52\pm .03$ & $.42\pm .02$ & $.05\pm .02$ &
$.012\pm .006$ & $ .67\pm .06$ & $-.99\pm .04$ & $-.54\pm .20$ \\
713 & $.52\pm .03$ & $.42\pm .02$ & $.04\pm .01$ &
$.020\pm .006$ & $ .65\pm .04$ & $-.98\pm .03$ & $-.48\pm .12$ \\
738 & $.47\pm .02$ & $.48\pm .01$ & $.03\pm .01$ &
$.019\pm .004$ & $ .72\pm .02$ & $-.99\pm .01$ & $-.63\pm .09$ \\
763 & $.50\pm .02$ & $.47\pm .01$ & $.02\pm .01$ &
$.019\pm .004$ & $ .63\pm .03$ & $-.99\pm .02$ & $-.51\pm .08$ \\
788 & $.44\pm .02$ & $.51\pm .01$ & $.03\pm .01$ &
$.016\pm .003$ & $ .62\pm .06$ & $-1.00\pm .01$ & $-.55\pm .09$ \\
813 & $.47\pm .02$ & $.52\pm .01$ & $.00\pm .01$ &
$.010\pm .003$ & $ .57\pm .04$ & $-1.00\pm .01$ & $-.53\pm .15$ \\
838 & $.44\pm .02$ & $.54\pm .01$ & $.01\pm .01$ &
$.008\pm .003$ & $ .59\pm .04$ & $-1.00\pm .02$ & $-.53\pm .18$ \\
863 & $.41\pm .02$ & $.57\pm .01$ & $.02\pm .01$ &
$.004\pm .003$ & $ .70\pm .04$ & $-.94\pm .11$ & $-.42\pm .28$ \\
888 & $.40\pm .02$ & $.59\pm .01$ & $.004\pm .006$ &
$.002\pm .002$ & $.82\pm .03$ & $-.96\pm .17$ & $-.64\pm .47$ \\
913 & $.37\pm .01$ & $.63\pm .01$ & $.000\pm .007$ &
$.002\pm .001$ & $.88\pm .02$ & $-.99\pm .07$ & $-.82\pm .34$ \\
938 & $.31\pm .02$ & $.68\pm .01$ & $.01\pm .01$ &
$.00\pm .01$ & $  .92\pm .02$ & $-.91\pm .8 $ & $-1.00\pm .33$ \\
963 & $.30\pm .01$ & $.70\pm .01$ & $.002\pm .004$ &
$.004\pm .001$ & $.97\pm .02$ & $-.96\pm .29$ & $-1.00\pm .03$ \\
988 & $.30\pm .01$ & $.70\pm .01$ & $.000\pm .004$ &
$.000\pm .001$ & $.99\pm .02$ & $-1.00\pm .57$ & $-1.00\pm 1.4$ \\
\hline
\end{tabular}}}
\rotl\tablefive
\end{table}

\end{document}